\documentclass[runningheads]{svmult}

\usepackage{makeidx}   
\usepackage{graphicx}  
\usepackage{subeqnar}  
\usepackage{multicol}  
\usepackage{physprbb}  
\makeindex             

\def\simge{%
    \mathrel{\rlap{\raise 0.511ex
        \hbox{$>$}}{\lower 0.511ex \hbox{$\sim$}}}}
\newcommand{\be}{\begin{equation}}
\newcommand{\ee}{\end{equation}}
\newcommand{\Msol}{\mbox{${\rm M}_{\odot}\;$}}

\newcommand{\Msun}{\mbox{${\rm M}_{\odot}\;$}}

\def\lsim{\mathrel{\rlap{
\lower3pt\hbox{\hskip-3pt$\sim$}}
\raise1pt\hbox{$<$}}}
\newcommand{\gsim}{\raisebox{-0.7ex}{$\stackrel{\textstyle >}{\sim}$ }}

\begin{document}
\title*{Evolution of a Neutron Star\protect\newline From its Birth to 
Old Age}
\toctitle{Evolution of a Neutron Star\protect\newline From its Birth to 
Old Age}
%
\titlerunning{Evolution of a Neutron Star}

%
\author{Madappa Prakash\inst{1}
\and James M. Lattimer\inst{1}
\and Jose A. Pons\inst{1}
\and Andrew W. Steiner\inst{1}
\and Sanjay Reddy\inst{2}}
\authorrunning{Madappa Prakash et al.}
%
\institute{Department of Physics \& Astronomy, \\ 
	State University of New York at Stony Brook, \\ 
	Stony Brook, NY-11794-3800, USA 
\and  	Institute for Nuclear Theory, \\ 
      	University of Washington, \\ 
      	Seattle, WA 98195, USA}

\maketitle  

\begin{abstract} The main stages in the evolution of a neutron star,
from its birth as a proto-neutron star, to its old age as a cold,
catalyzed configuration, are described.  A proto-neutron star is formed
in the aftermath of a successful supernova explosion and its evolution
is dominated by neutrino diffusion.  Its neutrino signal is a valuable
diagnostic of its internal structure and composition. During its
transformation from a hot, lepton-rich to a cold, catalyzed remnant, the
possibility exists that it can collapse into a black hole, which
abruptly terminates neutrino emissions.  The essential microphysics,
reviewed herein, that controls its evolution are the equation of state
of dense matter and its associated neutrino opacities. Several
simulations of the proto-neutron star evolution, involving different
assumptions about the composition of dense matter, are described.  After
its evolution into a nearly isothermal neutron star a hundred or so
years after its birth, it may be observable through its thermal emission
in X-rays during its life in the next million years.  
Its surface temperature will depend upon the rapidity of
neutrino emission processes in its core, which depends on the composition 
of dense matter and whether or not its constituents exhibit 
superfluidity and superconductivity.  Observations of thermal
emission offer the best hope of a determination of the radius of a
neutron star. The implications for the underlying dense matter equation
of state of an accurate radius determination are explored. 
\end{abstract}

\section{Introduction: The Tale}

A proto-neutron star (PNS) is born in the aftermath of a successful
supernova explosion as the stellar remnant becomes gravitationally
decoupled from the expanding ejecta. Initially, the PNS is optically
thick to neutrinos, that is, they are temporarily trapped within the star.
The subsequent evolution of the PNS is dominated by $\nu-$diffusion which
first results in deleptonization and subsequently in cooling.  After a much
longer time, photon emissions compete with neutrino emissions in neutron
star cooling.

In this paper, we will focus upon the essential microphysical ingredients
that govern the macrophysical evolution of neutron stars: the equation
of state (EOS) of dense matter and its associated neutrino opacity. 
Among the characteristics of matter that widely vary among EOS models
are their relative compressibilities (important in determining a neutron
star's maximum mass), symmetry energies (important in determining the
typical stellar radius and in the relative proton fraction) and specific
heats (important in determining the local temperature).  These
characteristics play important roles in determining the matter's
composition, in particular the possible presence of additional
components (such as hyperons, a pion or kaon condensate, or quark
matter), and also significantly affect calculated neutrino opacities and
diffusion time scales.

\begin{figure}[htb]
\begin{center}
\includegraphics[trim=0 0 0 0,scale=.52,angle=90]{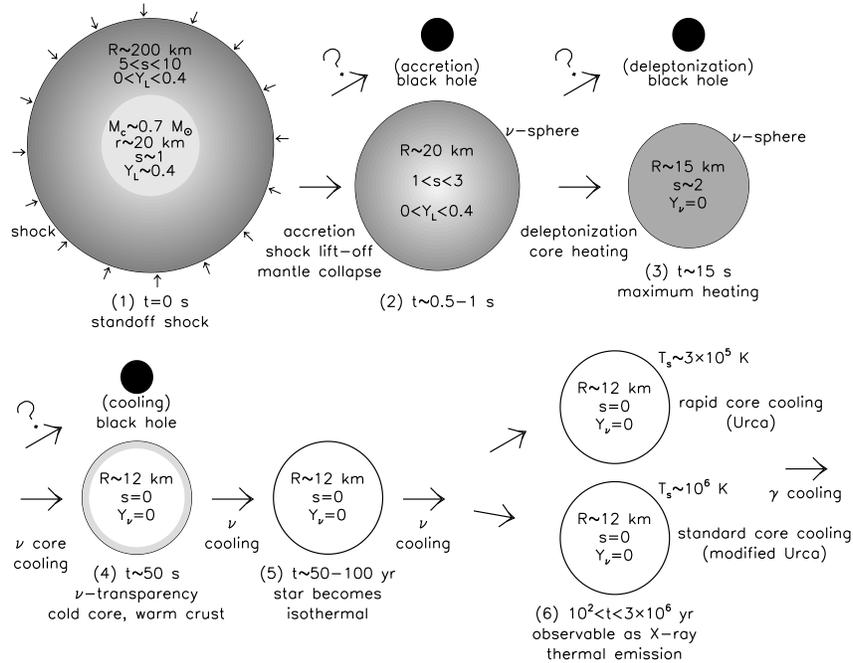}
\end{center}
\caption[]{The main stages of evolution of a neutron star.  Shading
indicates approximate relative temperatures.}
\label{pict1}
\end{figure}

The evolution of a PNS proceeds through several distinct stages
\cite{burr86,burr90} and with various outcomes, as shown schematically
in Fig.~\ref{pict1}. Immediately following core bounce and the passage
of a shock through the outer PNS's mantle, the star contains an
unshocked, low entropy core of mass $M_c\simeq0.7$ M$_\odot$ in which
neutrinos are trapped (the first schematic illustration, labelled (1) in
the figure). The core is surrounded by a low density, high entropy
($5<s<10$) mantle that is both accreting matter from the outer iron core
falling through the shock and also rapidly losing energy due to electron
captures and thermal neutrino emission. The mantle extends up to the
shock, which is temporarily stationary at a radius of about 200 km prior
to an eventual explosion.

After a few seconds (stage 2), accretion becomes less important if the
supernova is successful and the shock lifts off the stellar envelope. 
Extensive neutrino losses and deleptonization will have led to a  loss
of lepton pressure and the collapse of the mantle.  If enough accretion
occurs, however, the star's mass could increase beyond the maximum mass
capable of being supported by the hot, lepton-rich matter.  If this
occurs, the remnant collapses to form a black hole and its neutrino
emission is believed to quickly cease \cite{burr88}.

Neutrino diffusion deleptonizes the core on time scales of 10--15 s 
(stage 3).
The diffusion of high-energy (200--300 MeV) neutrinos from the core to
the surface where they escape as low-energy neutrinos (10--20 MeV)
generates a large amount of heat within the star (a process akin to
joule heating). The core's entropy approximately doubles, producing
temperatures in the range of 30--60 MeV, during this time, even as
neutrinos continue to be prodiguously emitted from the stars effective
surface, known as the $\nu-$sphere.

Strange matter, in the form of hyperons, a Bose condensate, or quark
matter, which is suppressed to extremely large densities when neutrinos
are trapped in matter, could appear at the end of the deleptonization. 
The appearance of strange matter leads to a decrease in the theoretical
maximum mass that matter is capable of supporting, leading to another
possibility for black hole formation~\cite{prak97a}.  This would occur
if the PNS's mass, which must be less than the maximum mass of hot,
lepton-rich matter (or else a black hole would already have formed), is
greater than the maximum mass of hot, lepton-poor matter.  However, if
strangeness does not appear, the theoretical maximum mass instead
increases during deleptonization and the appearance of a black hole
would be unlikely unless accretion in this stage remains significant.

The PNS is now lepton-poor, but it is still hot.  While the star has
zero net neutrino number, thermally produced neutrino pairs of all
flavors are abundant and dominate the emission.  Neutrino diffusion
continues to cool the star, but the average neutrino energy decreases,
and the neutrino mean free path increases.  After approximately 50
seconds (stage 4), the mean free path becomes comparable to the stellar
radius, and star finally becomes transparent to neutrinos.  Since the
threshold density for the appearance of strange matter decreases with
decreasing temperature, a delayed collapse to a black
hole is still possible during this epoch.

Neutrino observations from a galactic supernova will illuminate these
stages.  The observables will constrain time scales for
deleptonization and cooling and the star's binding energy.
Dimensionally, diffusion time scales are proportional to
$R^2(c\lambda)^{-1}$, where $R$ is the star's radius and $\lambda$ is
the effective neutrino mean free path.  This generic relation
illustrates how both the EOS and the composition, which determine both
$R$ and $\lambda$, influence evolutionary time scales.  The total
binding energy, which is primarily a function of stellar mass and
radius (Lattimer \& Prakash \cite{Lattimer00}), should be one of the
most accurately measured quantities from neutrino observatories.
Currently, Super-Kamiokande and SNO are capable of detecting thousands
of neutrinos from a galactic supernova (distance less than 10
kpc). Exciting possibilities lie ahead with many other existing and
planned new facilities \cite{AIP}.

Following the onset of neutrino transparency, the core continues to
cool by neutrino emission, but the star's crust remains warm and cools
less quickly. The crust is an insulating blanket which prevents the
star from coming to complete thermal equilibrium and keeps the surface
relatively warm ($T\approx3\times10^6$ K) for up to 100 years (stage
5).  This timescale is primarily sensitive to the neutron star's
radius and the thermal conductivity of the mantle~\cite{LvRPP94}, as
can be noted from the approximate diffusive relationship $\tau\propto
\Delta R^2/\lambda$, where $\Delta R$ is the thickness of the crust.
If the rapid decrease in the star's surface temperature predicted to
occur when thermal equilibrium is ultimately achieved (see
Fig.~\ref{fig:cooling-N-NQ} in Section 6), a valuable constraint on
the thickness of the crust, and hence the neutron star radius, could
be obtained. The temperature of the surface after the interior of the
star becomes isothermal (stage 6) is determined by the rate of
neutrino emission in the star's core.  The magnitude of the rate is
primarily determined by the question of whether or not one or more of
the so-called direct Urca processes can occur.  The basic Urca process
\begin{equation} n\rightarrow p+e^-+\bar\nu_e;\qquad p\rightarrow
n+e^++\nu_e \end{equation} operates even in degenerate matter because at
finite temperature some of the nucleons are in excited states.  In
addition, direct Urca process involving hyperons, Bose condensates and
quarks are also possible.  In general, the direct Urca rate is
proportional to $T^4$, and is so large that the surface temperatures
fall to just a few times $10^5$ K, which becomes very difficult to
observe in X-rays except for very nearby stars.  A relatively high
surface temperature, closer to $10^6$ K, will persist, however, if an
Urca process can only occur indirectly with the participation of a
spectator nucleon -- the modified Urca process, which in the case of
nucleons is
\begin{equation}
n+(n,p)\rightarrow p+(n,p)+e^-+\bar\nu_e;\qquad p+(n,p)\rightarrow
n+(n,p)+e^++\nu_e\,,
\end{equation}
and leads to the so-called standard cooling scenario.

However, there are two circumstances that could prevent the direct Urca
process from occuring.  First, if the composition of the matter is such
that the momentum triangle involving the non-neutrino particles cannot
be closed, momentum conservation disallows this process.  This occurs,
in the case of $n, p, e$, for example, if the $p$ and $e^-$ abundances,
which must be equal, are less than 1/8 the $n$ abundance. This would be
the case if the nuclear symmetry energy has a relatively weak density
dependence.  In addition, direct Urca processes involving hyperons, a
Bose condensate, or quarks would not occur, of course, if they are not
present.  Second, direct Urca processes are suppressed if one of the
reactants becomes superfluid.  In this case, when the core temperature
falls below the superfluid's critical temperature, the rapid cooling is
terminated.   In the case of a superfluid, the core cooling, and
therefore the surface temperature, will be intermediate between those
predicted by standard and rapid cooling models.  Neutrino emission
continues to dominate until neutron stars are approximately 1 million
years old, at which point photon cooling from the surface begins to
dominate.  Unless the interiors cool very rapidly, X-ray emissions from
stars remain relatively high until the photon cooling epoch.

Several neutron stars have been suggested to have observable thermal
emissions in X-rays. In addition, the nearby neutron star RX
J185635-3754, which is the closest known neutron star with a distance of
approximately 60 pc, has detectable UV and optical thermal emissions as
well.  Such objects represent the best chance for measuring a neutron
star's radius, especially if the redshift of spectral lines can be
determined.   Just-launched or proposed X-ray satellites, such as
Chandra and XMM offer abundant prospects of observations of photon
observations of neutron stars.

The organization of this article is as follows.  Section 2 contains a
summary of the basic equations of evolution for proto-neutron stars,
including a discussion of the equilibrium diffusion approximation.
Section 3 details the equation of state of dense matter, taking into
account the possibility that neutrinos are trapped in the matter.  The
possibilities of hyperon-rich matter, kaon condensation and strange
quark matter are also discussed here.  In the event of a transition to
matter containing kaons or quarks, we also consider the possibility
that matter could be inhomogeneous with droplets of the strange matter
embedded in normal matter.  Neutrino-matter interactions are
considered in Section 4, which includes discussions of the effects of
composition, in-medium dispersion relations, and correlations, in both
homogeneous and inhomogeneous phases.
In Section 5, we
present several simulations of the evolution of proto-neutron stars
with different assumptions about the composition of dense matter and
highlighting the role of the neutrino opacities.  Focus is placed upon
predicted neutrino signals and the differences anticipated for varying
assumptions about the matter's composition.  Section 6 describes the
long-term cooling epoch, with a special emphasis on the role of direct
Urca processes and superfluidity and a comparison with observations.
A discussion of the possibility of detecting superfluidity, including
quark color superfluidity, is included.  In Section 7, the dependence
of the structure of neutron stars on the underlying dense matter
equation of state is explored.  The relation between the matter's
symmetry energy and the radii of neutron stars is highlighted.  In
addition, the moments of inertia and binding energies of neutron stars
are discussed, and observational constraints on the mass and radius of
the Vela pulsar from these considerations are elaborated.  Section 8
contains our outlook.

\section{Short-Term Neutrino Cooling: The First Minutes}

The cooling of PNSs can be divided into two main regimes:  the
short-term, lasting perhaps one minute, during which the potential to
observe the neutrino signal in terrestrial detectors exists, and the
longer term period, lasting perhaps one million years, in which
neutrino emissions dominate the cooling but the star is observable
only through its thermal, photonic, emissions.  This section
summarizes the evolution equations relevant for the short-term
Kelvin-Helmholtz phase and the estimation of  its neutrino signature.

\subsection{PNS Evolution Equations}
The equations that govern the transport of energy and lepton number are
obtained from the Boltzmann equation  for massless particles 
\cite{burr86,lind66,thor81,pons99}.
We will focus on the  non-magnetic, spherically symmetric situation.
For the PNS problem, fluid velocities
are small enough so that hydrostatic equilibrium is nearly fulfilled. 
Under these conditions, the neutrino transport equations  
in a stationary metric
\begin{eqnarray}
\label{metric}
ds^2=-e^{2\phi}dt^2+e^{2\Lambda}dr^2+r^2d\theta^2+r^2\sin^2\theta \,d\Phi^2\,
\end{eqnarray}
are:
\begin{eqnarray}
\label{number}
\frac{\partial ({N_{\nu}/n_B})}{\partial t} &+&
 {\frac{\partial (e^{\phi} 4 \pi r^2 F_{\nu})}{\partial a}}
= e^\phi \frac{S_N}{n_B} \\
\label{energy}
\frac{\partial ({J_{\nu}/n_B})}{\partial t} &+& P_{\nu} \frac{\partial
({1/n_B})}{\partial t} + e^{-\phi} {\frac{\partial (e^{2 \phi} 4 \pi
r^2 H_{\nu})}{\partial a}} = e^\phi \frac{S_E}{n_B} \,,
\end{eqnarray}
where $n_B$ is the baryon number density and $a$ is the enclosed baryon number
inside a sphere of radius $r$. The quantities 
$N_{\nu}$, $F_{\nu}$, and $S_N$ are the number density, number flux and number
source term, respectively, while $J_{\nu}$, $H_{\nu}$, $P_{\nu}$, and $S_E$ are
the neutrino energy density, energy flux, pressure, and the energy
source term, respectively.

In the absence of accretion,   the enclosed baryon number
$a$ is a convenient Lagrangian variable.  The equations to be solved
split naturally into a transport part, which has a strong time
dependence, and a structure part, in which evolution is much slower.
Explicitly, the structure equations are
\begin{eqnarray}
\label{a-struc}
{{\partial r}\over{\partial a}} = \frac{1}{4 \pi r^2 n_B e^{\Lambda}}
\,&,& \quad
\frac{\partial m}{\partial a} = \frac{\rho}{n_B e^{\Lambda}} \\
\label{b-struc}
\frac{\partial \phi}{\partial a} = \frac{e^{\Lambda}}{4\pi r^4 n_B}
{\left( m + 4\pi r^3 P \right)} \,&,& \quad
\frac{\partial P}{\partial a} = - (\rho + P)
\frac{e^{\Lambda}}{4\pi r^4 n_B}
{\left( m + 4\pi r^3 P \right)} \,.
\end{eqnarray}
The quantities $m$ (enclosed gravitational mass), $\rho$ 
(mass-energy density), and $P$ (pressure) include
contributions from the leptons.  To
obtain the equations employed in the transport, 
Eq.~(\ref{number}) may be combined with 
the corresponding equation for the electron
fraction
\begin{eqnarray}
\frac{\partial Y_e}{\partial t}=-e^\phi\frac{S_N}{n_B} \
\end{eqnarray} to obtain \begin{eqnarray}
\label{a-number}
\frac{\partial Y_L}{\partial t} +
e^{-\phi} {\frac{\partial (e^{\phi} 4 \pi r^2 F_{\nu})}{\partial a}}
= 0 \,.
\end{eqnarray}
Similarly, Eq.~(\ref{energy}) may be combined with the matter energy equation
\begin{eqnarray}
\frac {dU}{dt} + P \frac{d({1/n_B})}{dt} = - e^\phi \frac{S_E}{n_B} \,,
\end{eqnarray}
where $U$ is the specific internal energy
and use of the first law of thermodynamics yields
\begin{eqnarray}
e^\phi T\frac{\partial s}{\partial t} + e^\phi\mu_{\nu} \frac{\partial
Y_L}{\partial t} + e^{-\phi} {\frac{\partial e^{2 \phi} 4 \pi r^2
H_{\nu}}{\partial a}} = 0 \,,
\label{a-energy}
\end{eqnarray}
where $s$ is the entropy per baryon.

\subsection{The Equilibrium Diffusion Approximation  }
At high density and for temperatures above several MeV, the source terms
in the Boltzmann equation 
are sufficiently strong to ensure that neutrinos are in thermal and
chemical equilibrium with the ambient matter. Thus, the neutrino
distribution function in these regions is both nearly Fermi-Dirac and
isotropic.  We can approximate the distribution function as an expansion in
terms of Legendre polynomials
to $O(\mu)$, which is known as the diffusion approximation.  Explicitly,
\begin{eqnarray}
f(\omega,\mu)= f_0(\omega) +  \mu f_1(\omega) \,, \quad
f_0 = [1+e^{\left(\frac{\omega-\mu_\nu}{kT}\right)}]^{-1} \,,
\end{eqnarray}
where $f_0$ is the Fermi--Dirac distribution function at equilibrium
($T=T_{mat}$, $\mu_{\nu}=\mu_{\nu}^{eq}$), with
$\omega$ and $\mu_\nu$ being the neutrino energy and chemical potential,
respectively.
The main goal is to obtain a relation for $f_1$ in terms of $f_0$.
In the diffusion approximation, one obtains \cite{pons99} 
\begin{eqnarray}
\label{f1}
f_1 = - D(\omega)
\left[ {e^{-\Lambda}} \frac{\partial f_0}{\partial r}
- {\omega} {e^{-\Lambda} \frac{\partial \phi}{\partial r}}
{\frac{\partial f_0}{\partial \omega}} \right] \,.
\end{eqnarray}
\noindent
The explicit form of the diffusion coefficient $D$ appearing above is given by
\begin{eqnarray}
D(\omega) = {\left( j+\frac{1}{\lambda_a}+\kappa^s_1 \right)}^{-1} \,.
\end{eqnarray}
The quantity $j=j_a+j_s$, where $j_a$ is the emissivity and 
$j_s$ is the scattering contribution to the source term. 
The absorptivity is denoted by 
$\lambda_a$ and $\kappa_1^s$ is the scattering contribution to the 
transport opacity.  
Substituting 
\begin{eqnarray}
\frac{\partial f_0}{\partial r} =
-\left( T \frac{\partial \eta_{\nu}}{\partial r} + \frac{\omega}{T}
\frac{\partial T}{\partial r}\right)
\frac{\partial f_0}{\partial \omega}~,
\end{eqnarray}
where $ \eta_{\nu}=\mu_{\nu}/T $ is the neutrino degeneracy parameter, 
in Eq.~(\ref{f1}), we obtain
\begin{eqnarray}
f_1 = - D(\omega) e^{-\Lambda} \left[
T \frac{\partial \eta}{\partial r}
+ \frac{\omega}{T e^{\phi}} \frac{\partial (T e^{\phi})}{\partial r} \right]
\left(- \, \frac{\partial f_0}{\partial \omega} \right)\,.
\end{eqnarray}
Thus, the energy-integrated lepton and energy fluxes are
\begin{eqnarray}
F_{\nu}&=&- \, \frac{e^{-\Lambda} e^{-\phi}T^2}{6 \pi^2}
\left[ D_3 \frac{\partial (T e^{\phi})}{\partial r} +
(T e^{\phi}) D_2 \frac{\partial \eta}{\partial r}  \right] \nonumber \\
H_{\nu}&=&- \, \frac{e^{-\Lambda} e^{-\phi}T^3}{6 \pi^2}
\left[ D_4 \frac{\partial (T e^{\phi})}{\partial r} +
(T e^{\phi}) D_3 \frac{\partial \eta}{\partial r}  \right] \,.
\label{fluxes}
\end{eqnarray}
The coefficients $D_2$, $D_3$, and $D_4$ are related to the energy-dependent
diffusion coefficient $D(\omega)$ through
\begin{eqnarray}
D_n = \int_0^\infty dx~x^n D(\omega)f_0(\omega)(1-f_0(\omega))~,
\label{d2d3}
\end{eqnarray} 
where $x=\omega/T$.  These diffusion coefficients 
depend only on the
microphysics of the neutrino-matter interactions (see \S4 for
details). The fluxes appearing in the above equations are for one
particle species. To include all six neutrino types, we redefine the
diffusion coefficients in Eq.~(\ref{fluxes}):
\begin{eqnarray}
D_2=D_2^{\nu_e}+D_2^{\bar{\nu}_e}\,, \quad
D_3=D_3^{\nu_e}-D_3^{\bar{\nu}_e}\,, \quad
D_4=D_4^{\nu_e}+D_4^{\bar{\nu}_e}+4 D_4^{\nu_\mu}\,.
\end{eqnarray}

\subsection{Neutrino Luminosities}
A fair representation of the
signal in a terrestrial detector can be found from the time dependence
of the total neutrino luminosity and average neutrino energy together
with an assumption of a Fermi-Dirac spectrum with zero chemical potential.
We will return to discuss the  improvements necessary to obtain 
more accurate information about the spectra. 

The total neutrino luminosity is
the time rate of change of the star's gravitational mass, and is
therefore primarily a global property of the evolution.  This
luminosity, due to energy conservation, must also equal
\begin{eqnarray}
L_\nu=e^{2 \phi} 4 \pi r^2 H_{\nu}\,
\end{eqnarray}
at the edge of the star.  This relation serves as a test of energy
conservation, at least for 
all times greater than about 5 ms, when the star comes into
radiative equilibrium.  For times greater than about 5 ms, initial
transients become quite small and the predicted luminosities should be
relatively accurate compared to full transport simulation.
Estimate of the average
energy of neutrinos is made from the temperature $T_\nu$ of the matter at the
neutrinosphere $R_\nu$, defined to be the location in the star where
the flux factor $\xi_H=0.25$. However, since the spectrum may not be
Fermi-Dirac at the neutrinosphere, a diffusion scheme cannot give a
very precise value for the average energy.  We use the average energy
$<E_\nu>\approx3T_\nu$, where $T_\nu$ is a mass average in the
outermost zone.  Because it is a globally determined quantity, the
luminosity $L_\nu$ is necessarily more accurately determined than
either $R_\nu$ or $T_\nu$.  

\section{The Equation of State of Neutrino Trapped Matter}

The rationale for considering different possibilities for the
composition of dense matter is largely due to the fact that QCD at
finite baryon density remains unsolved.  Effective QCD based models
have raised intriguing possibilites concerning the composition of
dense matter including the presence of hyperons, pion or kaon
condensates, and quark matter (see \cite{prak97a} for extensive
references).  It is also important to have predictions for the
plain-vanilla case of nucleons alone.  The contrast can be dramatic,
since additional components offer the possibility of BH formation
during the evolution of a PNS.  In what follows, the symbols $np$
refer to matter with nucleons alone, $npH$ to matter including
hyperons, $npK$ to matter with nucleons and kaons, and $npQ$ to matter
with nucleons and quarks.  In all cases, leptons in beta equilibrium
are included.

\subsection{Matter with Nucleons and Hyperons}
The masses and radii of neutron stars depend upon the matters'
compressibility, the composition of matter at high density, and the
nuclear symmetry energy (e.g., \cite{prak97a}).  In the PNS problem,
the finite temperature aspects of the EOS also play an important
role. During the early evolution the entropy in the central regions is
moderately high, $s\sim 1-2$ (in units of Boltzmann's constant), which
correspond to temperatures in the range $T=20-50$ MeV. These features
may be explored by employing a finite temperature field-theoretical
model in which the interactions between baryons are mediated by the
exchange of $\sigma,\omega$, and $\rho$ mesons\footnote{Note that the
couplings in these models may be chosen to reproduce the results of
numerically more intensive microscopic potential models, such as that
of Akmal and Pandharipande \cite{Akmal}, so that the gross features of
the zero temperature thermodynamics can be reproduced.  Additional
advantages to this approach are that the effects of finite temperature
and arbitrary proton fraction may be incorporated more easily.}.  The
hadronic Lagrangian density is given by~\cite{sero92}
\begin{eqnarray}
L_H &=& 
      \sum_{i} \overline{B_i}(-i\gamma^{\mu}\partial_{\mu}-g_{\omega i}
\gamma^{\mu}\omega_\mu
-g_{\rho i}\gamma^{\mu}{\bf{b}}_{\mu}\cdot{\bf t}-M_i+g_{\sigma i}\sigma)B_i
\nonumber \\
&-& \frac{1}{4}W_{\mu\nu}W^{\mu\nu}+\frac{1}{2}m_{\omega}^2\omega_{\mu}\omega^
{\mu} - \frac{1}{4}{\bf B_{\mu\nu}}{\bf
B^{\mu\nu}}+\frac{1}{2}m_{\rho}^2 b_{\mu}b^{\mu} \nonumber \\
&+& \frac{1}{2}\partial_{\mu}\sigma\partial^{\mu}\sigma -\frac{1}{2}
m_{\sigma}^2\sigma^2-U(\sigma)
\end{eqnarray}
Here, $B$ are the Dirac spinors for baryons and $\bf t$ is the isospin
operator. The sums include baryons $i=n,p,\Lambda,\Sigma$, and $\Xi$.
The field strength tensors for the $\omega$ and
$\rho$ mesons are $W_{\mu\nu} = \partial_\mu\omega_\nu-\partial_\nu\omega_\mu$
and ${\bf B}_{\mu\nu} =  \partial_\mu{\bf b}_\nu-\partial_\nu{\bf b}_\mu$,
respectively.  The potential $U(\sigma)$ represents the self-interactions of
the scalar field and is taken to be of the 
form~\cite{glen91}
\begin{eqnarray}
U(\sigma) =  \frac{1}{3}bM_n(g_{\sigma N}\sigma)^3 + \frac{1}{4}c(g_{\sigma
N}\sigma)^4\,.
\end{eqnarray}
The partition function $Z_H$ for
the hadronic degrees of freedom is evaluated  
in the mean field approximation. 
The total partition function $Z_{total}=Z_HZ_L$, where $Z_L$ is the standard
noninteracting partition function of the leptons.
Using $Z_{total}$, the thermodynamic quantities can be obtained in the 
standard way.  
The additional conditions needed to obtain a solution are provided by the
charge neutrality requirement, and, when neutrinos are 
trapped, the set of equilibrium chemical
potential relations required by the general condition
\begin{eqnarray}
\mu_i = b_i\mu_n - q_i(\mu_l-\mu_{\nu_\ell})\,.
\label{tbeta}
\end{eqnarray}
where $b_i$ is the baryon number of particle $i$ and $q_i$ is its
charge.  The introduction of additional variables, the neutrino
chemical potentials, requires additional constraints, which we supply
by fixing the lepton fractions, $Y_{L\ell}$, appropriate for
conditions prevailing in the evolution of the PNS.  In addition to
models containing only nucleonic degrees of freedom (GM1np \& GM3np)
we investigate models that allow for the presence of hyperons (GM1npH \& 
GM3npH). 
 For the determination of the various coupling constants
appearing in $Z_H$ see \cite{prak97a}.

The lepton chemical potentials influence the deleptonization epoch. 
For np models
a lower nuclear symmetry energy favors a larger $\nu_e$ fraction
and has little effect on the $e^-$ fraction at $Y_L=0.4$. Models with
hyperons lead to significantly larger $\mu_{\nu_e}$ and lower $\mu_e$, both of
which influence the diffusion of electron neutrinos. The electron
chemical potentials in neutrino free matter are reduced to a greater extent
by changes in composition and symmetry energy as there are no neutrinos to
compensate for changes in $\hat{\mu}=\mu_n-\mu_p$. 

\subsection{Matter with a Kaon Condensate}

The contents of this section are extracted from Pons et al. \cite{Pon00b}.
For the kaon sector, we use a Lagrangian that contains the usual
kinetic energy and mass terms along with the meson interactions 
\cite{GS99}.
Kaons are coupled to the meson fields through  minimal coupling;
specifically,
\begin{eqnarray}
L_K &=& {\cal D}^*_\mu K^+ {\cal D}^\mu K^- - m_K^{*2} K^+ K^- \,,
\end{eqnarray}
where the vector fields are coupled via the standard form \be {\cal
D}_\mu = \partial_\mu + i g_{\omega K} \omega_\mu + i g_{\rho K}
\gamma^{\mu}{\bf{b}}_{\mu}\cdot{\bf t} \ee and $m_K^{*} = m_K -
\frac{1}{2} g_{\sigma K} \sigma$ is the effective kaon mass.

In the mean field approach, the thermodynamic potential per unit volume 
in the kaon sector is \cite{Pon00b}
\begin{eqnarray}
\frac{\Omega_K}{V} &=& \textstyle{\frac{1}{2}}
(f\theta)^2(m_K^{*2}-(\mu+X_0)^2)\nonumber \\ &+&
T\int\limits_0^{\infty}\frac{d^3p}{(2\pi)^3}\left[
\ln(1-e^{-\beta(\omega^--\mu)})+
\ln(1-e^{-\beta(\omega^++\mu)})\right]\;,\label{zkexch}
\end{eqnarray}
where $ X_0 = g_{\omega K} \omega_0 + g_{\rho K}b_0$, 
the Bose occupation probability \newline 
$f_B(x)=(e^{\beta x}-1)^{-1}$, $\omega^{\pm} = 
{\sqrt {p^2+m_K^{*^2}}} \pm X_0$, 
$f=93$ MeV is the pion decay constant and the condensate amplitude,
$\theta$, can be found by extremization of the partition function.
This yields the solution $\theta=0$ (no condensate) or, if a
condensate exists, the equation \be m_K^{*} = \mu_K + X_0
\label{cond}
\ee
where $\mu_K$ is the kaon chemical potential.  In beta-stable stellar
matter the conditions of charge neutrality
\be
\label{cons1}
\sum_B q_B n_B - n_e - n_K = 0
\ee
and chemical equilibrium
\begin{eqnarray}
\mu_i &=& b_i\mu_n - q_i(\mu_l-\mu_{\nu_\ell}) \\
\mu_K &=& \mu_n - \mu_p
\label{tkbeta}
\end{eqnarray}
are also fulfilled.

The kaon condensate is assumed to appear by forming a mixed phase with
the baryons satisfying Gibbs' rules for phase equilibrium
\cite{Gibbs}.  Matter in this mixed phase is in mechanical, thermal
and chemical equilibrium, so that \be p^I=p^{II} \,, \quad
T^I=T^{II}\,, \quad \mu_i^I=\mu_i^{II}\,, \ee where the superscripts I
and II denote the nucleon and kaon condensate phases, respectively.
The conditions of global charge neutrality and baryon number
conservation are imposed through the relations \begin{eqnarray} \chi
q^I + (1-\chi) q^{II} &=& 0 \nonumber \\ \chi n_B^I + (1-\chi)
n_B^{II} &=& n_B \,, \end{eqnarray} where $\chi$ denotes the volume
fraction of nucleonic phase, $q$ the charge density, and $n_B$ the
baryon density.  We ignore the fact that the phase with the smallest
volume fraction forms finite-size droplets; in general, this would
tend to decrease the extent of the mixed phase region. Further general
consequences of imposing Gibbs' rules in a multicomponent system are
that the pressure varies continuously with density in the mixed phase
and that the charge densities must have opposite signs in the two
phases to satisfy global charge neutrality.  We note, however, that
not all choices of nucleon-nucleon and kaon-nucleon interactions
permit the Gibbs' rules to be satisfied (for an example of such an
exception, see \cite{Pon00b}).  The models chosen in this work {\it
do} allow the Gibbs' rules to be fulfilled at zero and finite
temperatures and in the presence of trapped neutrinos.

The nucleon-meson couplings are determined by adjusting them to
reproduce the properties of equilibrium nucleonic matter at $T=0$.  We
use the numerical values used by \cite{glen91},
i.e., equilibrium density $n_0=0.153$ fm$^{-3}$, equilibrium energy
per particle of symmetric nuclear $E/A=-16.3$ MeV, effective mass
$M^*=0.78M$, compression modulus $K_0=240$ MeV, and symmetry energy
$a_{sym}=32.5$ MeV. These values yield the coupling constants
$g_\sigma/m_\sigma = 3.1507~{\rm fm},~ g_\omega/m_\omega = 2.1954~{\rm
fm},~g_\rho/m_\rho = 2.1888,~b=0.008659$, and $c=-0.002421$.

The kaon-meson couplings $g_{\sigma K}$ and $g_{\omega K}$ are related
to the magnitude of the kaon optical potential $U_K$ at the saturation
density $n_0$ of isospin symmetric nuclear matter: \be U_K(n_0) = -
g_{\sigma K} {\sigma(n_0)} - g_{\omega K} \omega_0(n_0).  \ee Fits to
kaonic atom data have yielded values in the range $-(50-200)$ MeV
\cite{FGB94,Fri99,WW97,RO00,BGN00}. We use
$g_{\omega K} = g_{\omega N}/3$ and $g_{\rho K} = g_{\rho N}/2$ on the
basis of simple quark and isospin counting.  Given the uncertainty in
the magnitude of $|U_K|$, consequences for several values of $|U_K|$
were explored in \cite{Pon00b}.  
Moderate values of $|U_K|$ generally produce a
second order phase transition and, therefore, lead to moderate effects on
the gross properties of stellar structure.  Values  in excess of
100 MeV were found necessary for a first order phase transition to
occur; in this case kaon condensation occurs at a relatively low
density with an extended mixed phase region, which leads to more
pronounced effects on the structure due to a significant softening of
the EOS.

\begin{figure}
\begin{center}
\includegraphics[scale=0.4,angle=0]{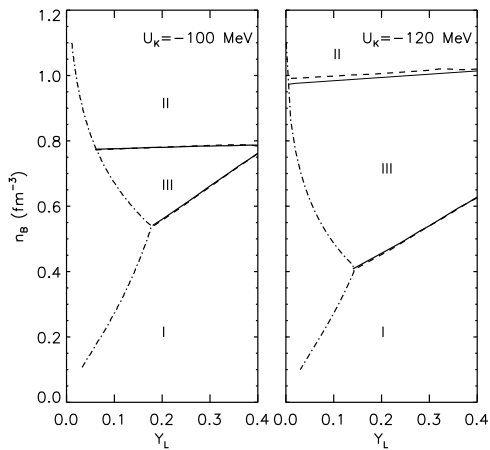}
\end{center}
\caption{Phase boundaries between pure nucleonic
matter (I), pure kaon condensed matter (II) and a mixed phase (III) in
the $Y_L$--$n_B$ plane for $U_K=-100$ MeV (left panel) and $U_K=-120$
MeV (right panel). The solid line corresponds to $s=0$ and the dashed
line to $s=1$. The dashed-dotted line shows the baryon density as a
function of the lepton fraction for $s=0$, neutrino-free ($Y_L=Y_e$)
matter}
\label{kaon-bdy}
\end{figure}

The phase boundaries of the different phases are displayed in 
Fig.~\ref{kaon-bdy} in a $Y_L$--$n_B$ plane for an optical potential $U_K$
of --100 MeV (left) and --120 MeV (right), respectively.  The
nucleonic phase, the pure kaon matter phase, and the mixed phase are
labelled I, II, and III, respectively. Solid lines mark the phase
transition at zero temperature and dashed lines mark the phase
transition at an entropy per baryon of $s=1$.  Note that finite
entropy effects are small and do not affect significantly the phase
transition density.  The dash-dotted line shows the electron fraction
$Y_e$ as a function of density in cold, catalyzed matter (for which
$Y_L=Y_e$), which is the final evolutionary state. The region to the
left of this line corresponds to negative neutrino chemical potentials
and cannot be reached during normal evolutions.  The solid and dashed
lines, which separate the pure phases from the mixed phase, vary
roughly linearly with the lepton fraction.  Also notice the large,
and nearly constant, densities of the boundary between the mixed phase III
and the pure kaon phase II.  These densities, for the cases shown, lie
above the central densities of the maximum mass stars, so that region
II does not generally exist in proto-neutron stars (see 
\cite{Pon00b}).  The effect of increasing the lepton number is
to reduce the size of the mixed phase (which in fact shrinks to become
a second order phase transition for $Y_L>0.4$ and $U_K=-100$ MeV) and
to shift the critical density to higher densities.  A similar effect
is produced by decreasing the magnitude of the optical potential.

\subsection{Matter with Quarks}

The discussion in this section follows the work of Steiner, Prakash \&
Lattimer \cite{SPL00}.  The thermodynamic potential of the quark phase
is $\Omega =\Omega_{\mathrm{FG}} + \Omega_{\mathrm{Int}}$, where
\begin{eqnarray}
\frac{\Omega_{\mathrm{FG}}}{V} = 2 N_c T \sum_{i=u,d,s}
\int \frac{d^3 p}{\left(2 \pi\right)^3}
\left[ \ln{(1-f_i)} + \ln{(1-{\bar f_i})} \right]\,
\label{FG}\end{eqnarray}
denotes the Fermi gas contribution arising from quarks.  We
consider three flavors, $i=u,d,s$ and three colors, $N_c=3$ of quarks.
The distribution functions of fermions and anti-fermions are $f_i=[1 +
\exp(\beta(E_i-\mu_i))]^{-1}$ and \newline ${\bar f_i} = [1 +
\exp(\beta(E_i+\mu_i))]^{-1}$, where $E_i$ and $\mu_i$ are the single
particle energy and chemical potential, respectively, of quark species
$i$.  To explore the sensitivity of the quark model, we contrast the
results of the MIT bag and the Nambu Jona-Lasinio (henceforth NJL)
models for $\Omega_{\mathrm{Int}}$.

In the MIT bag model, the Fermi gas contribution is calculated using
current, as opposed to
dynamical, quark masses. We will restrict ourselves to the
simplest bag model and keep only the constant cavity pressure term.
The results are qualitatively similar to what is obtained by including
perturbative corrections, if the bag constant $B$ is slightly altered
\cite{PBP}.

Several features of the Lagrangian of Quantum Chromo-Dynamics
(QCD), including the spontaneous breakdown of chiral
symmetry, are exhibited by the Nambu Jona-Lasinio (NJL) model, which
shares many symmetries with QCD.  In its commonly used form, the NJL
Lagrangian reads

\begin{eqnarray}
{\cal L} &=& \bar q ( i \partial{\hskip-2.0mm}/ - {\hat m_0}) q \;+\;
G \sum_{k=0}^8 [\,({\bar q}\lambda_k q)^2 + ({\bar q}
i\gamma_5\lambda_k q)^2\,] \nonumber \\ &-& K \,[ \,{\rm det}_f ({\bar
q}(1+\gamma_5) q) + {\rm det}_f ({\bar q}(1-\gamma_5) q) \,] \ . \label{L3}
\end{eqnarray}
The determinant operates over flavor space, ${\hat m_0}$ is the 3
$\times$ 3 diagonal current quark mass matrix, $\lambda_k$ represents
the 8 generators of SU(3), and $\lambda_0$ is proportional to the
identity matrix.  The four-fermion interactions stem from the original
formulation of this model \cite{NJL}, while the flavor mixing,
determinental interaction is added to break $U_A(1)$ symmetry
\cite{tHooft}.  Since the coupling constants $G$ and $K$ are
dimensionful, the quantum theory is non-renormalizable.  Therefore, an
ultraviolet cutoff $\Lambda$ is imposed, and results are considered
meaningful only if the quark Fermi momenta are well below this cutoff.
The coupling constants $G$ and $K$, the strange quark mass $m_{s,0}$,
and the three-momentum ultraviolet cutoff parameter $\Lambda$, are
fixed by fitting the experimental values of $f_\pi$, $m_\pi$, $m_K$
and $m_{\eta'}$.  We use the values of Ref.~\cite{Rehberg}, namely
$\Lambda = 602.3$ MeV, $G\Lambda^2 = 1.835$, $K\Lambda^5 = 12.36$, and
$m_{0,s}=140.7$ MeV, obtained using $m_{0,u}=m_{0,d}=5.5$ MeV.  The
subscript ``$0$'' denotes current quark masses.  Results of the gross
properties of PNSs obained by the alternative parameter sets of
Refs. \cite{parms2} and \cite{Hatsuda} are similar to the results
quoted below.

In the mean field approximation at finite temperature and at finite baryon
density, the
thermodynamic potential due to interactions among quarks
is given by \cite{Hatsuda}:
\begin{eqnarray}
\frac{\Omega_{\mathrm{Int}}}{V} &=& - 2 N_c \sum_{i=u,d,s}
\int \frac {d^3p}{(2\pi^3}
\left( {\sqrt{m_i^2 + p^2}} - {\sqrt{m_{0,i}^2 + p^2}} \right) \nonumber \\
&+& 2 G \langle\bar{q}_i q_i \rangle^2
- 4 K \langle \bar{q}_u q_u \rangle \langle \bar{q}_d q_d \rangle
\langle \bar{q}_s q_s \rangle\,.
\label{omegint}
\end{eqnarray}
In both Eqs.~(\ref{FG}) and (\ref{omegint}) for the NJL model, the
quark masses are dynamically generated as solutions of the gap
equation obtained by requiring that the potential be stationary with
respect to variations in the quark condensate $\langle {\bar{q}_i
q_i}\rangle$:
\begin{equation}
   m_i = m_{0,i} - 4 G \langle {\bar{q}_i q_i}\rangle +
     2 K \, \langle{\bar{q}_j q_j}\rangle \langle{\bar{q}_k q_k}\rangle \ ,
\label{gap}
\end{equation}
$(q_i,q_j,q_k)$ representing any permutation of $(u,d,s)$.
The quark condensate
$\langle {\bar{q}_i q_i}\rangle$ and the quark number density
$n_i=\langle {q_i^{\dagger} q_i}\rangle$ are given by:
\begin{eqnarray}
\langle{\bar{q}_i q_i}\rangle & = & - 2 N_c  \int
{ \frac{d^3 p}{\left(2 \pi \right)^3} \frac {m_i}{E_i}
\left[1-f_i-{\bar f_i}\right]  }  \nonumber \\
n_i=\langle {q^{\dagger}_i q_i}\rangle & = & 2 N_c \int
{ \frac{d^3 p}{\left(2 \pi \right)^3} \left[f_i-{\bar f_i}\right]}
\,.
\end{eqnarray}

A comparison between the MIT bag and NJL models is facilitated by
defining an effective bag pressure in the NJL model to be \cite{Buballa}
$B_{eff}=\Omega_{\mathrm{int}}/V-B_0$ with $B_0 V =
\Omega_{\mathrm{int}}|_{n_u=n_d=n_s=0}$ a constant value which makes
the vacuum energy density zero.  In this way, the thermodynamic
potential can be expressed as $\Omega= B_{eff}V + \Omega_{\rm FG}$ which
is to be compared to the MIT bag result $\Omega=BV+\Omega_{\rm
FG}$.  Note, however, that $\Omega_{FG}$ in the NJL model is
calculated using the dynamical quark masses from Eq.~(\ref{gap}).

\begin{figure}[htb]
\begin{center}
\includegraphics[scale=0.4]{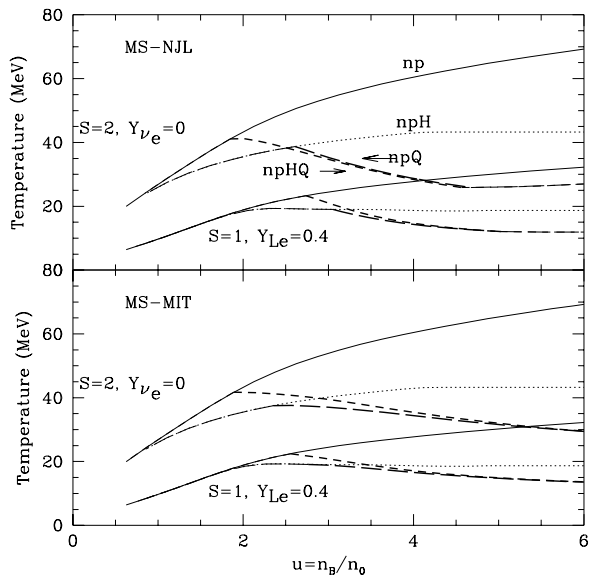}
\end{center}
\caption
{Temperature versus density in units of $n_0$
for two PNS evolutionary snapshots. The upper (lower) panel displays
results for the NJL (MIT bag) Lagrangian.
The parameters $\zeta=\xi=0$ in the M\"uller-Serot (MS)
hadronic Lagrangian \cite{MS} are chosen.  Results are compared for
matter containing only nucleons (np), nucleons plus hyperons (npH),
nucleons plus quarks (npQ) and nucleons, hyperons and quarks (npHQ).
Bold curves indicate the mixed phase region}
\label{spl:temp}
\end{figure}

The temperature as a function of baryon density for fixed entropy and
net lepton concentration is presented in Fig.~\ref{spl:temp}, which
compares the cases ($s=1, Y_{L_e}=0.4$) and ($s=2, Y_{\nu_e}=0$)
both including and ignoring quarks. The temperature for a
multicomponent system in a pure phase can be analyzed with the
relation for degenerate Fermi particles
\begin{eqnarray}
T=\frac{s}{\pi^2}\left(
\frac{\sum_i{ p_{F_i} \sqrt{p_{F_i}^2 + \left( m_i^{*} \right)^2} }}
{\sum_i{ p_{F_i}^3 }}
\right)^{-1}\,,
\end{eqnarray}
where $m^*_i$ and $p_{F_i}$ are the effective mass and the Fermi
momentum of component $i$, respectively.  This formula is quite
accurate since the hadron and quark Fermi energies are large compared
to the temperature.  The introduction of hyperons or quarks lowers the
Fermi energies of the nucleons and simultaneously increases the
specific heat of the matter, simply because there are more components.
In the case of quarks, a further increase, which is just as
significant, occurs due to the fact that quarks are rather more
relativistic than hadrons.  The combined effects for quarks are so
large that, in the case $M^{*}_0=0.6M$ shown in Fig.~\ref{spl:temp}, an
actual reduction of temperature with increasing density occurs along
an adiabat. The effect is not necessarily as dramatic for other
choices of $M^{*}_0$, but nevertheless indicates that the temperature
will be smaller in a PNS containing quarks than in stars without
quarks.  The large reduction in temperature might also influence
neutrino opacities, which are generally proportional to $T^2$.
However, the presence of droplet-like structures in the mixed phase,
not considered here, will modify the specific heat.  In addition,
these structures may dominate the opacity in the mixed phase
\cite{RBP}.  However, a PNS simulation is necessary to consistently
evaluate the thermal evolution, since the smaller pressure of
quark-containing matter would tend to increase the star's density and
would oppose this effect.

\begin{figure}[htb]
\begin{center}
\includegraphics[scale=0.4]{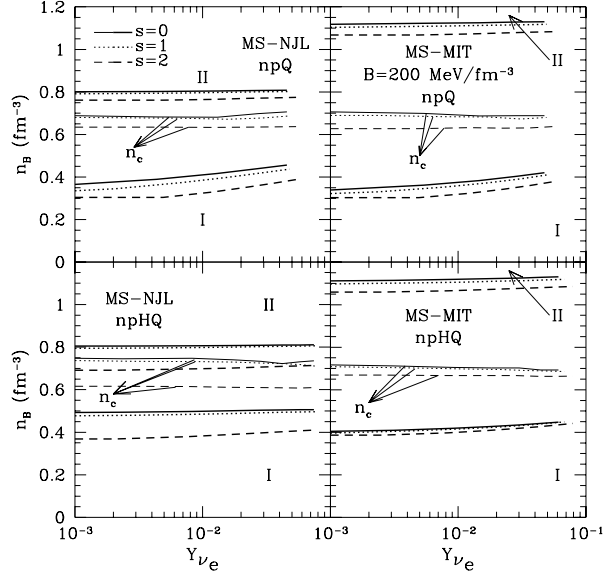}
\end{center}
\caption{The phase diagram of the quark-hadron
transition in the baryon number density - neutrino concentration plane
for three representative snapshots during the evolution of a
proto-neutron star.  The left (right) panels are for the NJL (MIT bag)
quark EOS, and hyperons are (are not) included in the bottom (top)
panels.
The parameters $\zeta=\xi=0$ in the M\"uller-Serot (MS)
hadronic Lagrangian are chosen.
The lower- and upper-density boundaries of the mixed phase
are indicated by bold curves.  The central densities of maximum mass
configurations are shown by thin curves}
\label{spl:phase}
\end{figure}

This last point is highlighted in Fig.~\ref{spl:phase} which shows phase
diagrams for the mixed phase in the baryon density-neutrino fraction
plane.  The upper and lower boundaries of the mixed phase region are
displayed as bold lines, while the central densities of the maximum
mass configurations are shown as light lines.  In no case, for either
quark model and whether or not hyperons are included, are pure quark stars
possible.  The high-density phase boundaries are always well
above the central densities. While
in the optimum case,  
in which the parameters of both the hadronic and quark 
EOSs are fine-tuned,
it is possible for a pure quark core to form if
$B<150$ MeV fm$^{-3}$, the maximum mass decreases below 1.44
$M_{\odot}$ if $B<145$ MeV fm$^{-3}$.  This narrow window, which
further decreases or disappears completely if the hadronic EOS is
altered, suggests that pure quark configurations may 
be unlikely.

\subsection{Inhomogeneous Phases}

It is widely believed that some type of phase transition will occur in
nuclear matter at high densities.  For example, a transition to
deconfined quark matter should exist at sufficiently high density, and
at lower densities, a first-order transition to a Bose condensate
phase might exist.  Such phase transitions are expected to soften the
equation of state, leading to changes in the mass-radius relation and
lowering the maximum mass.  Phase transitions can also influence
transport and weak interaction rates in  matter.  

Glendenning has shown that, due to the existence of two conserved
charges (baryon number and charge) instead of just one, first order
phase transitions can lead to a large mixed phase region in the
neutron star interior \cite{glen1}.  The mixed phase consists of high
baryon density, negatively charged, matter coexisting with lower
density, positively charged, baryonic matter.  The situation is
entirely analogous to the well-known situation involving the mixed
phase consisting of nuclei and a surrounding nucleonic vapor that
occurs below nuclear saturation density \cite{LR78}.  The occurence of
a mixed phase, as opposed to a Maxwell construction, results in a
wider transition in which bulk thermodynamic properties such as
pressure vary less rapidly but are softer over a wider density range.
In addition, the propagation of neutrinos whose wavelength is greater
than the typical droplet size and less than the inter-droplet spacing
will be greatly affected by the heterogeneity of the mixed phase, as a
consequence of the coherent scattering of neutrinos from the matter in
the droplet.  The thermodynamics and the effect on neutron star
structure of two situations have been studied in some detail: first
order kaon condensation \cite{GS99,GS98,size} and the quark-hadron
transition \cite{GP,PCL}.

\section{Neutrino-Matter Interaction Rates}
One of the important microphysical inputs in PNS simulations is the
neutrino opacity at supra-nuclear density~
\cite{burr86,brue85,mezz93,wils89,suzu92,keil95a}
Although it was
realized over a decade ago that the effects due to
degeneracy and strong interactions significantly alter the neutrino
mean free paths, it is only recently that detailed calculations have
become available 
\cite{RBP,redd97a,redd97b,prak97b,redd98,redd99a,burr98,burr99a}.
The scattering and absorption reactions that contribute to the neutrino
opacity are
\begin{eqnarray}
\nu_e+B&\rightarrow& e^-+B' \,,\quad \quad
\bar{\nu}_e+B\rightarrow e^++B' \,,\\
\nu_X+B&\rightarrow& \nu_X+B' \,,\quad \quad
\nu_X+e^-\rightarrow \nu_X +e^-  \,,
\end{eqnarray}
where the scattering reactions are common to all neutrino species and the
dominant source of opacity for the electron neutrinos is due to the charged
reaction. 
The weak interaction rates in hot
and dense matter are modified due to many in-medium effects.  The most
important of these are: \\

\noindent (1) {\it Composition}:   The neutrino mean free paths depend
sensitively on the composition which is sensitive to the nature of strong
interactions.  First, the different
degeneracies of the different Fermions determines the single-pair
response due to Pauli blocking. Second, neutrinos couple differently
to different baryonic species; consequently, the net rates will depend
on the individual concentrations. \\
\noindent (2) {\it In-medium dispersion relations}:
At high density, the single-particle spectra are
significantly modified from their noninteracting forms due to effects
of strong interactions.  Interacting matter features
smaller effective baryon masses and energy shifts relative to non-interacting
matter. \\
\noindent (3) {\it Correlations}: Repulsive particle-hole interactions
and Coulomb interactions
generally result in a screened dielectric response and also lead to
collective excitations in matter.  These effects may be
calculated using the Random Phase Approximation (RPA),  in which ring
diagrams are summed to all orders. Model calculations
\cite{redd97a,prak97b,redd99a,sawy75,sawy89,iwam82,horo91,raff95,sigl96}
indicate that at high density the neutrino
cross sections are suppressed relative to  the case in which these
effects are ignored.   In addition, these correlations enhance the
average energy transfer in neutrino-nucleon collisions.  Improvements in
determining the many-body dynamic form factor and assessing the role
of particle-particle interactions in dense matter at finite
temperature are  necessary before the full effects of many-body
correlations may be ascertained. \\

The relative importance of the various effects described above on
neutrino transport is only beginning to be studied systematically. As
a first step, we will focus on effects due to modifications (1)
through (3) above.  To see how this is accomplished, we start with a
general expression for the differential cross section~\cite{FW,DS}
\begin{eqnarray}
\frac {1}{V} \frac {d^3\sigma}{d^2\Omega_3 dE_3} &=&  -\frac {G_F^2}{128\pi^2}
\frac{E_3}{E_1}~
\left[1-\exp{\left(\frac{-q_0-(\mu_2-\mu_4)}{T}\right)}\right]^{-1}~ 
\nonumber \\ &\times &
(1-f_3(E_3))~{\rm Im}~~(L^{\alpha\beta}\Pi^R_{\alpha\beta}) \,,
\label{dcross}
\end{eqnarray}
where the incoming neutrino energy is $E_{1}$ and the outgoing
electron energy is $E_{3}$. The factor
$[1-\exp((-q_0-\mu_2+\mu_4)/T)]^{-1}$ maintains detailed balance, for
particles labeled `2' and '4' which are in thermal equilibrium at
temperature $T$ and in chemical equilibrium with chemical potentials
$\mu_2$ and $\mu_4$, respectively. The final state blocking of the
outgoing lepton is accounted for by the Pauli blocking factor
$(1-f_3(E_3))$. The lepton tensor $L_{\alpha\beta}$ is given by
\begin{equation}
L^{\alpha\beta}= 8[2k^{\alpha}k^{\beta}+(k\cdot q)g^{\alpha\beta}
-(k^{\alpha}q^{\beta}+q^{\alpha}k^{\beta})\mp i\epsilon^{\alpha\beta\mu\nu}
k^{\mu}q^{\nu}]
\end{equation}
The target particle retarded polarization tensor is
\begin{equation}
{\rm Im} \Pi^R_{\alpha\beta} =
\tanh{\left(\frac{q_0+(\mu_2-\mu_4)}{2T}\right)} {\rm Im}~\Pi_{\alpha\beta}
\,,\\
\end{equation}
where $\Pi_{\alpha\beta}$ is the time ordered or causal polarization and is
given by
\begin{equation}
\Pi_{\alpha\beta}=-i \int
\frac{d^4p}{(2\pi)^4} {\rm Tr}~[T(G_2(p)J_{\alpha} G_4(p+q)J_{\beta})]\,.
\end{equation}
Above, $k_{\mu}$ is the incoming neutrino four-momentum and $q_{\mu}$
is the four-momentum transfer. In writting the lepton tensor, we have
neglected the electron mass term, since typical electron energies are
of the order of a few hundred MeV.  The Greens' functions $G_i(p)$
(the index $i$ labels particle species) describe the propagation of
baryons at finite density and temperature.  The current operator
$J_{\mu}$ is $\gamma_{\mu}$ for the vector current and
$\gamma_{\mu}\gamma_5$ for the axial current.  Effects of strong and
electromagnetic correlations may be calculated by utilizing the RPA
polarization tensor 
\be 
\Pi^{RPA} = \Pi + \Pi^{RPA} D \Pi~, 
\ee 
where $D$ denotes the interaction matrix, in Eq.~(\ref{dcross}) (see 
\cite{redd99a} for more details).

\begin{figure}
\begin{center}
\includegraphics[scale=.4]{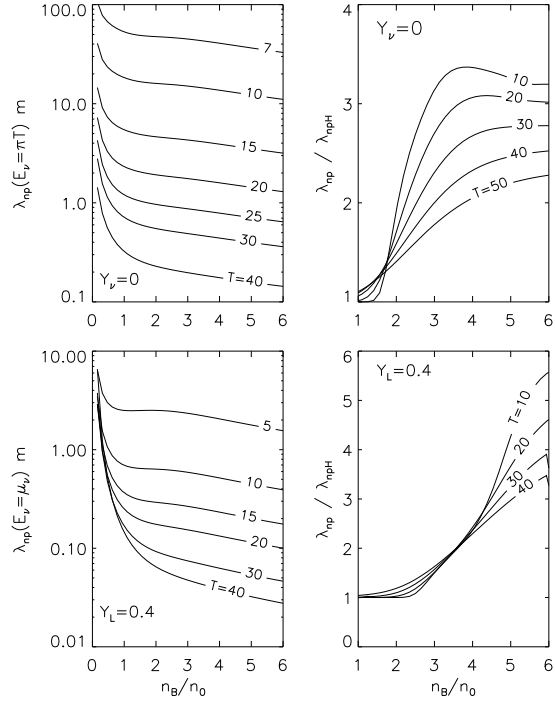}
\end{center}
\caption{
Neutrino mean free paths in matter with nucleons only (left panels).   
Right panels show ratios of mean free paths  in matter without and
with hyperons. Abscissa is  baryon density $n_B$ ($n_0$ is the nuclear
equilibrium density). Top panels show scattering mean free paths  common 
to all
neutrino species.  The bottom panels show results for electron neutrino mean
free paths where absorption reactions are included. The neutrino  content is
labelled in the different panels}
\label{sig}
\end{figure}

\subsection{Neutrino Mean Free Paths}

The differential cross section (Eq.~(\ref{dcross})) is needed in
multi-energy 
group neutrino transport codes. However, more approximate neutrino
transport algorithms (as in Section 2) require the total cross section
as a function of the neutrino energy for the calculation of diffusion
coefficients. The cross section per unit volume of
matter (or equivalently the inverse mean free path) is obtained by 
integrating $E_3$ and $\Omega_3$ in Eq.~(\ref{dcross}).

Under degenerate conditions even modest changes to the composition
significantly alter the neutrino scattering and absorption mean free
paths.  In Fig.~\ref{sig}, the neutrino scattering and absorption mean
free paths are shown for models GM3np and GM3npH relevant to the
deleptonization and cooling epochs. The top panels show the scattering
mean free paths common to all neutrino species in neutrino free
matter. The scattering mean free paths for thermal neutrinos
($E_{\nu}=\pi T$) is shown in the left panel for various
temperatures. To study the influence of hyperons, the ratio of the
$\lambda_{np}/ \lambda_{npH}$ is shown in the right panels.  The
presence of hyperons significantly increase the scattering cross
sections, by a factor $\sim (2- 3)$. Similar results for the
absorption cross sections are shown in the lower panels for $Y_L=0.4$.
Again we notice a significant enhancement (right panel) when hyperons
appear, the factor here could be as large as 5.

\begin{figure}
\begin{center}
\leavevmode
\includegraphics[scale=.3]{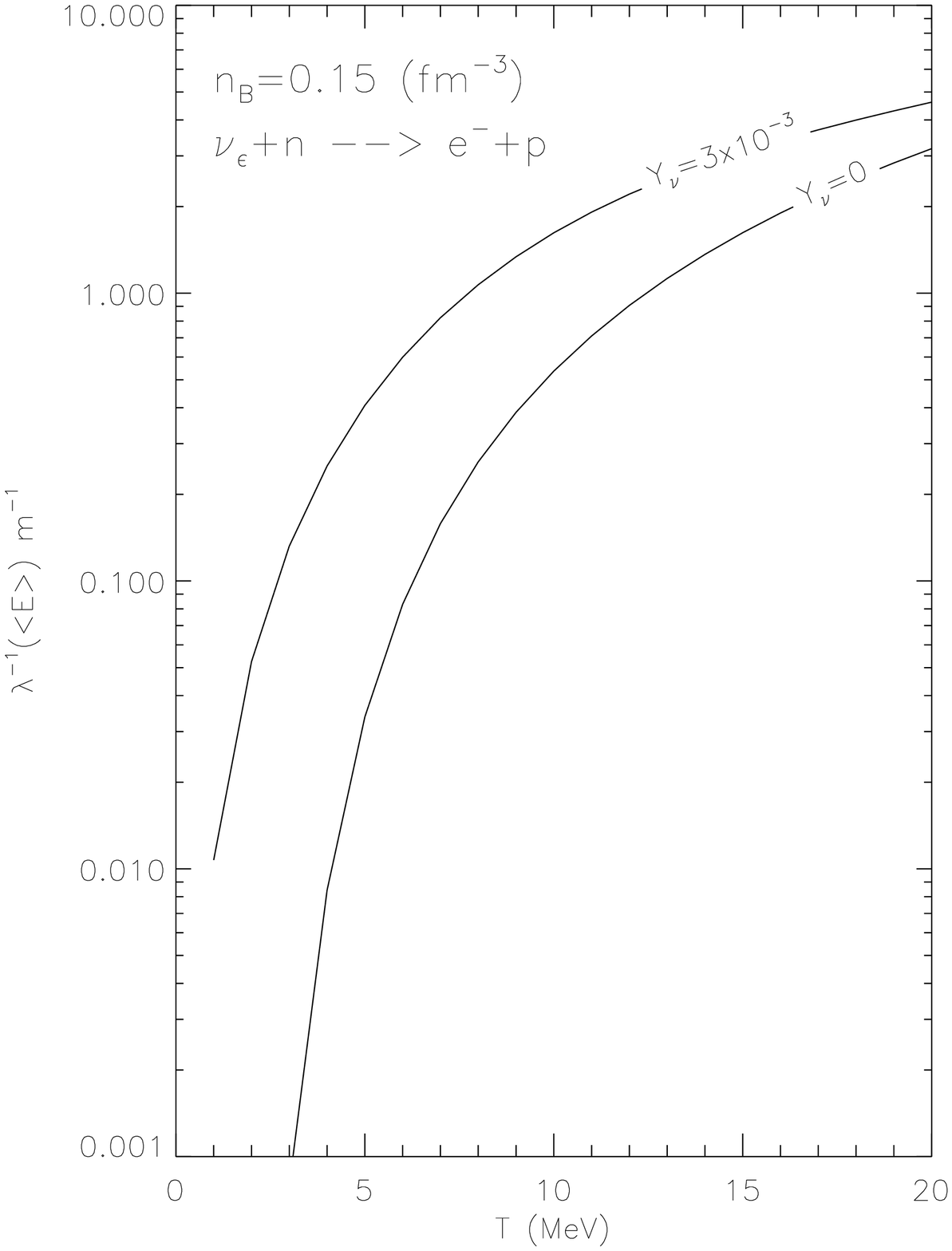}
\includegraphics[scale=.3]{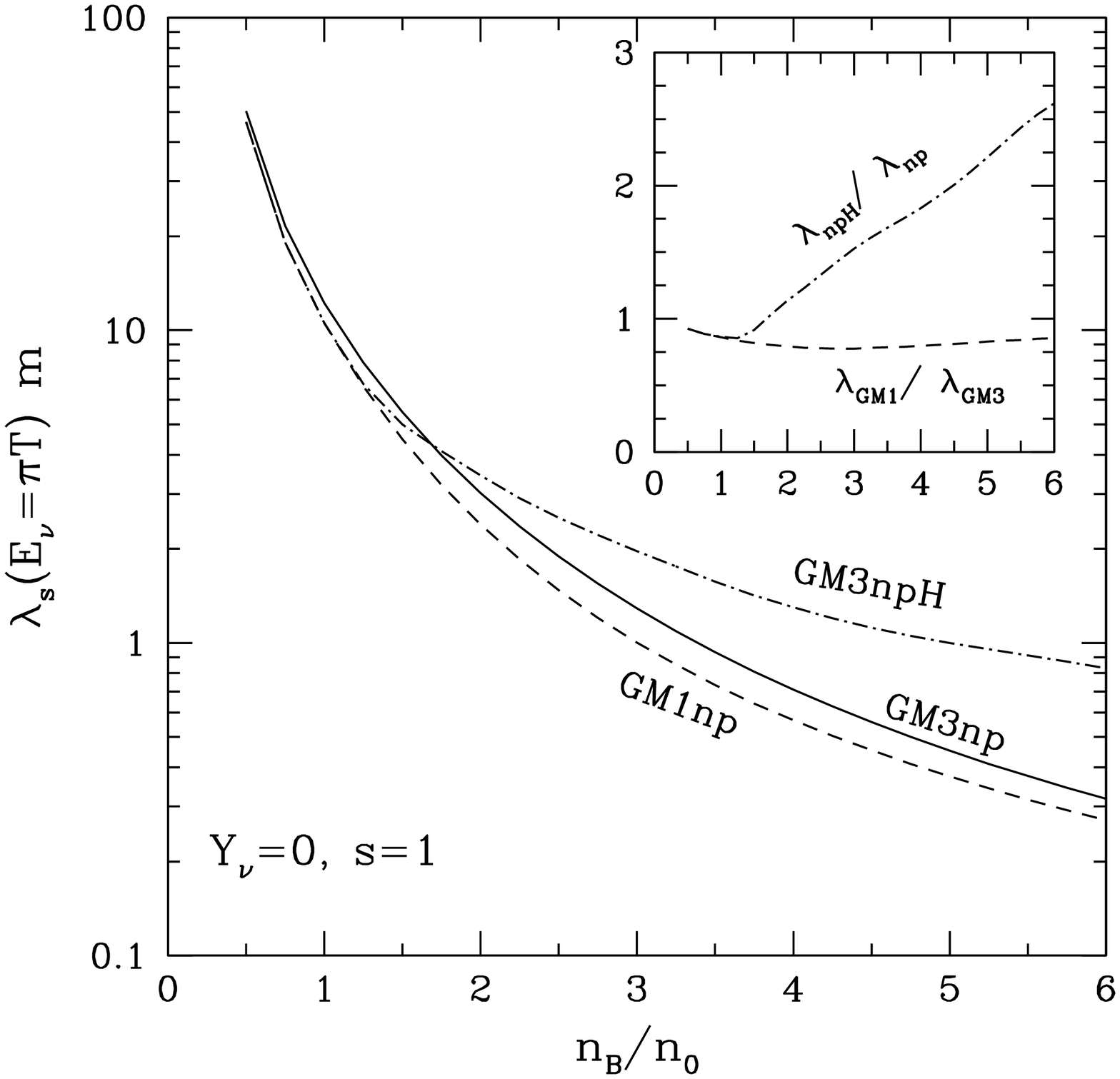}
\end{center}
\caption{ Left: Charged current inverse neutrino mean free paths
versus temperature. Right: Comparison of scattering mean free paths in
neutrino poor matter at fixed entropy for different EOSs in matter
containing nucleons and also hyperons}
\label{csig}
\end{figure}

During the deleptonization stage, lepton number transport is sensitive
to charged current reactions which dominate scattering reactions. At
zero temperature, charged current reactions $\nu + n \leftrightarrow e
+ p$ depend sensitively on the proton fraction $Y_p$~\cite{LPPH91}.
  Kinematic restrictions require $Y_p$ to be larger than
$11-14\%$ (direct Urca threshold).  At early times, a finite neutrino
chemical potential favors a large $Y_p$ throughout the star, which
enables these reactions to proceed without any hindrance. Toward the
late stages, however, $Y_p$ decreases with decreasing $\mu_{\nu}$ and
charged current reactions may be naively expected to become
inoperative. The threshold density for the charged current reaction
when $\mu_{\nu}=0$ and $T=0$ depends sensitively on the density
dependence of the nuclear symmetry energy.  In field-theoretical
models, in which the symmetry energy is largely given by contributions
due to $\rho$-meson exchange, the critical density is typically
$n_B=2\sim3 n_0$. However, finite temperatures favor larger $Y_p$'s
and increase the average neutrino energy enabling the charged current
reactions to proceed even below these densities.  Fig.~\ref{csig}
shows that this is the case even at relatively low temperatures
($T\sim 3-5$) MeV for a baryon density $n_B=0.15~{\rm fm}^{-3}$. The
sharp rise with temperature, which occurs even for $Y_{\nu}=0$,
clearly indicates that this reaction dominates the $\nu_e$ opacity
even during the late deleptonization era.  Thus, charged current
reactions cannot be simply turned off when the neutrino chemical
 potential becomes small enough as was done in prior PNS simulations
\cite{burr86}.

The EOS and neutrino mean free paths are intimately related, which is best
illustrated by  comparing the results shown in Fig.~\ref{sig} with those
shown in Fig.~\ref{csig}.  Composition and the baryon effective masses
influence both the neutrino mean free paths and the matter's specific heat.
Hyperons decrease the neutrino mean free paths at constant temperature
Fig.~\ref{sig}. This trend is reversed at constant entropy due to the
significantly lower temperatures favored in npH matter. Similar effects are
apparent  when we compare np models with different baryon effective  masses. At
a constant temperature, the larger effective mass in model GM3np favors
larger cross sections, while at constant entropy this trend is again
reversed due to the lower temperatures favored by the larger specific heat.

The diffusion coefficients are calculated using Eq.~(\ref{d2d3}) with the
cross sections discussed above.   The diffusion coefficients $D_2,D_3$,
and $D_4$ are functions of $n_B$, $T$, and $Y_{\nu_e}$.

\subsection{Inhomogeneous Phases: Effects of First Order Transitions}

The thermodynamics of the two situations, first order kaon
condensation \cite{GS99,GS98} and the quark-hadron transition
\cite{PCL}, has been previously considered. Reddy, Bertsch and Prakash
\cite{RBP} have studied the effects of inhomogeneous phases on
$\nu$-matter interactions. Based on simple estimates
of the surface tension between nuclear matter and the exotic phase,
typical droplet sizes range from $5-15$ fm \cite{size}, and
inter-droplet spacings range up to several times larger.  The
propagation of neutrinos whose wavelength is greater than the typical
droplet size and less than the inter-droplet spacing, i.e.,
$2{\rm~MeV}\lsim E_\nu\lsim40{\rm~MeV}$, will be greatly affected by the
heterogeneity of the mixed phase, as a consequence of the coherent
scattering of neutrinos from the matter in the droplet.

The Lagrangian that describes the neutral current coupling of
neutrinos to the droplet is
\begin{equation}
{\mathcal{L}}_W = \frac{G_F}{2\sqrt{2}} ~\bar{\nu}\gamma_\mu(1-\gamma_5) \nu
~J^{\mu}_D \,,
\label{nuD}
\end{equation}
where $J^{\mu}_D$ is the neutral current carried by the droplet and $G_F=1.166
\times 10^{-5}$ GeV$^{-2}$ is the Fermi weak coupling constant. For
non-relativistic droplets, $J^{\mu}_D = \rho_W(x)~\delta^{\mu 0}$ has
only a time like component. Here, $\rho_W(x)$ is the excess weak
charge density in the droplet. The total weak charge enclosed in a
droplet of radius $r_d$ is $N_W=\int_0^{r_d} d^3x ~\rho_W(x)$
and the form factor is $F(q)=(1/N_W)\int_0^{r_d} d^3x ~\rho_W(x)~
\sin{qx}/qx$. The differential cross section for neutrinos scattering
from an isolated droplet is then 
\begin{equation}
\frac{d\sigma}{d\cos{\theta}}= \frac{E_\nu^2}{16\pi}
G_F^2 N^2_W(1+\cos{\theta}) F^2(q) \,.
\label{diff}
\end{equation}
In the above equation, $E_\nu$ is the neutrino energy and $\theta$ is the
scattering angle. Since the droplets are massive, we consider only elastic
scattering for which the magnitude of the momentum transfer is 
$q=\sqrt{2}E_\nu(1-\cos{\theta})$.

We must embed the droplets into the medium to evaluate the neutrino
transport parameters.  The droplet radius $r_d$ and the inter-droplet
spacing are determined by the interplay of surface and Coulomb
energies.  In the Wigner-Seitz approximation, the cell radius is
$R_W=(3/4\pi N_D)^{1/3}$ where the droplet density is $N_D$. Except
for one aspect, we will neglect coherent scattering from more than one
droplet. If the droplets form a lattice, Bragg scattering will
dominate and our description would not be valid. But for low density
and a liquid phase, interference from multiple droplets affects
scattering only at long wavelengths. If the ambient temperature is not
small compared to the melting temperature, the droplet phase will be a
liquid and interference effects arising from scattering off different
droplets are small for neutrino energies $E_\nu \gsim
(1/R_W)$. However, multiple droplet scattering cannot be neglected for
$E_\nu \lsim 1/R_W$. The effects of other droplets is to cancel
scattering in the forward direction, because the interference is
destructive except at exactly zero degrees, where it produces a change
in the index of refraction of the medium. These effects are usually
incorporated by multiplying the differential cross section
Eq.~(\ref{diff}) by the static form factor of the medium. The static
form factor, defined in terms of the radial distribution function of
the droplets, $g(r)$, is
\begin{equation}
S(q)= 1 + N_D \int d^3r \exp{i \vec{q}.\vec{r}}~(g(r)-1) \,.
\end{equation}
The droplet correlations, which determine $g(r)$, arise due to the
Coulomb force and is measured in terms of the dimensionless Coulomb
number \newline $\Gamma=Z^2e^2/(8\pi R_W kT)$. Due to the long-range character
of the Coulomb force, the role of screening and the finite droplet
size, $g(r)$ cannot be computed analytically. We use a simple ansatz
for the radial distribution function $g(r< R_W) = 0$ and
$g(r>R_W)=1$. For this choice, the structure factor is independent of
$\Gamma$. Monte Carlo calculations \cite{CJH} of the liquid structure
function of a simple one component plasma indicate that our
choice of $S(q)$ is conservative for typical neutrino energies of
interest.

The simple ansatz for $g(r)$ is equivalent to subtracting, from the weak
charge density $\rho_W$, a uniform density which has the same total weak
charge $N_W$ as the matter in the Wigner-Seitz cell. Thus, effects due
to $S(q)$ may be incorporated by replacing the form factor $F(q)$ by
\begin{eqnarray}
F(q) \rightarrow \tilde{F}(q) = F(q) - 3~
\frac{\sin{qR_{W}} - (qR_{W})\cos{qR_{W}}}{(q R_{W})^3}  \,.
\label{formc}
\end{eqnarray}
The neutrino--droplet differential cross section per unit volume
then follows:
\begin{equation}
\frac{1}{V}\frac{d\sigma}{d\cos{\theta}}=
N_D~\frac{E_\nu^2}{16\pi} G_F^2 N^2_W(1+\cos{\theta}) \tilde{F}^2(q) \,.
\label{diff1}
\end{equation}
Note that even for small droplet density $N_D$, the factor $N_W^2$ acts to
enhance the droplet scattering. To quantify the importance of droplets as a
source of opacity, we compare with the standard scenario in which matter is
uniform and composed of neutrons. The dominant source of opacity
is then due to scattering from thermal fluctuations and 
\begin{eqnarray}
\frac {1}{V}\frac{d\sigma}{d\cos{\theta}} &=&
\frac{G_F^2}{8\pi}\left(c_V^2(1+\cos{\theta})
+(3-\cos{\theta})c_A^2\right) ~ E_{\nu}^2 \nonumber \\
&\times& \frac{3}{2}~ n_n ~\left[\frac{k_BT}{E_{fn}}\right]\,,
\label{diff2}
\end{eqnarray}
where $c_V$ and $c_A$ are respectively the vector and axial coupling
constants of the neutron, $n_n$ is the neutron number density,
$E_{fn}$ is the neutron Fermi energy and $T$ is the matter temperature
\cite{IP}.

The transport cross sections that are employed in studying the
diffusive transport of neutrinos in the core of a neutron star are
differential cross sections weighted by the angular factor
$(1-\cos{\theta})$. The transport mean free path $\lambda(E_\nu)$ for
a given neutrino energy $E_\nu$ is given by
\begin{eqnarray}
\frac{1}{\lambda(E_\nu)}=\frac{\sigma_T(E_\nu)}{V} =
\int d\cos{\theta}~ (1-\cos{\theta})
\left[\frac{1}{V} \frac{d\sigma}{d\cos{\theta}}\right] \,.
\label{fint}
\end{eqnarray}

\begin{figure}[t]
\begin{center}
\leavevmode
\includegraphics[scale=0.28,angle=0]{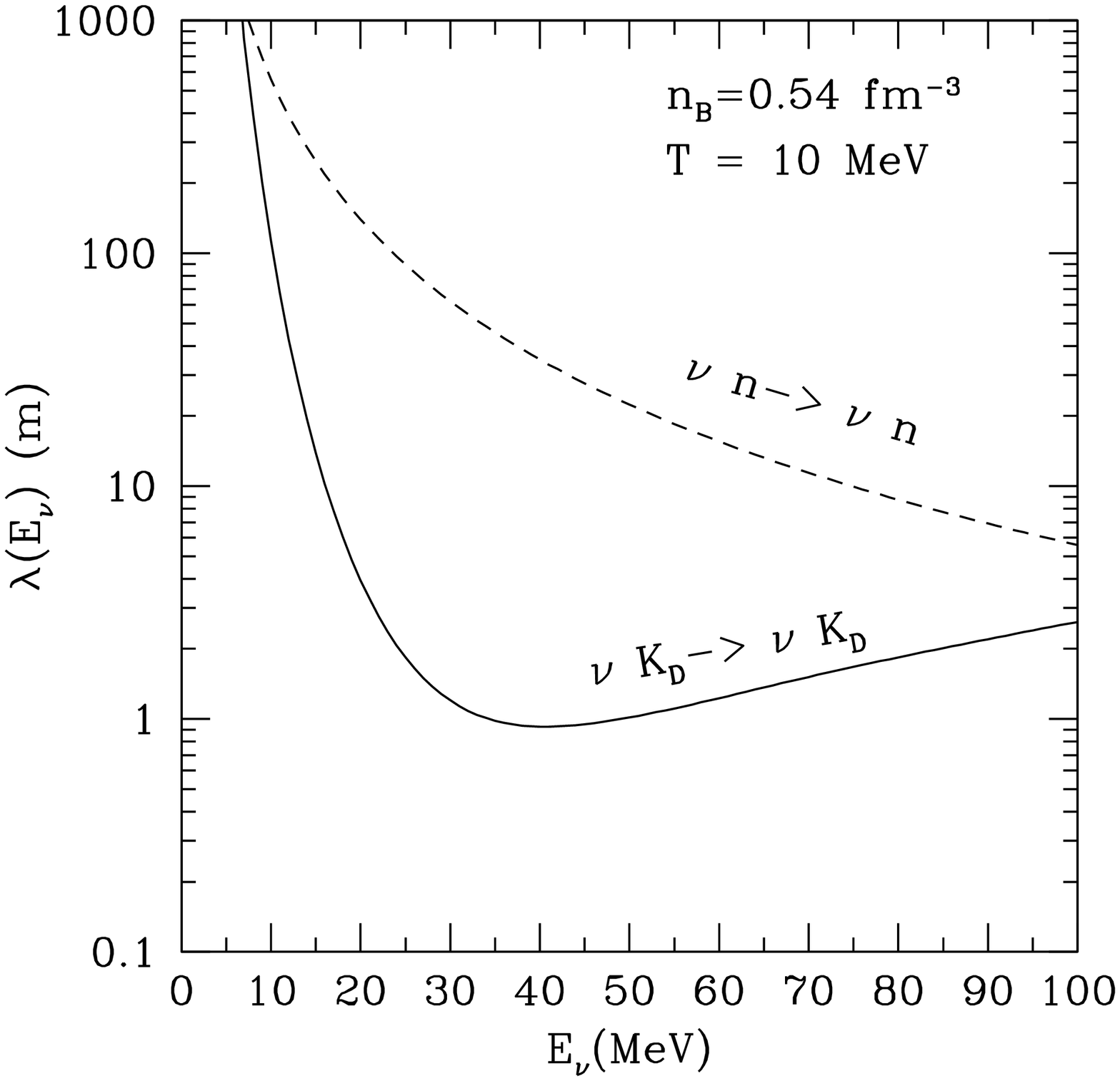}
\includegraphics[scale=0.28,angle=0]{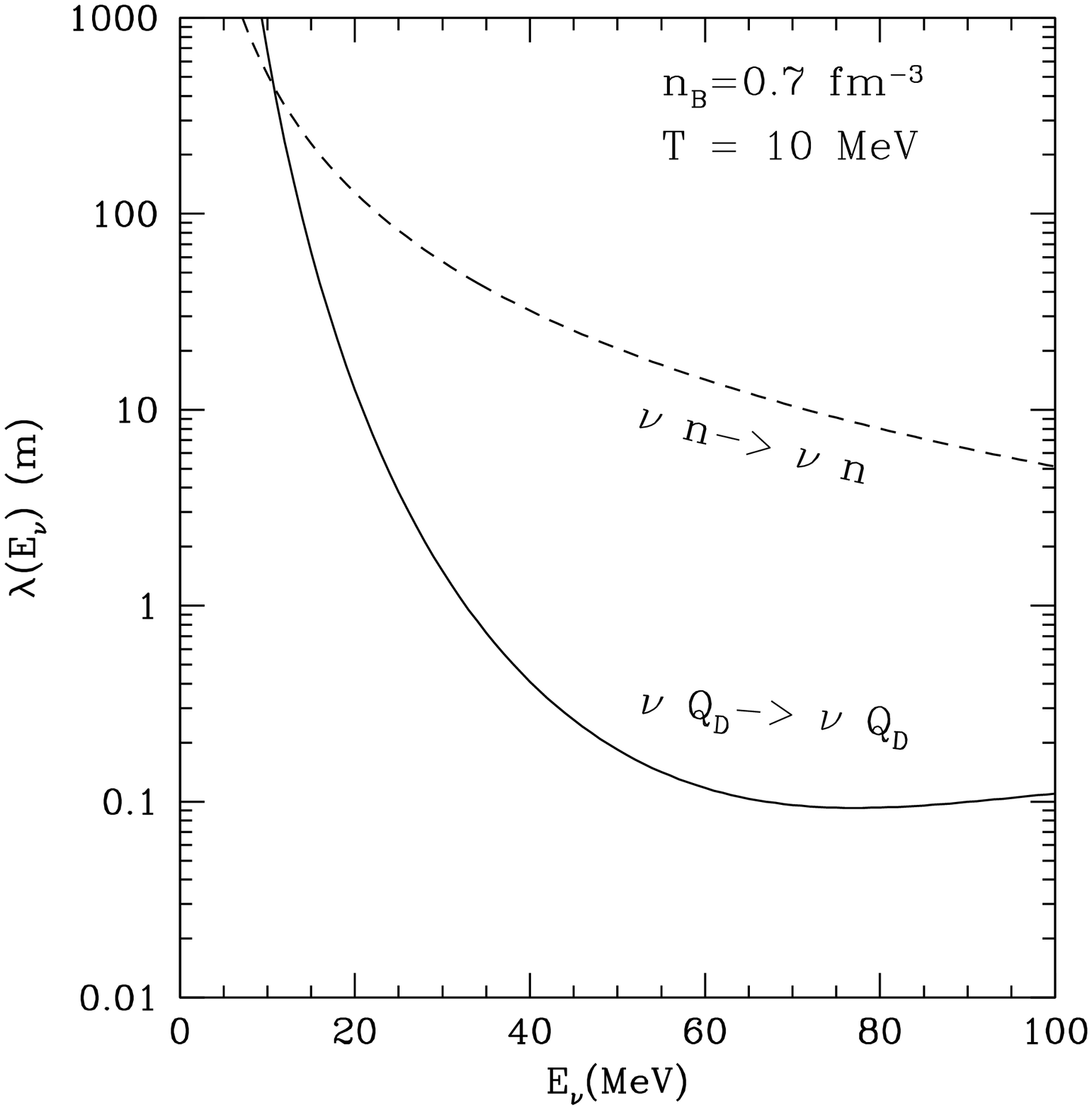}
\end{center}
\caption{Neutrino mean free paths as a function of neutrino energy.
Solid lines are for matter in a mixed phase containing kaons (left
panel) and quarks (right panel), and dashed curves are for uniform matter.}
\label{mpaths}
\end{figure}

Models of first order phase transitions in dense matter provide the
weak charge and form factors of the droplets and permit the evaluation
of $\nu$--droplet scattering contributions to the opacity of the mixed
phase. For the models considered, namely the first order kaon
condensate and the quark-hadron phase transition, the neutrino mean
free paths in the mixed phase are shown in the left and right panels of
Fig.~\ref{mpaths}, respectively.  The results are shown for the
indicated values of the baryon density $n_B$ and temperature $T$ where
the model predicts a mixed phase exists. The kaon droplets are
characterized by radii $r_d\sim7$ fm and inter-droplet spacings
$R_W\sim20$ fm, and enclose a net weak vector charge $N_W \sim
700$. The quark droplets are characterized by $r_d\sim5$ fm and
$R_W\sim11$ fm, and an enclosed weak charge $N_W \sim 850$.  For
comparison, the neutrino mean free path in uniform neutron matter at
the same $n_b$ and $T$ are also shown.  It is apparent that there is a
large coherent scattering-induced reduction in the mean free path for
the typical energy $E_\nu \sim \pi T$. At much lower energies, the
inter-droplet correlations tend to screen the weak charge of the
droplet, and at higher energies the coherence is attenuated by the
droplet form factor.

The large reduction in neutrino mean free path found here implies that the
mixed phase will cool significantly slower than homogeneous
matter. Consequently, the observable neutrino luminosity at late times might be
affected as it is driven by the transport of energy from the deep interior.
The reduced mean free path in the interior will tend to
prolong the late time neutrino emission phase.

\subsection{Effects of Quark Superconductivity and Superfluidity}

Recent theoretical works \cite{gap,qsf0} suggest that quarks form
Cooper pairs in medium, a natural consequence of attractive
interactions destabilizing the Fermi surface. Although the idea of
quark pairing in dense matter is not new \cite{gap,Bailin84}, it has
recently seen renewed interest in the context of the phase diagram of
QCD \cite{qsf0}.  Model calculations, mostly based on four-quark
effective interactions, predict the restoration of spontaneously
broken chiral symmetry through the onset of color superconductivity at
low temperatures.  They predict an energy gap of $\Delta \sim 100$ MeV
for a typical quark chemical potential of $\mu_q \sim 400 $ MeV.  As
in BCS theory, the gap will weaken for $T > 0$, and at some critical
temperature $T_c$ there is a (second-order) transition to a
``standard'' quark-gluon plasma.  During cooling from an initial
temperature in excess of $T_c$, the formation of a gap in the
fermionic
excitation spectrum in quark matter will influence
various transport properties of the system. Carter and Reddy
have studied its
influence on the transport of neutrinos \cite{CR00}.

The differential neutrino scattering cross section per unit volume in an
infinite and homogeneous system of relativistic fermions as calculated in
linear response theory is given by Eq.~(\ref{dcross}).
The medium is characterized by the quark polarization
tensor $\Pi_{\alpha\beta}$.  
In the case of free quarks, each flavor contributes a term of the form
\begin{equation}
\Pi_{\alpha\beta}(q)=-i {\rm Tr}_c \int
\frac{d^4p}{(2\pi)^4} {\rm Tr}~[S_0(p)\Gamma_{\alpha} 
S_0(p+q)\Gamma_{\beta}] \,, 
\label{pi_free}
\end{equation}
where $S_0(p)$ is the free quark propagator at finite chemical potential and
temperature.  
The outer trace is over color and simplifies to a $N_c = 3$ degeneracy.
The inner trace is over spin, and the
$\Gamma_\alpha$ are the neutrino-quark vertex functions which determine
the spin channel.
Specifically, the vector polarization is computed by choosing
$(\Gamma_{\alpha}, \Gamma_{\beta}) = ( \gamma_{\alpha}, \gamma_{\beta} )$.
The axial and mixed vector-axial polarizations are similarly obtained from 
$(\Gamma_{\alpha}, \Gamma_{\beta}) = (\gamma_{\alpha}\gamma_5,
\gamma_{\beta}\gamma_5)$ and $(\Gamma_{\alpha}, \Gamma_{\beta}) =
(\gamma_{\alpha}, \gamma_{\beta}\gamma_5)$, respectively.

The free quark propagators in Eq.~(\ref{pi_free}) are naturally
modified in a superconducting medium.  As first pointed out by
Bardeen, Cooper, and Schrieffer several decades ago, the
quasi-particle dispersion relation is modified due to the presence of
a gap in the excitation spectrum.  In calculating these effects, we
will consider the simplified case of QCD with two quark flavors which
obey SU(2)$_L \times$ SU(2)$_R$ flavor symmetry, given that the light
$u$ and $d$ quarks dominate low-energy phenomena.  Furthermore we will
assume that, through some unspecified effective interactions, quarks
pair in a manner analogous to the BCS mechanism \cite{BCS}.  The
relevant consequences of this are the restoration of chiral symmetry
(hence all quarks are approximately massless) and the existence of an
energy gap at zero temperature, $\Delta_0$, with approximate
temperature dependence,
\begin{equation}
\Delta(T) = \Delta_0 \sqrt{ 1 - \left(\frac{T}{T_c}\right)^2 }.
\end{equation}
The critical temperature $T_c \simeq 0.57 \Delta_0$ is likewise taken from 
BCS theory; this relation has been shown to hold for perturbative QCD 
\cite{tc} and is thus a reasonable assumption for non-perturbative physics.

Breaking the SU$_c$(3) color group leads to complications not found in
electrodynamics.  In QCD the superconducting gap is equivalent to a diquark
condensate, which can at most involve two of the three fundamental quark
colors.  The condensate must therefore be colored.  Since the scalar diquark
(in the $\bar{\bf 3}$ color representation) appears to always be the most
attractive channel, we consider the anomalous (or Gorkov) propagator
\cite{propagators}
\begin{eqnarray}
F(p)_{a b f g} &=&  
\langle q_{f a}^T(p) C\gamma_5 q_{g b}(-p) \rangle \nonumber\\
&=& -i \epsilon_{a b 3} \epsilon_{fg} 
\Delta \left(\frac{\Lambda^+(p)}{p_o^2 - \xi_p^2} + 
\frac{\Lambda^-(p)}{p_o^2 - \bar{\xi}_p^2}\right) \gamma_5~C\,. 
\label{a_bcs}
\end{eqnarray}
Here, $a,b$ are color indices, $f,g$ are flavor indices,
$\epsilon_{abc}$ is the usual anti-symmetric tensor and we have
conventionally chosen 3 to be the condensate color.  This propagator
is also antisymmetric in flavor and spin, with $C=-i\gamma_0\gamma_2$
being the charge conjugation operator.

The color bias of the condensate forces a splitting of the normal
quark propagator into colors transverse and parallel to the diquark.
Quarks of color 3, parallel to the condensate in color space, will be
unaffected and propagate freely, with
\begin{equation}
S_0(p)^{b g}_{a f}= i\delta_a^b \delta^g_f~
\left(\frac{\Lambda^+(p)}{p_o^2 - E_p^2} + 
\frac{\Lambda^-(p)}{p_o^2 - \bar{E}_p^2}\right)
~(p_\mu\gamma^\mu -\mu \gamma_0)\,.
\label{s_0}
\end{equation}
This is written in terms of the particle and anti-particle projection
operators $\Lambda^+(p)$ and $\Lambda^-(p)$ respectively, where
$\Lambda^{\pm}(p)=(1 \pm \gamma_0\vec{\gamma} \cdot \hat{p})/2$.  The
excitation energies are simply $E_p = |\vec{p}|-\mu$ for quarks and
$E_p = |\vec{p}|+\mu$ for anti-quarks.

On the other hand, transverse quark colors 1 and 2 participate in the diquark
and thus their quasi-particle propagators are given as
\begin{equation}
S(p)^{b g}_{a f}= i\delta_a^b \delta^g_f~
\left(\frac{\Lambda^+(p)}{p_o^2 - \xi_p^2} + 
\frac{\Lambda^-(p)}{p_o^2 - \bar{\xi}_p^2}\right)
~(p_\mu\gamma^\mu -\mu \gamma_0)\,.
\label{s_bcs}
\end{equation}
The quasi-particle energy is $\xi_p = \sqrt{(|\vec{p}|-\mu)^2 +
\Delta^2}$, and for the anti-particle $\bar\xi_p =
\sqrt{(|\vec{p}|+\mu)^2 + \Delta^2}$.

\begin{figure}[t]
\begin{center}
\includegraphics[scale=1.3,angle=0]{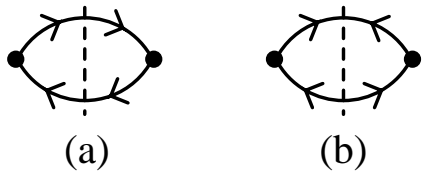}
\end{center}
\caption{Standard loop (a) and anomalous loop (b) diagrams contributing to 
the quark polarization operator}
\label{pol_fig}
\end{figure}

The appearance of an anomalous propagator in the superconducting phase
indicates that the polarization tensor gets contributions from both
the normal quasi-particle propagators Eq.~(\ref {s_bcs}) and anomalous
propagator Eq.~(\ref{a_bcs}).  Thus, to order $G_F^2$,
Eq.~(\ref{pi_free}) is replaced with the two contributions
corresponding to the diagrams shown in Fig.~\ref{pol_fig}, and written
\begin{eqnarray}
\Pi_{\alpha\beta}(q) &=& -i \!\int\! \frac{d^4p}{(2\pi)^4} \left\{
{\rm Tr}~[S_0(p)\Gamma_{\alpha} S_0(p+q)\Gamma_{\beta}] 
\right. \nonumber \\ 
& \qquad + & \left.
2 {\rm Tr}~[S(p)\Gamma_{\alpha} S(p+q)\Gamma_{\beta}] 
+ 2 {\rm Tr}~[F(p)\Gamma_{\alpha} \bar{F}(p+q)\Gamma_{\beta}]
\right\} \,.
\label{pi_bcs}
\end{eqnarray}
The remaining trace is over spin, as the color trace has been
performed.  Fig.~\ref{pol_fig}(a) corresponds to the first two terms,
which have been decomposed into one term with ungapped propagators
Eq.~(\ref{s_0}) and the other with gapped quasi-particle propagators
Eq.~(\ref{s_bcs}).  Fig.~\ref{pol_fig}(b) represents the third,
anomalous term.

For neutrino scattering we must consider vector, axial, and mixed
vector-axial channels, all summed over flavors.  
The full polarization, to be used in evaluating Eq.~(\ref{dcross}),
may be written
\begin{equation}
\Pi_{\alpha\beta} = \sum_f\,\left[ (C_V^f)^2 \Pi^V_{\alpha\beta} +
(C_A^f)^2 \Pi^A_{\alpha\beta} - 2 C_V^f C_A^f\Pi^{VA}_{\alpha\beta}\right]\,.
\label{polsum}
\end{equation}
The coupling constants for up quarks are $C_V^u =
\textstyle{\frac{1}{2}} - {\textstyle{\frac{4}{3}}} \sin^2\theta_W$
and $C_A^u = \textstyle{\frac{1}{2}}$ , and for down quarks, $C_V^d =
-\textstyle{\frac{1}{2}}+\textstyle{\frac{2}{3}}\sin^2\theta_W$ and
$C_A^d = -\textstyle{\frac{1}{2}}$, where $\sin^2\theta_W \simeq 0.23$
is the Weinberg angle.

\begin{figure}[t]
\begin{center}
\leavevmode
\includegraphics[scale=0.28,angle=0]{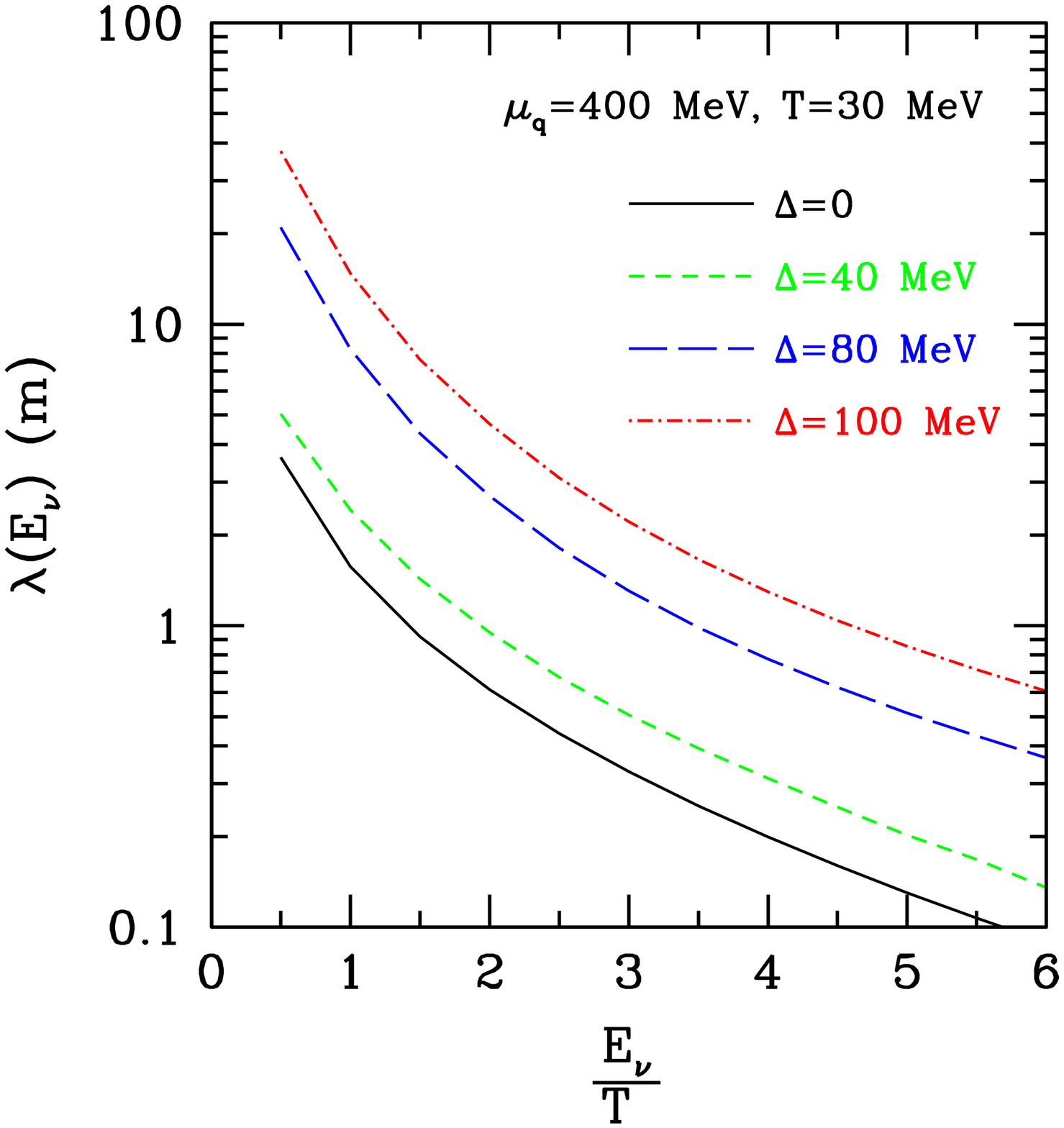}
\includegraphics[scale=0.28,angle=0]{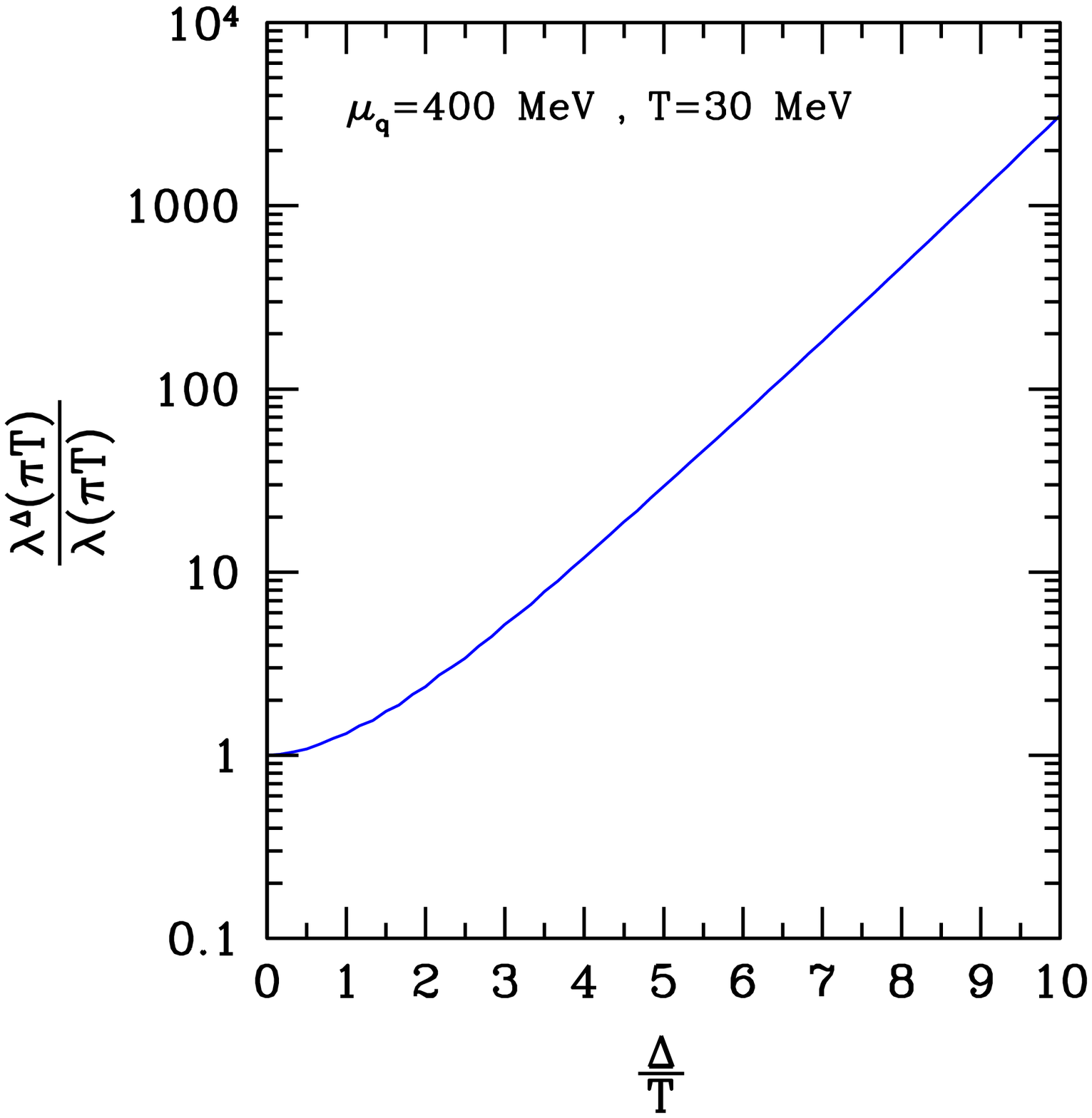}
\end{center}
\caption{Left panel: Neutrino mean free path as a function of neutrino
energy $E_\nu$.  Right panel: Neutrino mean free paths for $E_\nu=\pi
T$ as a function of $\Delta/T$.  These results are virtually
independent of temperature for $T \lsim 50$ MeV.}
\label{lambda}
\end{figure}

The differential cross section, Eq.~(\ref{dcross}) and the total cross
section are obtained by integrating over all neutrino energy transfers
and angles.  Results for the neutrino mean free path,
$\lambda=V/\sigma$, are shown in Fig.~\ref{lambda} as a function of
incoming neutrino energy $E_\nu$ (for ambient conditions of
$\mu_q=400$ MeV and $T=30$ MeV).  They show the same energy dependence
found previously for free relativistic and degenerate fermionic matter
\cite{redd98}; $\lambda\propto1/E_\nu^2$ for $E_\nu \gg T$ and
$\lambda\propto1/E_\nu$ for $E_\nu \ll T$.  The results indicate that
this energy dependence is not modified by the presence of a gap when
$\Delta \sim T$.  Thus, the primary effect of the superconducting phase
is a much larger mean free path.  This is consistent with the
suppression found in the vector-longitudinal response function, which
dominates the polarization sum Eq.~(\ref{polsum}), at $q_0<q$.

We now consider the cooling of a macroscopic sphere of quark matter, a
toy approximation for the core of a neutron star with a mixed quark
phase, as it becomes superconducting.  As in the preceding calculation, we
consider the relatively simple case of two massless flavors with
identical chemical potentials and disregard the
quarks parallel in color to the condensate; {\it i.e.} we consider a
background comprised exclusively of quasi-quarks.

The cooling of a spherical system of quark matter from $T \sim T_c \sim 50$ 
MeV is driven by neutrino diffusion, for
the neutrino mean free path is much smaller than the dimensions of system
of astrophysical size, and yet several orders of magnitude larger than
the mean free path of the quarks.
The diffusion equation for energy transport by neutrinos in a spherical
geometry is
\begin{eqnarray}
C_V\, \frac{dT}{dt}=-\frac{1}{r^2} \frac{\partial L_{\nu}}{\partial r} \,,
\label{ediff}
\end{eqnarray}
where $C_V$ is the specific heat per unit volume of quark matter, $T$
is the temperature, and $r$ is the radius.  The neutrino energy
luminosity for each neutrino type, $L_\nu$, depends on the neutrino
mean free path and the spatial gradients in temperature and is
approximated by an integral over neutrino energy $E_\nu$
\begin{equation}
L_\nu \cong 6\int dE_\nu\,\frac{c}{6\pi^2}\, E_\nu^3 r^2 \lambda(E_\nu)
\left| \frac{\partial f(E_\nu)}{\partial r} \right| \,.
\label{eflux}
\end{equation}
We assume that neutrino interactions are dominated by the neutral current
scattering which is common to all neutrino types, accounting for
the factor 6 in Eq.~(\ref{eflux}).

\begin{figure}[t]
\begin{center}
\includegraphics[scale=0.4,angle=0]{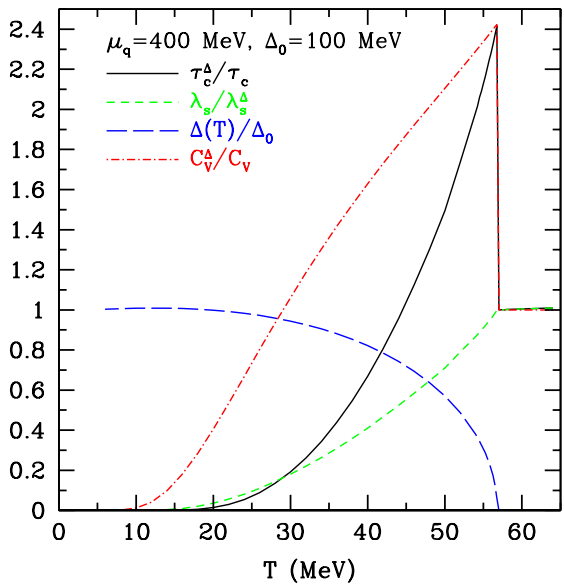}
\end{center}
\caption{The extent to which different physical quantities are
affected by a superconducting transition.  Ratios of the cooling time
scale (solid curve), the inverse mean free path (short-dashed curve)
and the matter specific heat (dot-dashed curve) in the superconducting
phase to that in the normal phase is shown as a function of the matter
temperature. The ratio of the gap to its zero temperature value
$\Delta_0$ is also shown (long-dashed curve).}
\label{temp}
\end{figure}

The solution to the diffusion equation will depend on the initial
temperature gradients.  However, being primarily interested in a
qualitative description of cooling through a second-order phase
transition to superconducting matter, the temporal behavior can be
characterized by a time scale $\tau_c$ which is proportional to the
inverse cooling rate.  This characteristic time is
\begin{equation}
\tau_c(T)= C_V(T) \frac{R^2}{c\langle\lambda(T)\rangle} \,,
\label{tauc}
\end{equation}
and is a strong function of the temperature since it depends on both
the specific heat $C_V$ and the energy-averaged neutrino mean free
path, $\langle\lambda(T)\rangle$.  The latter quantity is here
approximated by $\lambda(E_\nu=\pi T)$ since the neutrinos are in
thermal equilibrium.  Using BCS theory, as described in the previous
section, $\langle\lambda(T)\rangle$ depends on the gap $\Delta$ as
shown in the right panel of Fig.~\ref{lambda}.  The results indicate
that for small $\Delta/T$, $\lambda$ is not strongly modified, but as
$\Delta/T$ increases so too does $\lambda$, non-linearly at first and
then exponentially for $\Delta/T \gsim 5$.  Also, in the BCS theory,
the temperature dependences of $C_V$ (dashed curve) and $\Delta$
(dot-dashed curve) are shown in Fig.~\ref{temp}.
Finally, the ratio $\tau^{\Delta}_c(T)/\tau_c(T)$, a measure of the extent to
which the cooling rate is changed by a gap, is shown by the solid line in
Fig.~\ref{temp}.   We note that the
diffusion approximation is only valid when $\lambda \ll R$ and will
thus fail for very low temperatures, when $\lambda \lsim R$.

These results are readily interpreted.  The cooling rate around $T_c$
is influenced mainly by the peak in the specific heat associated with
the second order phase transition, since the neutrino mean free path
is not strongly affected when $T\ge\Delta$.  Subsequently, as the
matter cools, both $C_V$ and $\lambda^{-1}$ decrease in a non-linear
fashion for $T\sim\Delta$.  Upon further cooling, when $T\ll\Delta$,
both $C_V$ and $\lambda^{-1}$ decrease exponentially.  Both 
effects accelerate the cooling process.

We conclude that if it were possible to measure the neutrino
luminosity from the hypothetical object described here, a second order
superconducting transition might be identified by the temporal
characteristics of the late time supernova neutrino signal from a PNS.
Specifically, there might be a brief interval during which the cooling
would slow when the core temperature falls below $T_c$, signified by a
period of reduced neutrino detection.  However, this effect might be obscured by $\nu-$opaque matter outside the star's core.  If the neutrino
opacity of these outer regions of the star is large, it is likely that
any sharp temporal feature associated with neutrino transport in the
core will be diluted as the neutrino diffuse through the outer
regions.  Nevertheless, the main finding, which is that phase
transitions in the core can have a discernible impact on the transport
of neutrinos and suggests that the late time supernova neutrino signal
is a promising probe of the high density and low temperature region of
the QCD phase diagram.

\section{Results of PNS Simulations}

\begin{figure}[hbt]
\begin{center}
\leavevmode
\includegraphics[angle=0,scale=0.3]{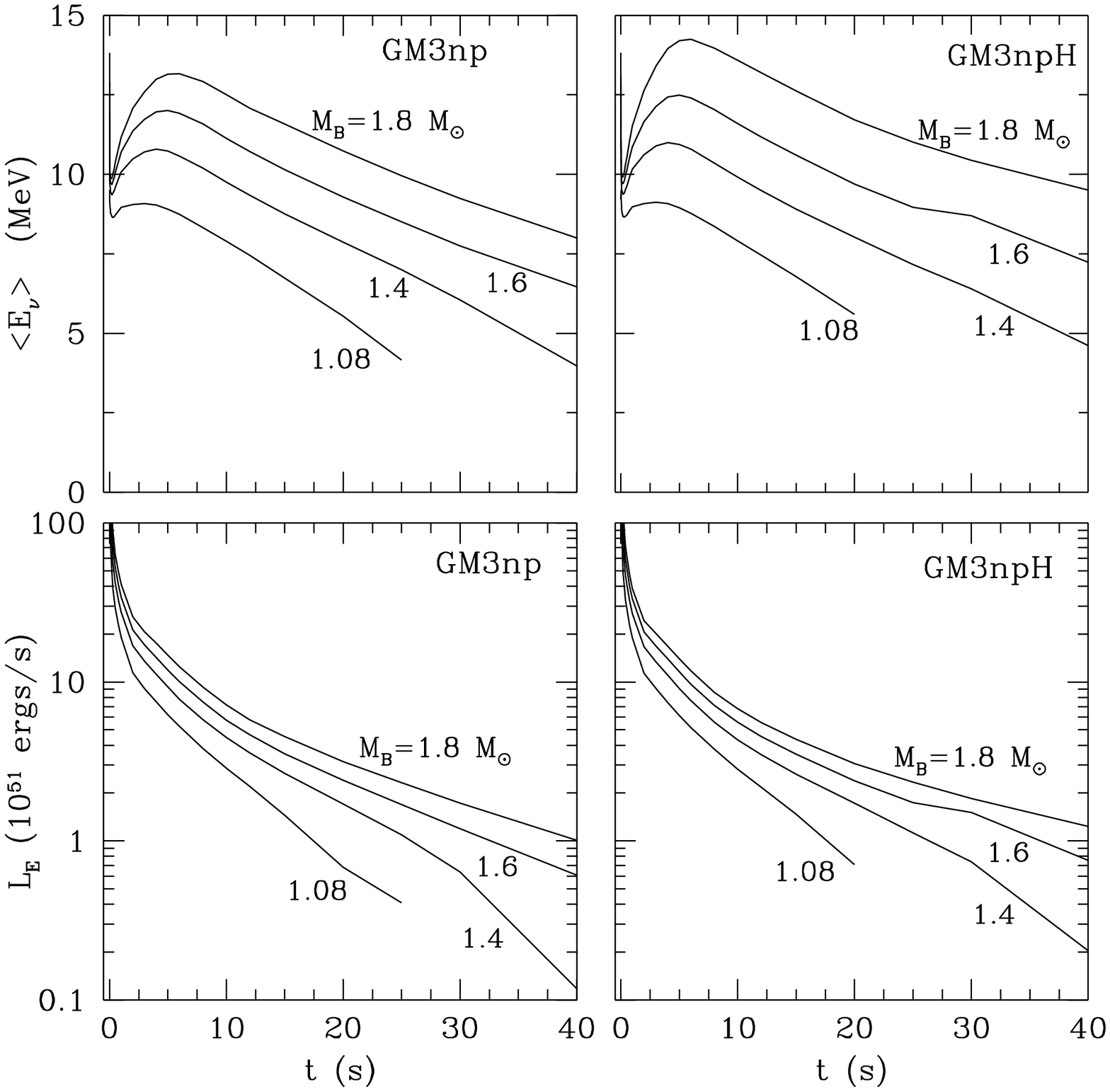}
\includegraphics[angle=0,scale=0.3]{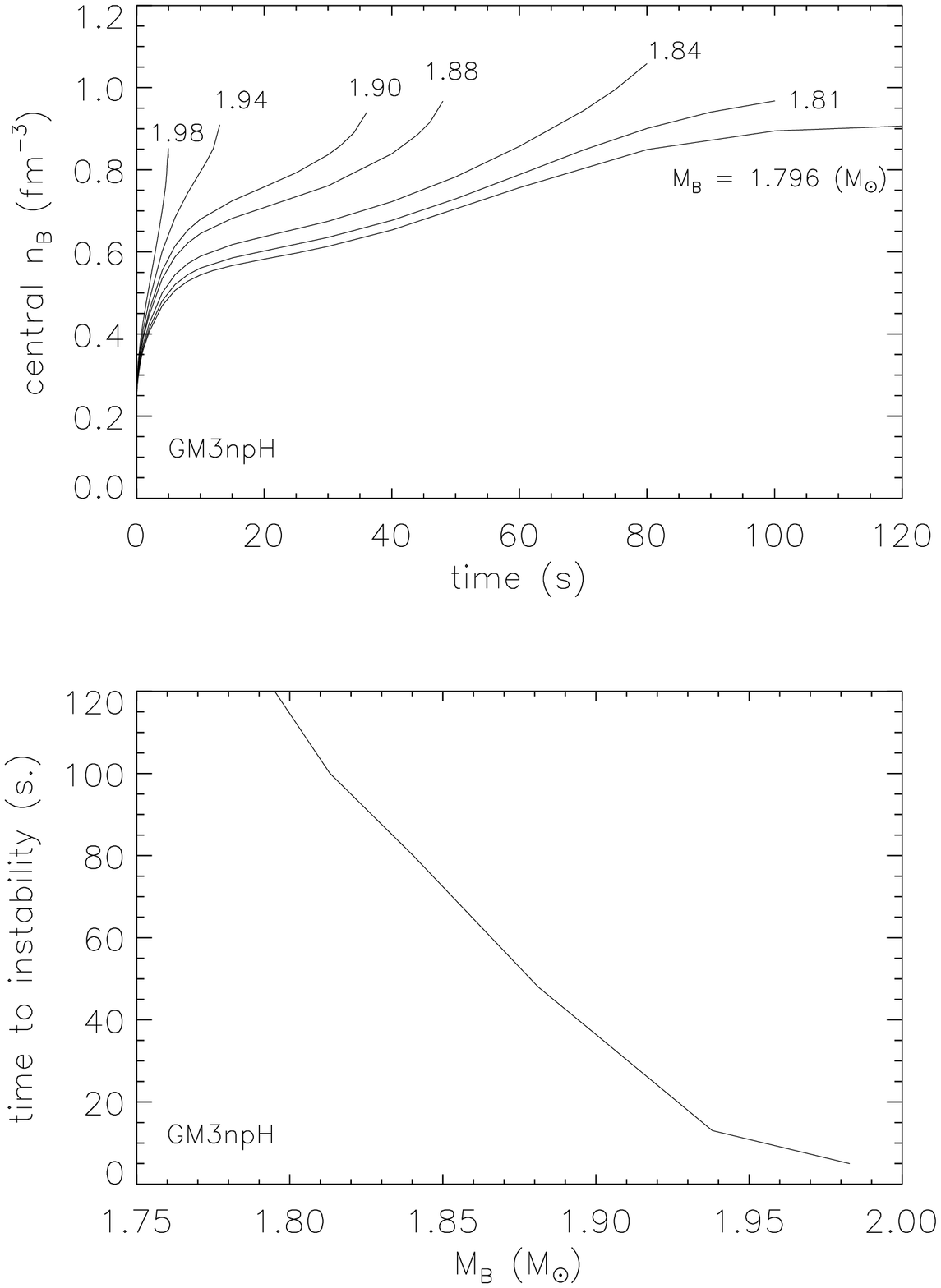}
\end{center}
\caption{
Left:  The evolution of the neutrino average energy and total
neutrino luminosity is compared for several assumed PNS 
baryon masses and EOSs.  The EOSs in the left panels contain only
baryons and leptons while those in the right panels also contain
hyperons.
Top right panel: Evolution of the central baryon number density
for different baryonic mass stars containing hyperons (model GM3npH) which
are metastable. Bottom right panel: Time required by stars shown in the top 
panel to reach the unstable configuration}
\label{res_mass}
\end{figure}

Neutrino signals from PNSs depends on many stellar
properties, including the mass; initial entropy, lepton fraction and
density profiles; and neutrino opacities. Pons et al. \cite{pons99}
carried
out a detailed study of the dependence of neutrino emission on
PNS characteristics.  They verified the generic results
of Burrows \& Lattimer \cite{burr86} that both neutrino luminosities and
average energies increased with increasing mass (see
Fig.~\ref{res_mass}).  In addition, they found that variations in
initial entropy and lepton fraction profiles in the outer regions of
the PNS caused only transient (lasting a few tenths of
a second) variations in neutrino luminosities and energies.
Variations in the central lepton fraction and entropy were found to
produce modest changes in neutrino luminosities that persisted to late
times.  The central values of lepton fraction and entropy are
established during core collapse, and will depend upon the initial
properties of the star as well as the EOS and neutrino transport
during the collapse.  

\subsection{Baseline Results}

Properties of the dense matter EOS that affect PNS evolution include
the compressibility, symmetry energy, specific heat, and composition.
Pons et al. \cite{pons99} employed a field theoretical EOS
\cite{prak97a}, with which the results due to some differences in
stellar size (due to variations in nuclear interactions) and
composition were studied.  Some results are summarized in
Fig.~\ref{res_mass}.  Overall, both the average energies and
luminosities of stars containing hyperons were larger compared to
those without.  In addition, for stars without hyperons, those stars
with smaller radii had higher average emitted neutrino energy,
although the predicted luminosities for early times ($t<10$ s) were
insensitive to radii.  This result only holds if the opacities are
calculated consistently with each EOS \cite{redd98,redd99a}; otherwise
rather larger variations in evolutions would have been found
\cite{keil95a,burr98}.  The same held true for models which allowed
for the presence of hyperons, except when the initial proto-neutron
star mass was significantly larger than the maximum mass for cold,
catalyzed matter.  Another new result was that the average emitted
neutrino energy of all flavors increased during the first 2-5 seconds
of evolution, and then decreased nearly linearly with time.  For times
larger than about 10 seconds, and prior to the occurrence of neutrino
transparency, the neutrino luminosities decayed exponentially with a
time constant that was sensitive to the high-density properties of
matter. Significant variations in neutrino emission occurred beyond 10
seconds: it was found that neutrino luminosities were larger during
this time for stars with smaller radii and with the inclusion of
hyperons in the matter.  Finally, significant regions of the stars
appeared to become convectively unstable during the evolution , as
several works have found \cite{burr93c,hera94,keil95b,mezz98,mpu00}.

The right panel of Fig.~\ref{res_mass} shows the time development of
the central baryon density (top panel) and also the time to the
collapse instability as a function of baryon mass (bottom panel).  The
larger the mass, the shorter the time to instability, since the PNS
does not have to evolve in lepton number as much.  Above 2.005
M$_\odot$, the metatstability disappears because the GM3npH initial
model with the lepton and entropy profiles we chose is already
unstable.  Below about 1.73 M$_\odot$, there is no metastability,
since this is the maximum mass of the cold, catalyzed npH star for
GM3.  The signature of neutrino emission from a metastable PNS should
be identifiable and it is discussed in Section 5.4.

\subsection{Influence of Many-Body Correlations}

\begin{figure}[hbt]
\begin{center}
\leavevmode
\includegraphics[scale=0.3]{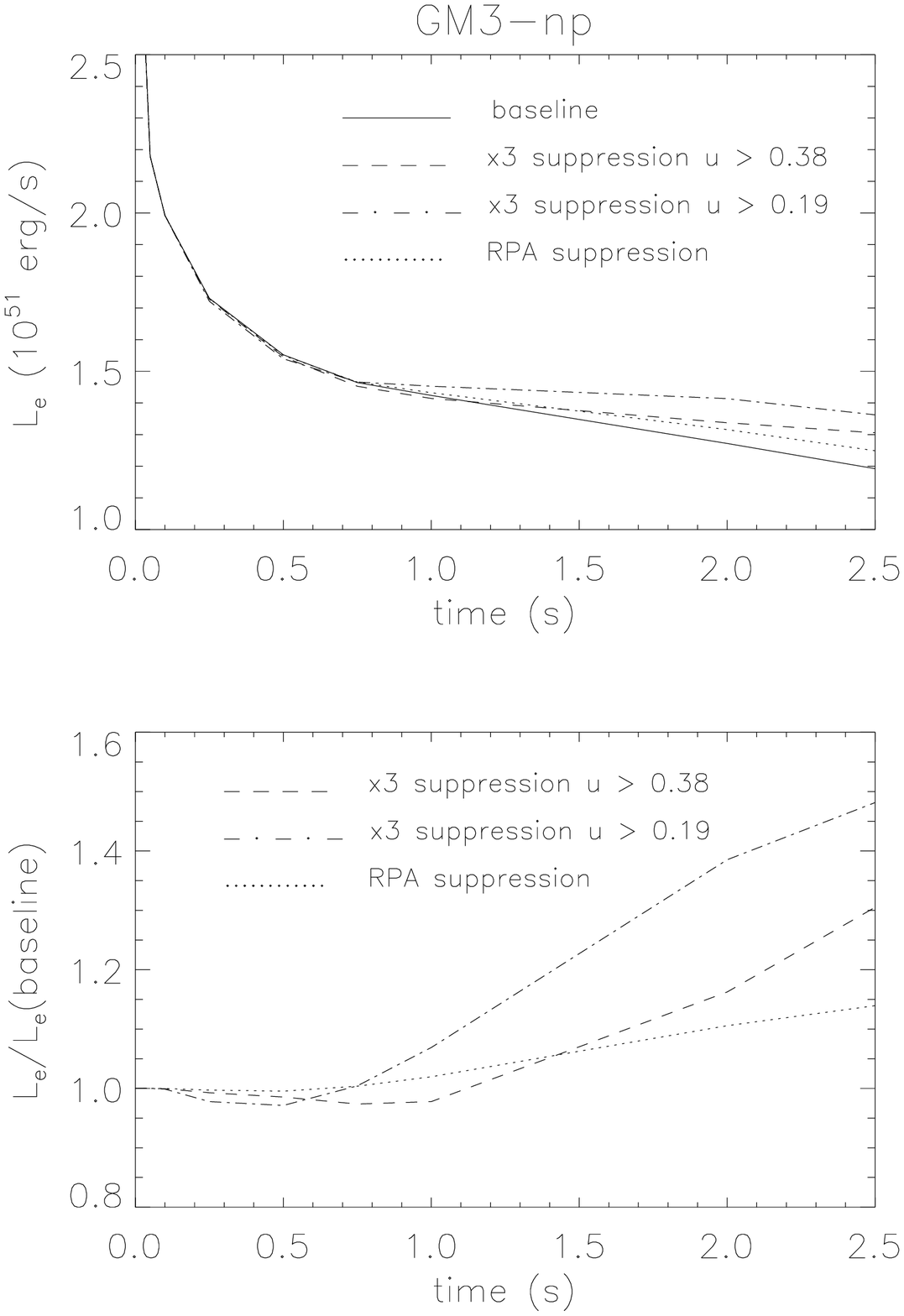}
\includegraphics[scale=0.3]{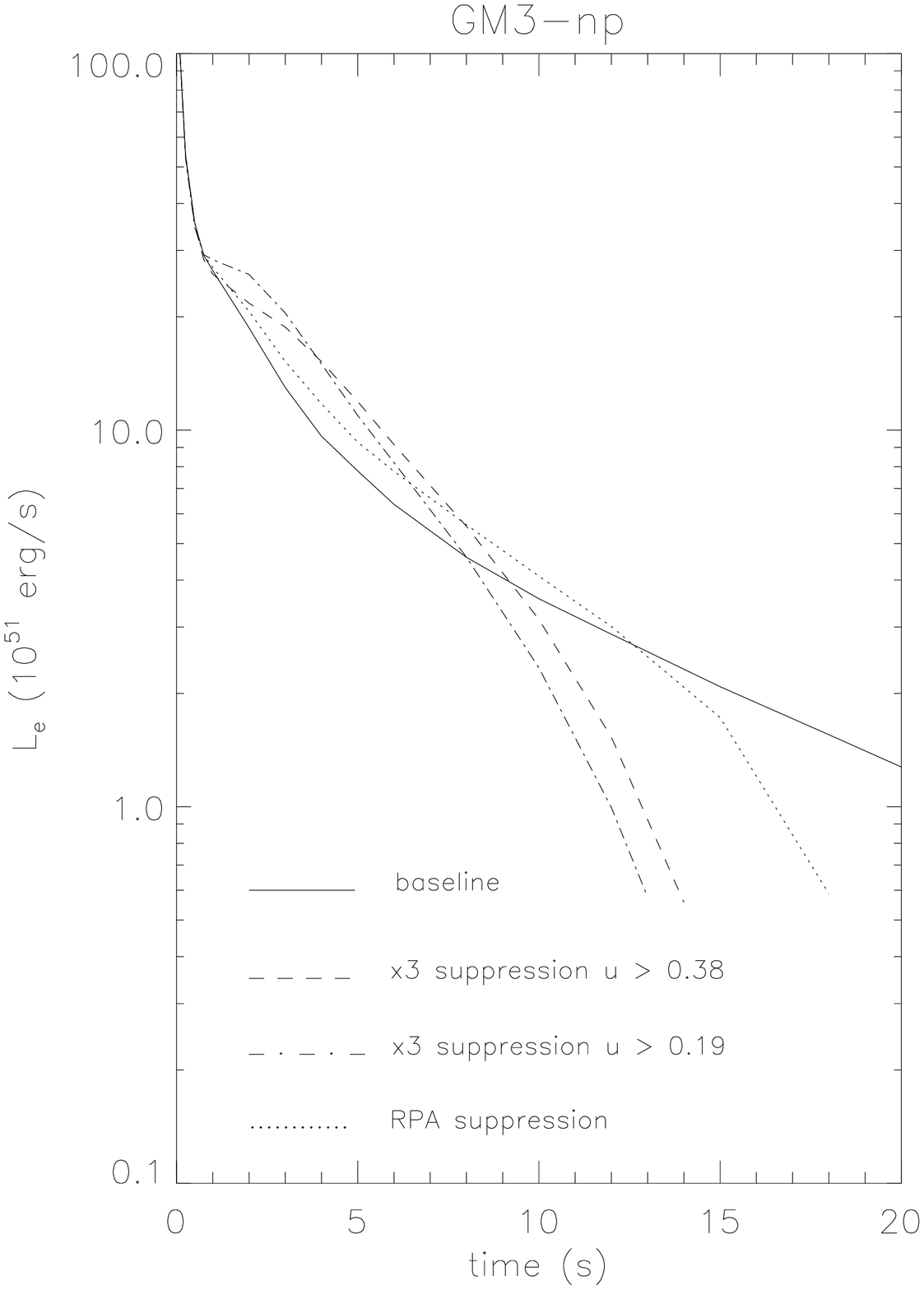}
\end{center}
\caption{
Left: The upper panel shows the total emitted neutrino luminosity for the
PNS evolutions described in Reddy et al. \cite{redd99a}. 
The lower panel shows the ratio of the luminosities
obtained in the three  models which contain corrections to the baseline
(Hartree approximation) model. 
Right: Emitted neutrino luminosity for long-term
PNS evolutions}
\label{lums}
\end{figure}

The main effect of the larger mean free paths produced by  RPA
corrections \cite{redd99a} 
is  that the inner core deleptonizes more quickly.  In
turn, the maxima in central temperature and entropy are reached on
shorter timescales. In addition, the faster increase in thermal
pressure in the core slows the compression associated with the
deleptonization stage, although after 10 s the net compressions of all
models converge.

The relatively large, early, changes in the central thermodynamic
variables do not, however, translate into similarly large effects on
observables such as the total neutrino luminosity and the average
radiated neutrino energy, relative to the baseline simulation.  The
luminosities for the different models are shown as a function of time
in Fig.~\ref{lums}.  The left panel  shows the
early time development in detail.  The exploratory models agree with
the results reported in \cite{burr98,burr99a}. 
 However, the
magnitude of the effects when  full RPA corrections are applied is
somewhat reduced compared to the exploratory models.  It is
especially important that at and below nuclear density, the
corrections due to correlations are relatively small.  Since
information from the inner core is transmitted only by the neutrinos,
the time scale to propagate any high density effect to the
neutrinosphere is the neutrino diffusion time scale. Since the
neutrinosphere is at a density approximately 1/100 of nuclear density,
and large correlation corrections occur only above 1/3 nuclear density
where nuclei disappear, we find that correlation corrections
calculated here have an effect at the neutrinosphere only after 1.5 s.
Moreover, the RPA suppresion we have calculated is considerably
smaller than those reported in \cite{burr98,burr99a}, 
reaching a maximum of
about 30\% after 5 s, compared to a luminosity increase of 50\% after
only 2 s.  However, the corrections are still very important during
the longer-term cooling stage (see Fig.~\ref{lums}), and result in
a more rapid onset of neutrino transparency compared to the Hartree results.

\subsection{Signals in Detectors}

\begin{figure}[htb]
\begin{center}
\includegraphics[scale=0.4]{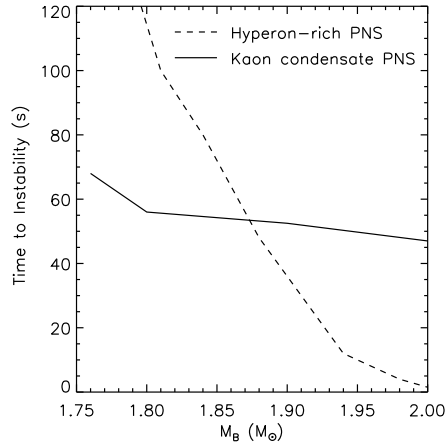}
\end{center}
\caption{Lifetimes of metastable stars as a function
of the stellar baryon mass.  Solid lines show results for PNSs
containing kaon-condensates and dashed lines show the results of Pons, 
et al.  \cite {pons99} for PNSs containing hyperons}
\label{fig:ttc}
\end{figure}

In Fig. \ref{fig:ttc} the lifetimes versus $M_B$ for stars containing
hyperons ($npH$) and $npK$ stars are compared \cite{Pon00a}.  In both
cases, the larger the mass, the shorter the lifetime.  For kaon-rich
PNSs, however, the collapse is delayed until the final stage of the
Kelvin-Helmholtz epoch, while this is not necessarily the case for
hyperon-rich stars.

\begin{figure}[htb]
\begin{center}
\includegraphics[scale=0.37]{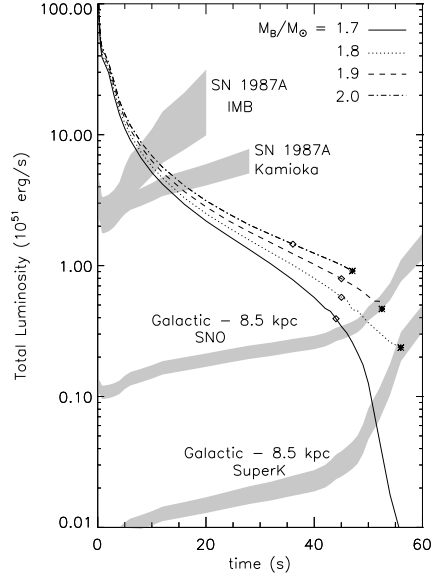}
\end{center}
\caption{The evolution of the total neutrino luminosity for stars of
various baryon masses.  Shaded bands illustrate the limiting
luminosities corresponding to a count rate of 0.2 Hz in all detectors,
assuming a supernova distance of 50 kpc for IMB and Kamioka, and 8.5
kpc for SNO and SuperK. The width of the shaded regions represents
uncertainties in the average neutrino energy from the use of a
diffusion scheme for neutrino transport}
\label{fig:lum1}
\end{figure}

In Fig.~\ref{fig:lum1} the evolution of the total neutrino energy
luminosity is shown for different models. Notice that the drop in the
luminosity for the stable star (solid line), associated with the end
of the Kelvin-Helmholtz epoch, occurs at approximately the same time
as for the metastable stars with somewhat higher masses.  In all
cases, the total luminosity at the end of the simulations is below
$10^{51}$ erg/s.  The two upper shaded bands correspond to SN 1987A
detection limits with KII and IMB, and the lower bands correspond to
detection limits in SNO and SuperK for a future galactic supernova at
a distance of 8.5 kpc.  The times when these limits intersect the
model luminosities indicate the approximate times at which the count
rate drops below the background rate $(dN/dt)_{BG}=0.2$ Hz.

The poor statistics in the case of SN 1987A precluded a precise
estimate of the PNS mass.  Nevertheless, had a collapse to a black
hole occurred in this case, it must have happened after the detection
of neutrinos ended. Thus the SN 1987A signal is compatible with a late
kaonization-induced collapse, as well as a collapse due to
hyperonization or to the formation of a quark core.  More information
would be extracted from the detection of a galactic SN with the new
generation of neutrino detectors.

In SNO, about 400 counts are expected for electron antineutrinos from
a supernova located at 8.5 kpc.  The statistics would therefore be
improved significantly compared to the observations of SN 1987A.  A
sufficiently massive PNS with a kaon condensate becomes metastable,
and the neutrino signal terminates, before the signal decreases below
the assumed background.  In SuperK, however, up to 6000 events are
expected for the same conditions (because of the larger fiducial mass)
and the effects of metastability due to condensate formation in lower
mass stars would be observable.

\subsection{What Can We Learn From Neutrino Detections?} 

The calculations of Pons, et al. \cite{Pon00a} show that the variations
in the neutrino light curves caused by the appearance of a kaon
condensate in a stable star are small, and are apparently insensitive
to large variations in the opacities assumed for them.  Relative to a
star containing only nucleons, the expected signal differs by an
amount that is easily masked by an assumed PNS mass difference of
$0.01-0.02$ M$_\odot$.  This is in spite of the fact that, in some
cases, a first order phase transition appears at the star's center.
The manifestations of this phase transition are minimized because of
the long neutrino diffusion times in the star's core and the Gibbs'
character of the transition.  Both act in tandem to prevent either a
``core-quake'' or a secondary neutrino burst from occurring during the
Kelvin-Helmholtz epoch.

Observable signals of kaon condensation occur only in the case of
metastable stars that collapse to a black hole.  In this case, the
neutrino signal for a star closer than about 10 kpc is expected to
suddenly stop at a level well above that of the background in a
sufficiently massive detector with a low energy threshold such as
SuperK.  This is in contrast to the signal for a normal star of
similar mass for which the signal continues to fall until it is
obscured by the background.  The lifetime of kaon-condensed metastable
stars has a relatively small range, of order 50--70 s for the models
studied here, which is in sharp contrast to the case of hyperon-rich
metastable stars for which a significantly larger variation in the
lifetime (a few to over 100 s) was found.  This feature of kaon
condensation suggests that stars that destabilize rapidly cannot do so
because of kaons.

Pons, et al. \cite{Pon00a} determined the minimum lifetime for
metastable stars with kaons to be about 40 s by examining the most
favorable case for kaon condensation, which is obtained by maximizing
the magnitude of the optical potential.  The maximum optical potential
is limited by the binary pulsar mass constraint, which limits the
star's maximum gravitational mass to a minimum value of 1.44
M$_\odot$.  Therefore, should the neutrino signal from a future
supernova abruptly terminate sooner than 40 s after the birth of the
PNS, it would be more consistent with a hyperon- or quark-induced
instability than one due to kaon condensation.

It is important to note that the collapse to a black hole in the case
of kaon condensation is delayed until the final stages of the
Kelvin-Helmholtz epoch, due to the large neutrino diffusion time in
the inner core.  Consequently, to distinguish between stable and
metastable kaon-rich stars through observations of a cessation of a
neutrino signal from a galactic supernovae is only possible using
sufficiently massive neutrino detectors with low energy thresholds and
low backgrounds, such as the current SNO and SuperK, and future
planned  detectors including the UNO. 

\subsection{Expectations From Quark Matter}

Strangeness appearing in the form of a mixed phase of strange quark
matter also leads to metastability.  Although quark matter is also
suppressed by trapped neutrinos \cite{PCL,SPL00}, the transition to
quark matter can occur at lower densities than the most optimistic
kaon case, and the dependence of the threshold density on $Y_L$ is
less steep than that for kaons.  Thus, it is an expectation that
metastability due to the appearance of quarks, as for the case of
hyperons, might be able to occur relatively quickly. Steiner, et
al. \cite{SPL00} have demonstrated that the temperature along adiabats
in the quark-hadron mixed phase is much smaller than what is found for
the kaon condensate-hadron mixed phase.  Calculations of PNS evolution
with a mixed phase of quark matter, including the possible effects of
quark matter superfluidity \cite{CR00} are currently in progress and
will be reported separately.

\section{Long-Term Cooling: The Next Million Years}

Following the transparency of the neutron star to neutrinos, the only
observational link with these objects is through photon emissions,
either as a pulsar or through thermal emissions or both.  Thermal
emissions of course are controlled by the temperature evolution of the
star, and this depends sensitively upon its internal composition.  The
tabulation of temperatures and ages for a set of neutron stars would
go a long way to deciding among several possibilities.

\subsection{Thermal Evolution}

The cooling of a young (age $<10^5$ yr) neutron star is mainly
governed by $\nu-$emission processes and the specific heat
\cite{P98a}.  Due to the extremely high thermal conductivity of
electrons, a neutron star becomes nearly isothermal within a time
$t_w\approx1-100$ years after its birth, depending upon the thickness
of the crust~\cite{LvRPP94}.  After this time its thermal evolution is
controlled by energy balance: 
\begin{eqnarray} \frac{dE_{th}}{dt} = C_V \frac{dT}{dt} = -L_{\gamma} -L_{\nu} + H \,,
\label{equ:balance}
\end{eqnarray} where $E_{th}$ is the total thermal energy and $C_V$ is the
specific heat.  $L_{\gamma}$ and $L_{\nu}$ are the total luminosities
of photons from the hot surface and $\nu$s from the interior,
respectively.  Possible internal heating sources, due, for example, to
the decay of the magnetic field or friction from differential
rotation, are included in $H$.  Our cooling simulations were performed
by solving the heat transport and hydrostatic equations including
general relativistic effects (see \cite{P98a}).  The surface's
effective temperature $T_e$ is much lower than the internal
temperature $T$ because of a strong temperature gradient in the
envelope.  Above the envelope lies the atmosphere where the emerging
flux is shaped into the observed spectrum from which $T_e$ can be
deduced.  As a rule of thumb $T_e/10^6$ K
$\approx\sqrt{T/10^8{\rm~K}}$, but modifications due to magnetic
fields and chemical composition may occur.

\subsection{Rapid vs. Slow Cooling}

The simplest possible $\nu$ emitting processes are the direct Urca
processes $f_1 + \ell \rightarrow f_2 + \nu_\ell\,, f_2 \rightarrow
f_1 + \ell + \overline{\nu_\ell}$, where $f_1$ and $f_2$ are either
baryons or quarks and $\ell$ is either an electron or a muon.  These
processes can occur whenever momentum conservation is
satisfied among $f_1, f_2$ and $\ell$ (within minutes of
birth, the $\nu$ chemical potential vanishes).
{If the unsuppressed 
direct Urca process for {\em any} component occurs, a neutron star
will rapidly cool because of enhanced emission:
the star's interior temperature $T$ will drop below 10$^9$ K in minutes
and reach 10$^7$ K in about a hundred years}.
$T_e$ will hence drop to less than 300,000 K after
the crustal diffusion time $t_w$ {\cite{LvRPP94,PA92,note}}.  This is the
so-called {\rm rapid cooling} paradigm.  If no direct Urca
processes are allowed, or they are all suppressed, cooling instead
proceeds through the significantly less rapid modified Urca process in
which an additional fermion enables momentum conservation.  This
situation could occur if no hyperons are present, or the nuclear
symmetry energy has a weak density dependence~\cite{LPPH91,PPLP92}.
The $\nu$ emisssion rates for the nucleon, hyperon, and quark Urca
and modified Urca processes can be found in~\cite{crevs}.

\subsection{Superfluid and Superconducting Gaps }

Pairing is unavoidable in a degenerate Fermi liquid if there is an
attractive interaction in {\it any} channel.  The resulting
superfluidity, and in the case of charged particles,
superconductivity, in neutron star interiors has a major effect on the
star's thermal evolution through suppressions of neutrino
($\nu$) emission processes and specific heats \cite{P98a,PA92}.  Neutron
($n$), proton ($p$) and $\Lambda$-hyperon superfluidity in the $^1S_0$
channel and $n$ superfluidity in the $^3P_2$ channel have been shown
to occur with gaps of a few MeV or less \cite{BEEHJS98,BB97}.
However, the density ranges in which gaps occur remain
uncertain.  At large baryon densities for which perturbative QCD
applies, pairing gaps for like quarks have been estimated to be a few
MeV \cite{Bailin84}.  However, the pairing gaps of unlike quarks ($ud,~
us$, and $ds$) have been suggested to be several tens to
hundreds of MeV through non-perturbative studies \cite{qsf0}
kindling interest in quark superfluidity and superconductivity
\cite{qsf,B99} and their effects on
neutron stars \cite{Blaschke00,PPLS}.

The effect of the pairing gaps on the emissivities and specific heats
for massive baryons are investigated in \cite{LY9496} and are here
generalized to the case of quarks.  The principal effects are severe
suppressions of both the emissivity and specific heat when
$T<<\Delta$, where $\Delta$ is the pairing gap.  In a system in which
several superfluid species exist the most relevant gap for these
suppressions is the smallest one.  The specific heat suppression is
never complete, however, because leptons remain unpaired.  Below the
critical temperature $T_c$, pairs may recombine, resulting in the
emission of $\nu\bar\nu$ pairs with a rate that exceeds the modified
Urca rate below $10^{10}$ K~\cite{FRS}; these processes are included
in our calculations.

\begin{figure}
\begin{center}
\includegraphics[scale=0.5]{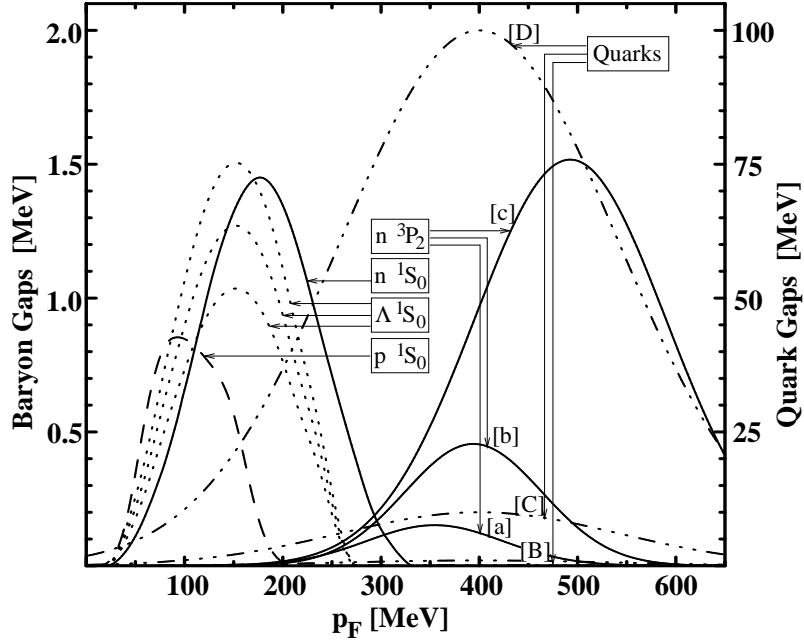}
\end{center}
\caption{Pairing gaps adopted for neutron $^1$S$_0$ and $^3$P$_2$,
     proton $^1$S$_0$, $\Lambda$ $^1$S$_0$, and quarks.
     The n $^3$P$_2$ gaps are anisotropic; plotted values are
     angle-averaged.    The $\Lambda$ gaps
correspond, in order of increasing $\Delta$, to background densities
$n_B=0.48, 0.64$ and 0.8 fm$^{-3}$, respectively. 
     The s-wave quark gaps are schematic; see text for details}
\label{fig:baryon_Tc}
\end{figure}

The baryon and quark pairing gaps we adopt are shown in
Fig.~\ref{fig:baryon_Tc}.  Note that gaps are functions of Fermi
momenta ($p_F(i)$, $i$ denoting the species) which translates into a
density dependence.  For $p_F(n,p) \lsim 200 - 300$ MeV/$c$, nucleons
pair in the $^1$S$_0$ state, but these momenta correspond to densities
too low for enhanced $\nu$ emission involving nucleons to occur. At
higher $p_F$'s, baryons pair in higher partial waves.  The $n~^3$P$_2$
gap has been calculated for the Argonne V$_{18}$, CD-Bonn and Nijmegen
I \& II interactions \cite{BEEHJS98}.  This gap is crucial since it
extends to large $p_F(n)$ and can reasonably be expected to occur at
the centers of neutron stars.  For $p_F(n)>350$ MeV/c, gaps are
largely uncertain because of both experimental and theoretical
uncertainties \cite{BEEHJS98}.  The curves [a], [b] and [c] in
Fig.~\ref{fig:baryon_Tc} reflect the range of uncertainty.  The $p$
$^3$P$_2$ gap is too small to be of interest.  Gaps for the $^1$S$_0$
pairing of $\Lambda$, taken from \cite{BB97} and shown as dotted
curves, are highly relevant since $\Lambda$s participate in direct
Urca emission as soon as they appear \cite{PPLP92}.  Experimental
information beyond the $^1$S$_0$ channel for $\Lambda$ is not
available.  $\Delta$s for $\Sigma-$hyperons remain largely
unexplored. The quark ($q$) gaps are taken to be Gaussians centered at
$p_F(q) = 400$ MeV/$c$ with widths of 200 MeV/$c$ and heights of 100
MeV [model D], 10 MeV [C], 1 MeV [B] and 0.1 MeV [A], respectively.
The reason for considering quark gaps much smaller than suggested in
\cite{qsf0,Bailin84} is associated with the multicomponent nature of
charge-neutral, beta-equilibrated, neutron star matter as will become
clear shortly.

\subsection{Effects of Composition}

We consider four generic compositions: charge-neutral, beta
equilibrated matter containing nucleons only ($np$), nucleons with
quark matter ($npQ$), nucleons and hyperons ($npH$), and nucleons,
hyperons and quarks ($npHQ$).  In the cases involving quarks, a mixed phase
of baryons and quarks is constructed by satisfying Gibbs' phase rules
for mechanical, chemical and thermal equilibrium \cite{glen1}.  The
phase of pure quark matter exists only for very large baryon
densities, and rarely occurs in our neutron star models.  Baryonic
matter is calculated using a field-theoretic model at the mean field
level \cite{ZM90}; quark matter is calculated using either a bag-like
model or the Nambu-Jona-Lasinio quark model \cite{PCL,SPL00}. 
The
equation of state (EOS) is little affected by the pairing phenomenon,
since the energy density gained 
is negligible compared to the ground state energy densities without pairing.

Additional particles, such as quarks or hyperons, have the
effect of softening the EOS and increasing the central densities of
stars relative to the $np$ case.  
For the $npQ$ model studied, a mixed phase appears at
the density $n_B=0.48$ fm$^{-3}$.  Although the volume fraction of
quarks is initially zero, the quarks themselves have a significant
$p_F(q)$ when the phase appears.  The $p_F$s of the three quark
flavors become the same at extremely high density, but for the densities
of interest they are different due to
the presence of negatively charged leptons. In particular, $p_F(s)$
is much smaller than $p_F(u)$ and $p_F(d)$ due to the
larger $s$-quark mass.  Use of the Nambu--Jona-Lasinio model, in which quarks
acquire density-dependent masses resembling those of constituent
quarks, exaggerates the reduction of $p_F(s)$.  This has
dramatic consequences since the pairing phenomenon operates at its
maximum strength when the Fermi momenta are exactly equal; even small
asymmetries cause pairing gaps to be severely reduced
\cite{B99,ARSW96}.  In addition, one may also expect p-wave
superfluidity, to date unexplored, which may yield gaps smaller than
that for the s-wave.  We therefore investigate pairing gaps that are
much smaller than those reported for the case of s-wave superfluidity
and equal quark $p_F$'s.  

The introduction of hyperons does not change these generic trends.  In
the case $npH$, the appearance of hyperons changes the lepton and
nucleon $p_F$'s similarly to the appearance of quarks 
although with less magnitude.  
While the appearance of quarks is delayed by the
existence of hyperons, at high densities the $p_F$'s of nucleons and
quarks remain similar to those of the $npQ$ case.  The existence of either
hyperons or quarks, however, does allow the possibility of additional
direct Urca processes involving themselves as well as those involving
nucleons by decreasing $p_F(n)-p_F(p)$.  For the $npQ$ and $npHQ$
models studied, the maximum masses are $\cong 1.5$\Msol, the central
baryon densities are $\cong 1.35$ fm$^{-3}$, and the volume fractions
of quarks at the center are $\cong 0.4$.

\subsection{Examples of Results}

\begin{figure}
\begin{center}
\includegraphics[scale=1.2]{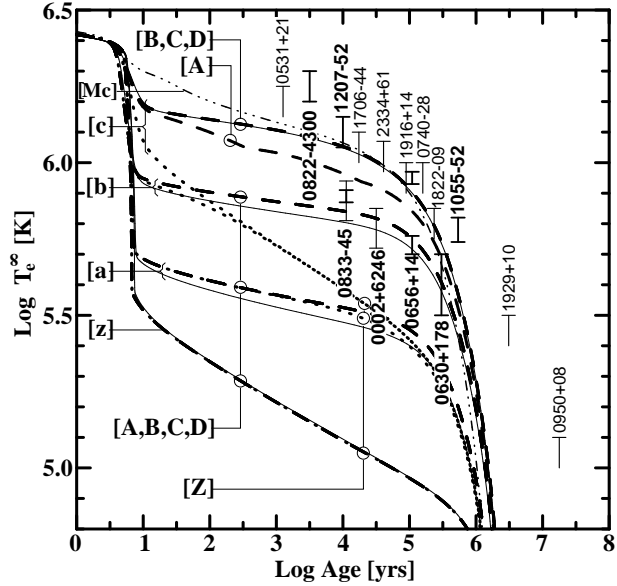}
\end{center}
\caption{Cooling of 1.4\Msun stars with $np$
matter (continuous curves) and npQ matter (dashed and dotted
curves).  The curves labelled as [a], [b], and [c] correspond to
$n~^3$P$_2$ gaps as in Fig.~\protect\ref{fig:baryon_Tc}; [z]
corresponds to zero $n$ gap.  Models labelled [A], [B], [C] and
[D] correspond to quark gaps as in
Fig.~\protect\ref{fig:baryon_Tc}; [Z] corresponds to zero quark
gap} 
\label{fig:cooling-N-NQ} 
\end{figure}

\begin{figure}
\begin{center}
\includegraphics[scale=1.2]{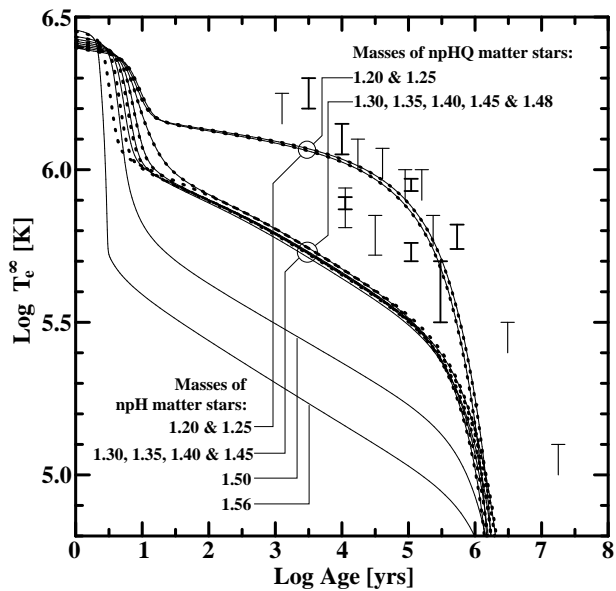} 
\end{center}
\caption{Cooling of stars
with $npH$ (continuous lines) and $npHQ$ matter (dotted lines)
for various stellar masses (in \Msun).  $n~^3$P$_2$ gaps are from
case [c] while quark gaps, when present, are from model [C] of
Fig.~\ref{fig:baryon_Tc}} 
\label{fig:cooling-NH-NHQ}
\end{figure}

Cooling simulations of stars without hyperons and with hyperons are
compared, in Figs.~\ref{fig:cooling-N-NQ} and
\ref{fig:cooling-NH-NHQ}, respectively, to available observations of
thermal emissions from pulsars.  Sources for the observational data
can be found in \cite{P97}.  However, at the present time, the
inferred temperatures must be considered as upper limits because the
total flux is contaminated, and in some cases dominated, by the
pulsar's magnetospheric emission and/or the emission of a surrounding
synchrotron nebula.  Furthermore, the neutron star surface may be
reheated by magnetospheric high energy photons and particles;
late-time accretion for non-pulsing neutron stars is also
possible.  Other uncertainties arise in the temperature estimates due
to the unknown chemical composition and magnetic field strength in the
surface layers, and in the age, which is based upon the observed
spin-down time.  In these figures, the bolder the data symbol the
better the data.

\subsection*{$np$ and $npQ$ Matter}

The $np$ case is considered in Fig.~\ref{fig:cooling-N-NQ}, in which
solid lines indicate the temperature evolution of a 1.4 M$_\odot$ star
for quarkless matter: case [z] is for no nucleon pairing at all, and
cases [a], [b] and [c] correspond to increasing values for the neutron
$^3P_2$ gap, according to Fig.~\ref{fig:baryon_Tc}.  The
field-theoretical model employed for the nucleon interactions allows
the direct nucleon Urca process, which dominates the cooling.  The
unimpeded direct Urca process carries the temperature to values well
below the inferred data.  Pairing suppresses the cooling for $T<T_c$,
where $T$ is the interior temperature, so $T_e$ increases with
increasing $\Delta$.  If the direct Urca process is not allowed, the
range of predicted temperatures is relatively narrow due to the low
emissivity of the modified Urca process. We show an example of such
cooling (curve [Mc]) using the $n~^3P_2$ gap [c] for a 1.4\Msun with
an EOS \cite{APR98} for which the direct Urca cooling is not allowed.

The other curves in the figure illustrate the effects of quarks upon
the cooling.  The dotted curves [Z] are for vanishingly small quark
gaps; the dashed curves ([A], [B], [C] and [D]) are for quark gaps as
proposed in Fig.~\ref{fig:baryon_Tc}.  For nonexistent ([z]) or small
([a]) nucleon gaps, the quark Urca process is irrelevant and the
dependence on the existence or the size of the quark gaps is very
small.  However, for large nucleon gaps ([b] and [c]), the quark
direct Urca process quickly dominates the cooling as the nucleon
direct Urca process is quenched.  It is clear that for quark gaps of
order 1 MeV or greater ([B], [C] or [D]) the effect of quarks is
again very small.  There is at most a slight increase in the stars
temperatures at ages between 10$^1$ to 10$^{5 - 6}$ years due to the
reduction of $p_F(n)$ and the consequent slightly larger gap
(Fig.~\ref{fig:baryon_Tc}).  Even if the quark gap is quite small
([A]), quarks have an effect only if the nucleon gap is very large
([b] or [c]), i.e., significantly larger than the quark gap: the
nucleon direct Urca process is suppressed at high temperatures and the
quark direct Urca process has a chance to contribute to the cooling.
We find that the effects of changing the stellar mass $M$ are similar
to those produced by varying the baryon gap, so that only combinations
of $M$ and $\Delta$ might be constrained by observation.

\subsection*{$npH$ and $npHQ$ Matter}

The thermal evolution of stars containing hyperons has been discussed
in \cite{P98b,SBSB98}, but we obtain qualitatively different results
here.  Hyperons open new direct Urca channels:
$\Lambda \rightarrow p + e + \overline{\nu}_e$ and
$\Lambda + e \rightarrow \Sigma^- + \nu_e$ if $\Sigma^-$'s are present,
with their inverse processes.
Previous results showed that the cooling is
naturally controlled by the smaller of the $\Lambda$ or $n$ gap.
However, this is significantly modified when the
$\Lambda$ gap is more accurately treated.  At the $\Lambda$ appearance
threshold, the gap must vanish since $p_F(\Lambda)$ is vanishingly
small.  We find that a very thin layer, only a few meters thick, of
unpaired or weakly paired $\Lambda$'s is sufficient to control the
cooling.  This effect was overlooked in previous works perhaps because
they lacked adequate zonal resolution.

In Fig.~\ref{fig:cooling-NH-NHQ} we compare the evolution of stars of
different masses made of either $npH$ or $npHQ$ matter.  We find that
all stars, except the most massive $npH$ ones, follow two distinctive
trajectories depending on whether their central density is below or
above the threshold for $\Lambda$ appearance (= 0.54 fm$^{-3}$ in our
model EOS, the threshold star mass being 1.28\Msun).  In the case of
$npH$ matter, stars with $M>1.50\Msun$ are dense enough so that the
$\Lambda$ $^1$S$_0$ gap vanishes and hence undergo fast cooling, while
stars made of $npHQ$ matter do not attain such high densities.  The
temperatures of $npH$ stars with masses between 1.3 and 1.5 \Msun are
below the ones obtained in the models of Fig.~\ref{fig:cooling-N-NQ}
with the same $n$ $^3$P$_2$ gap [b], which confirms that the cooling
is dominated by the very thin layer of unpaired $\Lambda$'s (the
slopes of these cooling curves are typical of direct Urca processes).
Only if the $n~^3$P$_2$ gap $\lsim0.3$ MeV do the cooling curves fall
below what is shown in Fig.~\ref{fig:cooling-NH-NHQ}.  Notice,
moreover, that in the mass range 1.3 -- 1.48 \Msun the cooling curves
are practically indistinguishible from those with unpaired quark
matter shown in Fig.~\ref{fig:cooling-N-NQ}.  In these models with
$npH$ or $npHQ$ matter, there is almost no freedom to ``fine-tune''
the size of the gaps to attain a given $T_e$: stars with $\Lambda$'s
will all follow the same cooling trajectory, determined by the
existence of a layer of unpaired or weakly paired $\Lambda$'s, as
long as the $n~^3P_2$ gap is not smaller.  It is, in some sense, the
same result as in the case of $np$ and $npQ$ matter: the smallest gap
controls the cooling and now the control depends on how fast the
$\Lambda$ $^1$S$_0$ gap increases with increasing $p_F(\Lambda)$.

\subsection{Implications}

Our results indicate that observations could constrain combinations of
the smaller of the neutron and $\Lambda-$hyperon pairing gaps and the
star's mass.  Deducing the sizes of quark gaps from observations of
neutron star cooling will be extremely difficult.  Large quark gaps
render quark matter practically invisible, while vanishing quark gaps
lead to cooling behaviors which are nearly indistinguishable from
those of stars containing nucleons or hyperons.  Moreover, it also
appears that cooling observations by themselves will not provide
definitive evidence for the existence of quark matter itself.

\section{The Structure of Catalyzed Stars}

In this section, we explore from a theoretical perspective, how the
structure of neutron star depends upon the assumed EOS.  This study
is crucial if new observations of masses and radii are to lead to
effective constraints of the EOS of dense matter.  Two general classes
of stars can be identified: {\it normal} stars in which the density
vansihes at the stellar surface, and {\it self-bound} stars in which
the density at the surface is finite.  Normal stars originate from
nuclear force models which can be conveniently grouped into three
broad categories: nonrelativistic potential models, relativistic field
theoretical models, and relativistic Dirac-Brueckner-Hartree-Fock
models.  In each of these approaches, the presence of additional
softening components such as hyperons, Bose condensates or quark
matter, can be incorporated.  Details of these approaches have been
further considered in Lattimer et al. \cite{LPMY} and Prakash et
al. \cite{prak97a}.  A representative sample, and some general attributes,
including references and typical compositions, of equations of state
employed here are summarized in Table~\ref{eosname}.

For normal matter, the EOS is that of interacting nucleons above
a transition density of 1/3 to 1/2 $n_s$.  Below this density, the
ground state of matter consists of heavy nuclei in equilibrium with a
neutron-rich, low-density gas of nucleons.
However, for most of the purposes of this paper, the pressure in the
region $n<0.1$ fm$^{-3}$ is not relevant as it does not significantly
affect the mass-radius relation or other global aspects of the star's
structure.  Nevertheless, the value of the transition density, and the
pressure there, are important ingredients for the determination of the
size of the superfluid crust of a neutron star that is believed to be
involved in the phenomenon of pulsar glitches (Link, Epstein, \&
Lattimer \cite{Link}).

Four equations of state are taken from Akmal \& Pandharipande
\cite{Akmal}.  These are: AP1 (the AV18 potential), AP2 (the AV18 potential
plus $\delta v_b$ relativistic boost corrections), AP3 (the AV18
potential plus a three-body UIX potential ), and AP4 (the AV18
potential plus the UIX potential plus the $\delta v_b$ boost).  Three
equations of state from M\"uller \& Serot \cite{MS}, labelled MS1--3,
correspond to different choices of the parameters $\xi$ and $\zeta$
which determine the strength of the nonlinear vector and isovector
interactions at high densities.  The numerical values used are
$\xi=\zeta=0; \xi=1.5, \zeta=0.06$; and $\xi=1.5, \zeta=0.02$,
respectively.  Six EOSs come from the phenomenological non-relativistic
potential model of Prakash, Ainsworth \& Lattimer \cite{PAL}, labelled
PAL1--6, which have different choices of the symmetry
energy parameter at the saturation density, its density dependence,
and the bulk nuclear matter incompressibility parameter $K_s$.  The
incompressibilities of PAL1--5 were chosen to be $K_s=180$ or 240 MeV,
but PAL6 has $K_s=120$ MeV.  Three interactions denoted GM1--3 come from the
field-theoretical model of Glendenning \& Moszkowski \cite{glen91}.  Two
interactions come from the field-theoretical model of Glendenning \&
Schaffner-Bielich \cite{GS99}: GL78 with
$U_K(\rho_0)=-140$ MeV and TM1 with $U_K=-185$ MeV.  The labels
denoting the other EOSs in Table~\ref{eosname} are identical to those in the
original references.

The rationale for exploring a wide variety of EOSs, even some that are
relatively outdated or in which systematic improvements are performed,
is two-fold.  First, it provides contrasts among widely different
theoretical paradigms.  Second, it illuminates general relationships
that exist between the pressure-density relation and the macroscopic
properties of the star such as the radius.  For example, AP4
represents the most complete study to date of Akmal \& Pandharipande
\cite{Akmal}, in which many-body and special relativistic corrections are
progressively incorporated into prior models, AP1--3.  AP1--3 are
included here because they represent different pressure-energy
density-baryon density relations, and serve to reinforce correlations
between neutron star structure and microscopic physics observed using
alternative theoretical paradigms.  Similarly, several different
parameter sets for other EOSs are chosen.

\begin{table}
\caption{Approach refers to the
underlying theoretical technique.  Composition refers to strongly
interacting components (n=neutron, p=proton, H=hyperon, K=kaon,
Q=quark); all models include leptonic contributions.}
\begin{center}
\begin{tabular}{l|l|l|l} \hline
Symbol & Reference & Approach & Composition \\ \hline
FP  & Friedman \& Pandharipande \cite{FP} & Variational & np \\
PS & Pandharipande \& Smith \cite{PS} & Potential & n$\pi^0$ \\
WFF(1-3) & Wiringa, Fiks \& Fabrocine \cite{WFF} & Variational & np \\
AP(1-4) & Akmal \& Pandharipande \cite{Akmal} & Variational & np \\
MS(1-3) & M\"uller \& Serot \cite{MS} & Field Theoretical & np \\
MPA(1-2) & M\"uther, Prakash \& Ainsworth \cite{MPA} & Dirac-Brueckner
HF & np \\ 
ENG & Engvik et al. \cite{Engvik} & Dirac-Brueckner HF & np \\
PAL(1-6)  & Prakash, Ainsworth \& Lattimer \cite{PAL} & Schematic
Potential & np \\ 
GM(1-3) & Glendenning \& Moszkowski \cite{glen91} & Field Theoretical & npH \\
GS(1-2) & Glendenning \& Schaffner-Bielich \cite{GS98} & Field
Theoretical & npK\\
PCL(1-2) & Prakash, Cooke \& Lattimer \cite{PCL} & Field Theoretical &
npHQ \\
SQM(1-3) & Prakash, Cooke \& Lattimer \cite{PCL} & Quark Matter & 
Q$(u,d,s)$\\ 
\hline
\end{tabular}
\label{eosname}
\end{center}
\end{table}

In all cases, except for PS (Pandharipande \& Smith \cite{PS}), the
pressure is evaluated assuming zero temperature and beta equilibrium
without trapped neutrinos.  PS only contains neutrons among the
baryons, there being no charged components.  We chose to include this
EOS, despite the fact that it has been superceded by more
sophisticated calculations by Pandharipande and coworkers, because it
represents an extreme case producing large radii neutron stars.

\begin{figure}
\begin{center}
\includegraphics[scale=0.525,angle=90]{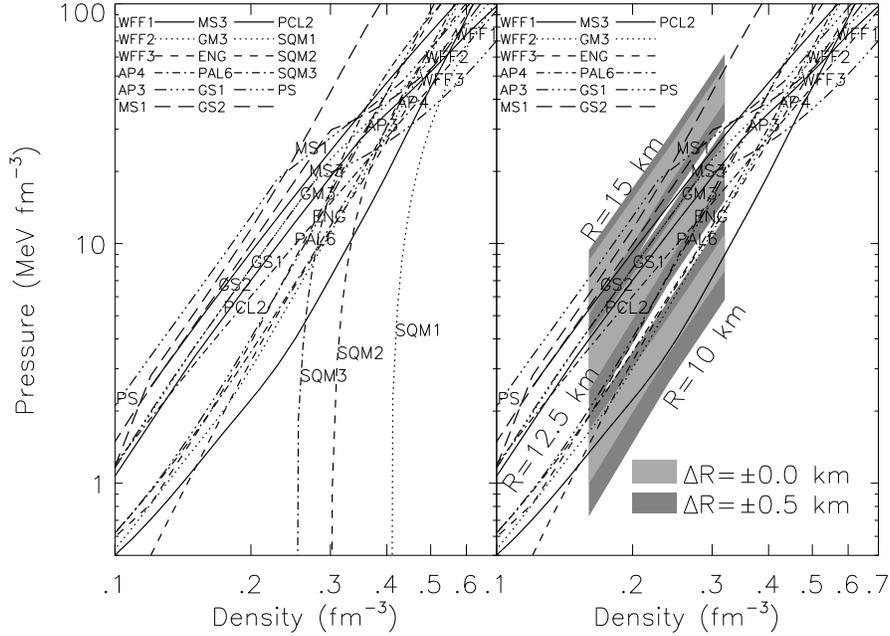}
\end{center}
\caption{Left: The pressure-density relation for a selected set of EOSs
contained in Table~\ref{eosname}.  The pressure is in units of MeV
fm$^{-3}$ and 
the density is in units of baryons per cubic fermi.  The nuclear
saturation density is approximately $0.16$ fm$^{-3}$.
Right: The pressures inferred from the empirical correlation
Eq.~(\ref{correl}), for three hypothetical radius values (10, 12.5
and 15 km) overlaid on the pressure-density relations shown on the left.
  The light shaded region takes into account only the
uncertainty associated with $C(n,M)$; the dark shaded region also
includes a hypothetical uncertainty of 0.5 km in the radius
measurement.  The neutron star mass was assumed to be 1.4 M$_\odot$}
\label{fig:P-rho}
\end{figure}

The pressure-density relations for some of the selected EOSs are shown
in Fig.~\ref{fig:P-rho} which displays three significant features to
note for normal EOSs.  First, there is a fairly wide range of
predicted pressures for beta-stable matter in the density domain
$n_s/2<n<2n_s$.  For the EOSs displayed, the range of pressures covers
about a factor of five, but this survey is by no means exhaustive.
That such a wide range in pressures is found is somewhat surprising,
given that each of the EOSs provides acceptable fits to
experimentally-determined nuclear matter properties.  Clearly, the
extrapolation of the pressure from symmetric matter to nearly pure
neutron matter is poorly constrained.  Second, the {\em slopes} of the
pressure curves are rather similar.  A polytropic index of $n\simeq1$,
where $P=Kn^{1+1/n}$, is implied.  Third, in the density domain below
$2n_s$, the pressure-density relations seem to fall into two groups.
The higer pressure group is primarily composed of relativistic
field-theoretical models, while the lower pressure group is primarily
composed of non-relativistic potential models.  It is significant that
relativistic field-theoretical models generally have symmetry energies
that increase proportionately to the density while potential models
have much less steeply rising symmetry energies.

A few of the plotted normal EOSs have considerable softening at high
densities, especially PAL6, GS1, GS2, GM3, PS and PCL2.  PAL6 has an
abnormally small value of incompressibility ($K_s=120$ MeV).  GS1 and
GS2 have phase transitions to matter containing a kaon condensate, GM3
has a large population of hyperons appearing at high density, PS has a
phase transition to a neutral pion condensate and a neutron solid, and
PCL2 has a phase transition to a mixed phase containing strange quark
matter.  These examples are representative of the kinds of softening
that could occur at high densities.

The best-known example of self-bound stars results from Witten's
\cite{Witten84} conjecture (also see Fahri \& Jaffe \cite{Fahri84}, Haensel, Zdunik \&
Schaeffer \cite{Haensel86}, Alcock \& Olinto \cite{Alcock88}, and Prakash et al. \cite{Prakash90}) that
strange quark matter is the ultimate ground state of matter.  In this
paper, the self-bound EOSs are represented by strange-quark matter
models SQM1--3, using perturbative QCD and an MIT-type bag model, with
parameter values given in Table~\ref{strangetable}. 
The existence of an energy ceiling
equal to the baryon mass, 939 MeV, for zero pressure matter requires
that the bag constant $B\le94.92$ MeV fm$^{-3}$.  This limiting value
is chosen, together with zero strange quark mass and no interactions
($\alpha_c=0$), for the model SQM1.  The other two models chosen, SQM2
and SQM3, have bag constants adjusted so that their energy ceilings
are also 939 MeV.

\begin{table}
\caption{Parameters for self-bound strange quark stars.
Numerical values employed in the MIT bag
model as described in Fahri \& Jaffe \cite{Fahri84}.}
\begin{center}
\begin{tabular}{l|c|c|c} \hline
Model & $B$ (MeV fm$^{-3})$ & $m_s$ (MeV) & $\alpha_c$ \\ \hline
SQM1 & 94.92 & 0 & 0 \\
SQM2 & 64.21 & 150 & 0.3 \\
SQM3 & 57.39 & 50 & 0.6 \\ \hline
\end{tabular}
\end{center}
\label{strangetable}
\end{table}

\subsection{Neutron Star Radii}

\begin{figure}
\begin{center}
\includegraphics[scale=0.52,angle=90]{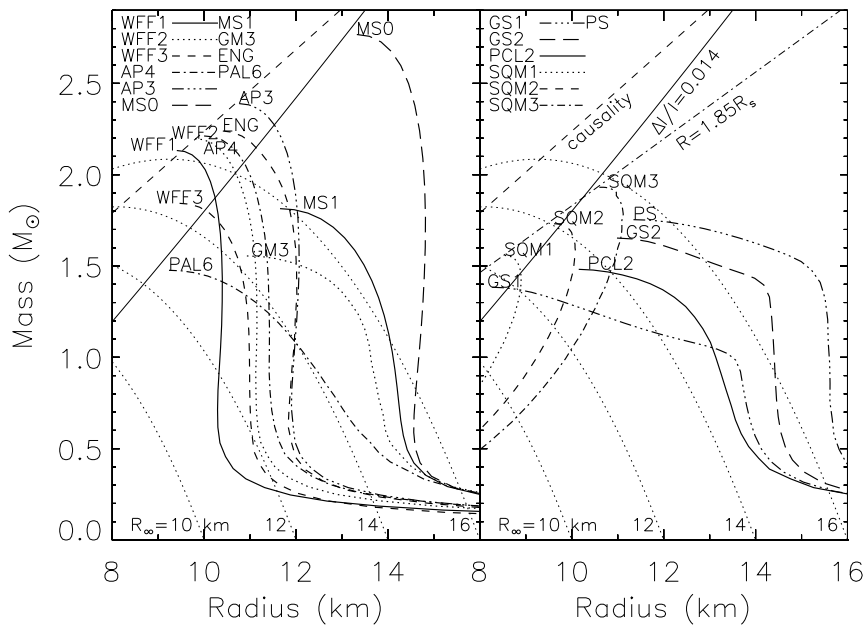} 
\end{center}
\caption{Mass-radius curves for several EOSs listed in
Table~\ref{eosname}.  The 
left panel is for stars containing nucleons and, in some cases,
hyperons.  The right panel is for stars containing more exotic
components, such as mixed phases with kaon condensates or strange
quark matter, or pure strange quark matter stars.  In both panels, the
lower limit causality places on $R$ is shown as a dashed line, a
constraint derived from glitches in the Vela pulsar is shown as the
solid line labelled $\Delta I/I=0.014$, and contours of constant
$R_\infty=R/\sqrt{1-2GM/Rc^2}$ are shown as dotted
curves. In the right panel, the theoretical trajectory of maximum
masses and radii for pure strange quark matter stars is marked by the
dot-dash curve labelled $R=1.85R_s$}
\label{fig:M-R}
\end{figure}

Fig.~\ref{fig:M-R} displays the resulting mass-radius relations for
catalyzed matter.  Rhoades \& Ruffini \cite{Rhoades74} demonstrated that the
assumption of causality beyond a fiducial density $\rho_f$ sets an
upper limit to the maximum mass of a neutron star:
$4.2\sqrt{\rho_s/\rho_f}$ M$_\odot$.  Lattimer et al. \cite{LPMY} have shown
that the causality constraint also sets a lower limit to the radius:
$R\simge1.52 R_s$, where $R_s=2GM/c^2$, which is shown in Fig.~\ref{fig:M-R}.  For a 1.4 M$_\odot$ star, this is about 4.5 km.  The
most reliable estimates of neutron star radii in the near future will
likely stem from observtins of thermal emission from their surfaces.
Such estimates yield the so-called ``radiation radius''
$R_\infty=R/\sqrt{1-R_s}$, a quantity resulting from redshifting the
star's luminosity and temperature.  A given value of $R_\infty$
imiplies that $R<R_\infty$ and $M<0.13 (R_\infty/{\rm
km}){\rm~M}_\odot$.  Contours of $R_\infty$ are also displayed.
With the exception of model GS1,
the EOSs used to generate Fig.~\ref{fig:M-R} result in maximum
masses greater than 1.442 M$_\odot$, the limit obtained from PSR
1913+16.  From a theoretical perspective, it appears that values of
$R_\infty$ in the range of 12--20 km are possible for normal neutron
stars whose masses are greater than 1 M$_\odot$.

One observes that {\em normal} neutron stars have minimum masses of
about 0.1 M$_\odot$ that are primarily determined by the EOS below
$n_s$.  At the minimum mass, the radii are generally in excess of 100
km.  Self-bound stars have no minimum mass and the maximum mass
self-bound stars have nearly the largest radii possible for a given
EOS.  If the strange quark mass $m_s=0$ and interactions are neglected
($\alpha_c=0$), the maximum mass is related to the bag constant $B$ in
the MIT-type bag model by
$M_{max}=2.033~(56{\rm~MeV~fm}^{-3}/B)^{1/2}~{\rm M}_\odot$.  Prakash
et al. \cite{Prakash90} and Lattimer et al. \cite{LPMY} showed that the addition of a
finite strange quark mass and/or interactions produces larger maximum
masses.  The constraint that $M_{max}>1.44$ M$_\odot$ is thus
automatically satisfied by the condition that the energy ceiling is
939 MeV, and non-zero values of $m_s$ and $\alpha_c$ yield larger
radii for every mass.  The locus
of maximum masses is given simply by $R\cong1.85
R_s$~(Lattimer et al. \cite{LPMY}) as shown in the right-hand panel of
Fig.~\ref{fig:M-R}.  Strange quark stars with electrostatically
supported normal-matter crusts~(Glendenning \& Weber
\cite{Glendenning92}) have larger 
radii than those with bare surfaces.  Coupled with the additional
constraint $M>1{\rm M}_\odot$ from protoneutron star models, MIT-model
strange quark stars cannot have $R<8.5$ km or $R_\infty<10.5$ km.
These values are comparable to the smallest possible radii for a Bose
(pion or kaon) condensate EOS.

One striking feature of Fig.~\ref{fig:M-R} is that in the mass range
from 1--1.5 M$_\odot$ or more the radius
has relatively little dependence upon the stellar mass.  The major
exceptions illustrated are the model GS1, in which a mixed phase
containing a kaon condensate appears at a relatively low density, and
the model PAL6, which has an extremely small nuclear incompressibility
(120 MeV).  Both of these have considerable softening and a large
increase in central density for $M>1$ M$_\odot$.

While it is generally assumed that a stiff EOS implies both a large
maximum mass and a large radius, many counter examples exist.  For
example, GM3, MS1 and PS have relatively small maximum masses but have
large radii compared to most other EOSs with larger maximum masses.
Also, not all EOSs with extreme softening have small radii for $M>1$
M$_\odot$ (e.g., GS2, PS).  Nonetheless, for stars with masses greater than
1 M$_\odot$, only models with a large degree of softening (including
strange quark matter configurations) can have
$R_\infty<12$ km.

To understand the relative insensitivity of the radius to the mass for
normal neutron stars, it is relevant that a Newtonian polytrope with
$n=1$ has the property that the stellar radius is independent of both
the mass and central density.  An $n=1$ polytrope also has the
property that the radius is proportional to the square root of the
constant $K$ in the polytropic pressure law $P=K\rho^{1+1/n}$.  This
suggests that there might be a quantitative relation between the
radius and the pressure that does not depend upon the EOS at the
highest densities, which determines the overall softness or stiffness
(and hence, the maximum mass).

\begin{figure}[hbt]
\includegraphics[scale=0.52,angle=90]{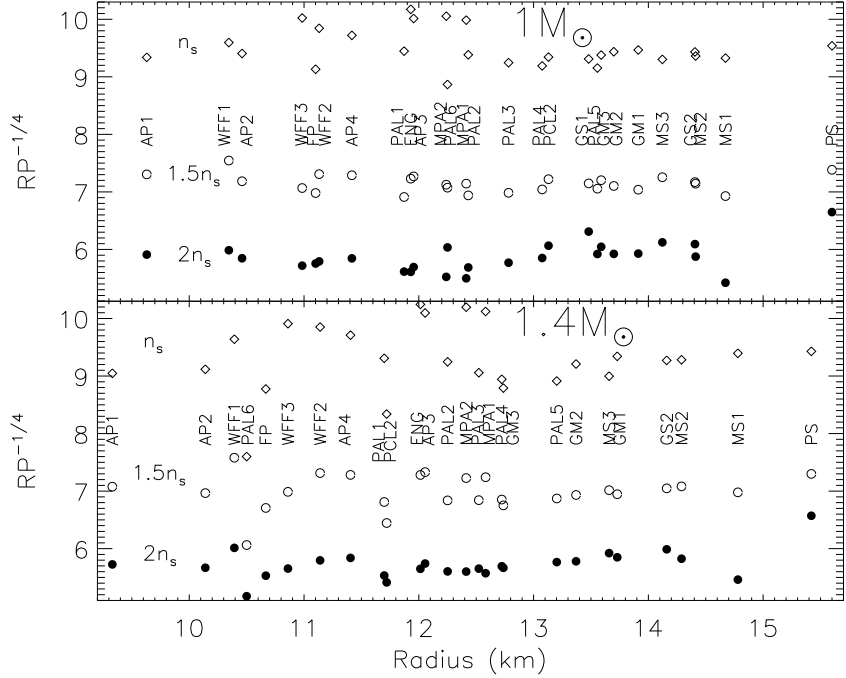} 
\caption{Empirical relation between pressure, in units of MeV fm$^{-3}$,
and $R$, in km, for EOSs listed in Table~\ref{eosname}.  The upper panel shows
results for 1 M$_\odot$ (gravitational mass) stars; the lower panel is
for 1.4 M$_\odot$ stars.  The different symbols show values of
$RP^{-1/4}$ evaluated at three fiducial densities}
\label{fig:P-R}
\end{figure}

In fact, this conjecture may be verified.  Fig.~\ref{fig:P-R} shows
a remarkable empirical correlation between the radii of 1 and 1.4
M$_\odot$ normal stars and the matter's pressure evaluated at fiducial
densities of 1, 1.5 and 2 $n_s$.  Numerically, the correlation has the
form of a power law between the radius $R_M$, defined as the radius at
a particular mass $M$, and the total pressure $P(n)$ evaluated at a
given density:
\begin{equation}
R_M \simeq C(n,M)~[P(n)]^{0.23-0.26}\,.
\label{correl}
\end{equation}
$C(n,M)$ is a number that depends on
the density $n$ at which the pressure was evaluated and the stellar
mass $M$.  An exponent of 1/4 was chosen for display in
Fig.~\ref{fig:P-R}, but the correlation holds for a small range of
exponents about this value.  Using an exponent of 1/4, and ignoring
points associated with EOSs with phase transitions in the density
ranges of interest, we find values for $C(n,M)$, in units of km
fm$^{3/4}$ MeV$^{-1/4}$, which are listed in Table~\ref{cnmtab}.
The error bars 
are taken from the standard deviations.
The correlation is seen to be somewhat tighter for the baryon density
$n=1.5 n_s$ and $2 n_s$ cases.
Despite the relative insensitivity of radius to mass for a particular
EOS in this mass range, the nominal radius $R_M$ has a variation $\sim
5$ km.  The largest deviations from this correlation occur for EOSs
with extreme softening or for configurations very near their maximum
mass.  This correlation is valid only for cold, catalyzed neutron
stars, i.e., not for protoneutron stars which have finite entropies
and might contain trapped neutrinos.  

\begin{table}
\caption{The quantity $C(n,M)$ of Eq.~(\ref{correl})
The quantity $C(n,M)$, in units of km
fm$^{3/4}$ MeV$^{-1/4}$, which relates the pressure (evaluated at density
$n$) to the radius of neutron stars of mass $M$.  The errors are
standard deviations}
\begin{center}
\begin{tabular}{l|l|l}\hline
$n$ & 1 M$_\odot$ & 1.4 M$_\odot$ \\ \hline
$n_s$ & $9.53\pm0.32$ & $9.30\pm0.60$ \\
$1.5n_s$ & $7.14\pm0.15$ & $7.00\pm0.31$ \\
$2n_s$ & $5.82\pm0.21$ & $5.72\pm0.25$ \\ \hline
\end{tabular}
\end{center}
\label{cnmtab}
\end{table}

If a measurement of $P$ near $n_s$ can be deduced in this way, an
important clue about the symmetry properties of matter will be
revealed.  The energy per particle and pressure of cold, beta stable
nucleonic matter is
\begin{eqnarray}
E(n,x) &\simeq& E(n,1/2) + S_v(n)(1-2x)^2   \,, \nonumber \\
P(n,x) &\simeq& n^2[E^\prime(n,1/2)+ S_v^\prime(n) (1-2x)^2] \,,
\label{enuc}
\end{eqnarray}
where $E(n,1/2)$ is the energy per particle of symmetric matter and
$S_v(n)$ is the bulk symmetry energy (which is density dependent).
Primes denote derivatives with respect to density. If only one term in
this expansion is important, as noted by Prakash, Ainsworth \&
Lattimer \cite{PAL}, then
\begin{equation}
S_v(n)\simeq{1\over2}{\partial^2E(n,x)\over\partial x^2}\simeq E(n,0)-E(n,1/2)\,.
\end{equation}
At $n_s$, the symmetry energy can be estimated from nuclear mass
systematics and has the value $S_v\equiv S_v(n_s) \approx 27-36~{\rm
MeV}$.  Attempts to further restrict this range from consideration of
fission barriers and the energies of giant resonances provide
constraints between $S_v$ and $S_v(n)$ primarily by providing
correlations between $S_v$ and $S_s$, the surface symmetry parameter.
Lattimer \& Prakash \cite{Lattimer01} detail how $S_s$ is basically a
volume integral of the quantity $1-S_v/S_v(n)$ through the nucleus.
However, both the magnitude of $S_v$ and its density dependence
$S_v(n)$ remain uncertain.  Part of the bulk symmetry energy is due to the
kinetic energy for noninteracting matter, which for degenerate
nucleonic matter is proportional to $n^{2/3}$, but the remainder of
the symmetry energy, due to interactions, is also expected to
contribute significantly to the overall density dependence.

Leptonic contributions must to be added to Eq.~(\ref{enuc}) to
obtain the total energy and pressure; the electron energy per baryon
is $(3/4)\hbar cx(3\pi^2nx)^{1/3}$.  Matter in neutron stars is in
beta equilibrium, i.e., $\mu_e = \mu_n - \mu_p = - \partial E/\partial
x$, which permits the evaluation of the equilibrium proton fraction.
The pressure at the saturation density becomes
\begin{eqnarray}
P_s=n_s(1-2x_s)[n_sS_v^\prime(1-2x_s)+S_v x_s]\,,
\end{eqnarray}
where $S_v^\prime\equiv S_v^\prime(n_s)$
and the equilibrium proton fraction at $n_s$ is
\begin{eqnarray}
x_s\simeq(3\pi^2 n_s)^{-1}(4S_v/\hbar c)^3 \simeq 0.04\,,
\end{eqnarray}
for $S_v=30$ MeV. Due to the small value of $x_s$, we find that
$P_s\simeq n_s^2 S_v^\prime$.

Were we to evaluate the pressure at a larger density, contributions
featuring other nuclear parameters, including the nuclear
incompressibility $K_s=9(dP/dn)|_{n_s}$ and the skewness
$K_s^\prime=-27n_s^3(d^3E/dn^3)|_{n_s}$, also contribute.  However,
the $K_s$ and $K_s^\prime$ terms largely cancel, up to $2n_s$, so the
symmetry term dominates.

At present, experimental guidance concerning the density dependence of
the symmetry energy is limited and mostly based upon the division of
the nuclear symmetry energy between volume and surface contributions.
Upcoming experiments involving heavy-ion collisions which might sample
densities up to $\sim (3-4)n_s$, will be limited to analyzing
properties of the nearly symmetric nuclear matter EOS through a study
of matter, momentum, and energy flow of nucleons.  However, the
parity-violating experiment \cite{PREX} to accurately determine the
thickness of the neutron skin in $^{208}$Pb at Jefferson Lab will
provide important constraints.  The neutron skin thickness is directly
proportional to $S_s/S_v$.  In addition, studies of heavy nuclei far
off the neutron drip lines using radioactive ion beams might also
provide useful constraints.

\subsection{Moments of Inertia}

Besides the stellar radius, other global attributes of neutron stars
are potentially observable, including the moment of inertia and the
binding energy.  These quantities depend primarily upon the ratio
$M/R$ as opposed to details of the EOS (Lattimer \& Prakash \cite{Lattimer00}).

The moment of inertia, for a star uniformly
rotating with angular velocity $\Omega$, is
\begin{equation}
I=(8\pi/3)\int_0^R r^4(\rho+P/c^2)e^{(\lambda-\nu)/2}
(\omega/\Omega) dr\,.
\label{inertia}
\end{equation}
The metric function $\omega(r)$ is a solution of the equation
\begin{equation}
d[r^4e^{-(\lambda+\nu)/2}\omega^\prime]/dr + 4r^3\omega
de^{-(\lambda+\nu)/2}/dr=0
\label{diffomeg}
\end{equation}
with the surface boundary condition
\begin{equation}\omega_R=\Omega-{R\over3}\omega^\prime_R
=\Omega\left[1-{2GI\over R^3c^2}\right].
\label{boundary}
\end{equation}
The second equality in the above follows from the definition of $I$ and the TOV
equation.  Writing $j=\exp[-(\nu+\lambda)/2]$, the
TOV equation becomes
\begin{equation}
j^\prime=-4\pi Gr(P/c^2+\rho)je^\lambda/c^2\,.
\end{equation}
Then, one has
\begin{equation}
I=-{2c^2\over3G}\int {\omega\over\Omega}r^3dj =
{c^2R^4\omega^\prime_R\over6G\Omega} \,. \end{equation}

\begin{figure}[hbt]
\includegraphics[scale=0.52,angle=90]{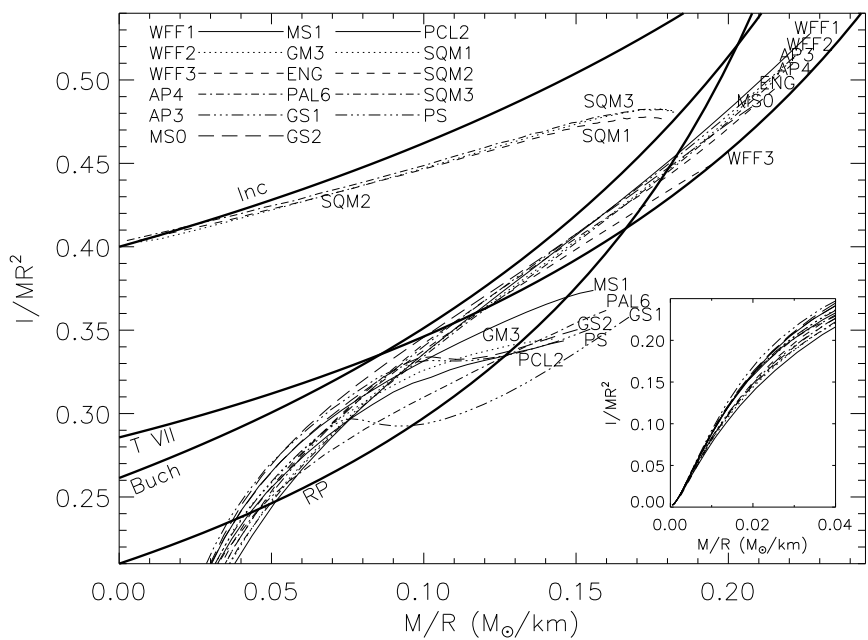} 
\caption{The moment of inertia $I$, in units of $MR^2$, for several
EOSs listed in Table~\ref{eosname}.  The curves labelled ``Inc'', ``T
VII'', ``Buch'' 
and ``RP''
are for an incompressible fluid, the Tolman \cite{Tolman} VII
solution, the Buchdahl \cite{Buchdahl} solution,
and an approximation of Ravenhall \& Pethick \cite{RP},
respectively.  The inset shows details of $I/MR^2$ for $M/R \rightarrow 0$}
\label{mominert}
\end{figure}

Useful approximations which are valid for three analytic, exact,
solutions to GR, the incompressible fluid (Inc), the Tolman VII
(Tolman \cite{Tolman}; VII) solution, and Buchdahl's \cite{Buchdahl} solution (Buch), are
\begin{eqnarray}
I_{Inc}/MR^2 &\simeq&  2(1-0.87\beta-0.3\beta^2)^{-1}/5\,,\label{iinc} \\
I_{Buch}/MR^2 &\simeq&
(2/3-4/\pi^2)(1-1.81\beta+0.47\beta^2)^{-1}\,,\label{ibuc} \\
I_{T VII}/MR^2 &\simeq& 2(1-1.1\beta-0.6\beta^2)^{-1}/7\,.\label{itol}
\end{eqnarray}
Fig.~\ref{mominert} indicates that the T VII
approximation is a rather good approximation to most EOSs without
extreme softening at high densities, for $M/R\ge0.1$ M$_\odot$/km.
The EOSs with softening fall below this trajectory.

Another interesting result from Fig.~\ref{mominert} concerns the
moments of inertia of strange quark matter stars.  Such stars are
relatively closely approximated by incompressible fluids, this
behavior becoming exact in the limit of $\beta\rightarrow0$.  This
could have been anticipated from the $M\propto R^3$ behavior of the
$M-R$ trajectories for small $\beta$ strange quark matter stars as
observed in Fig.~\ref{fig:M-R}.

\subsection{Crustal Fraction of the Moment of Inertia} 

A new observational constraint involving $I$ concerns pulsar
glitches.  Occasionally, the spin rate of a pulsar will suddenly
increase (by about a part in $10^6$) without warning after years of
almost perfectly predictable behavior.  However, Link, Epstein \&
Lattimer \cite{Link} argue that these glitches are not completely random:
the Vela pulsar experiences a sudden spinup about every three years,
before returning to its normal rate of slowing.  Also, the size of a
glitch seems correlated with the interval since the previous glitch,
indicating that they represent self-regulating instabilities for which
the star prepares over a waiting time.  The angular momentum
requirements of glitches in Vela imply that $\ge 1.4$\% of the
star's moment of inertia drives these events.

Glitches are thought to represent angular momentum transfer between
the crust and another component of the star. In this picture, as a
neutron star's crust spins down under magnetic
torque, differential rotation develops between the stellar crust and
this component. The more rapidly rotating component then acts as an
angular momentum reservoir which occasionally exerts a spin-up torque
on the crust as a consequence of an instability.  A popular notion at
present is that the freely spinning component is a superfluid flowing
through a rigid matrix in the thin crust, the region in which
dripped neutrons coexist with nuclei, of the star.  As the solid
portion is slowed by electromagnetic forces, the liquid continues to
rotate at a constant speed, just as superfluid He continues to spin
long after its container has stopped.  This superfluid is usually
assumed to locate in the star's crust, which thus must contain at least
1.4\% of the star's moment of inertia.

The high-density boundary of the crust is naturally set by the phase
boundary between nuclei and uniform matter, where the pressure is
$P_t$ and the density $n_t$.  The low-density boundary is the neutron
drip density, or for all practical purposes, simply the star's surface
since the amount of mass between the neutron drip point and the
surface is negligible.  One can utilize Eq.~(\ref{inertia}) to
determine the moment of inertia of the crust alone with the
assumptions that $P/c^2<<\rho$, $m(r)\simeq M$, and $\omega
j\simeq\omega_R$ and $P\propto\rho^{4/3}$ in the crust (Lattimer \&
Prakash \cite{Lattimer00}:

\begin{equation}{\Delta I\over I}\simeq{28\pi P_t
R^3\over3
Mc^2}{(1-1.67\beta-0.6\beta^2)\over\beta}\left[1+{2P_t(1+5\beta-14\beta^2)\over
n_t
m_bc^2\beta^2}\right]^{-1}.
\label{dii}
\end{equation}

In general, the EOS parameter $P_t$, in the units of MeV fm$^{-3}$,
varies over the range $0.25<P_t<0.65$ for realistic EOSs.  The
determination of this parameter requires a calculation of the
structure of matter containing nuclei just below nuclear matter
density that is consistent with the assumed nuclear matter EOS.
Unfortunately, few such calculations have been performed.  Like the
fiducial pressure at and above nuclear density which appears in
Eq.~(\ref{correl}), $P_t$ should depend sensitively upon the behavior
of the symmetry energy near nuclear density.

\begin{figure}[hbt]
\includegraphics[scale=0.52,angle=90]{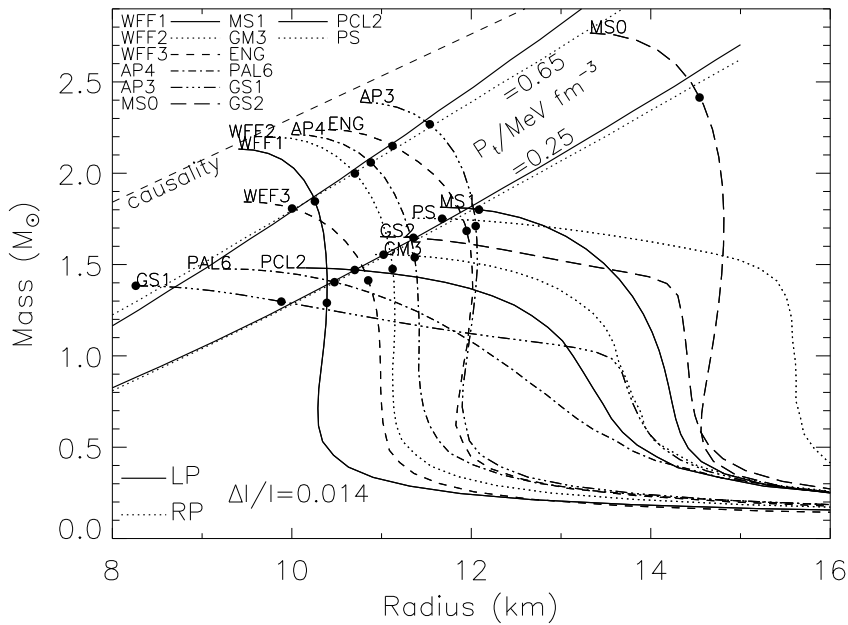}
\caption{Mass-radius curves for selected EOSs from
Table~\ref{eosname}, comparing 
theoretical contours of $\Delta I/I=0.014$ from approximations
developed in this paper, labelled ``LP'', and from Ravenhall \&
Pethick \cite{RP}, labelled ``RP'', to numerical results (solid dots).
Two values of $P_t$, the transition pressure demarking the
crust's inner boundary, which bracket estimates in the literature, are
employed.  The region to the left of the $P_t=0.65$ MeV fm$^{-3}$
curve is forbidden if Vela glitches are due to angular momentum transfers
between the crust and core, as discussed in Link, Epstein \& Lattimer
\cite{Link}.  For comparison, the region excluded by causality alone lies
to the left of the dashed curve labelled ``causality'' as determined
by Lattimer et al. \cite{LPMY} and Glendenning \cite{glen}}
\label{fig:M-R-2}
\end{figure}

Link, Epstein \& Lattimer \cite{Link} established a lower limit to the
radius of the Vela pulsar by using Eq.~(\ref{dii}) with $P_t$ at
its maximum value and the glitch constraint $\Delta I/I\ge0.014$:
\begin{equation}
R>3.9+3.5 M/{\rm M}_\odot-0.08 (M/{\rm M}_\odot)^2{\rm~km}\,.
\label{glitch}
\end{equation}
As shown in Fig.~\ref{fig:M-R-2}, this constraint is somewhat more
stringent than one based upon causality.  Better estimates of the
maximum value of $P_t$ should make this constraint more stringent.

\subsection{Binding Energies}

The binding energy formally represents the energy gained by assembling
$N$ baryons.  If the baryon mass is $m_b$, the binding energy is
simply $BE=Nm_b-M$ in mass units.  However, the quantity $m_b$ has
various interpretations in the literature.  Some authors take it to be
939 MeV/$c^2$, the same as the neutron or proton mass.  Others take it
to be about 930 MeV/$c^2$, corresponding to the mass of C$^{12}$/12 or
Fe$^{56}$/56.  The latter choice would be more appropriate if $BE$ was
to represent the energy released in by the collapse of a
white-dwarf-like iron core in a supernova explosion.  The difference
in these definitions, 10 MeV per baryon, corresponds to a shift of
$10/939\simeq0.01$ in the value of $BE/M$.  This energy, $BE$, can be deduced
from neutrinos detected from a supernova
event; indeed, it might be the most precisely determined aspect of the
neutrino signal.

Lattimer \& Yahil \cite{LY} suggested that the binding energy could be
approximated as
\begin{equation}
BE\approx 1.5\cdot10^{51} (M/{\rm M}_\odot)^2 {\rm~ergs} = 0.084
(M/{\rm M}_\odot)^2 {\rm~M}_\odot\,.
\label{lybind}
\end{equation}
Prakash et al. \cite{prak97a} also concluded that such a formula was a
reasonable approximation, based upon a comparison of selected
non-relativistic potential and field-theoretical
models, good to about $\pm20$ \%.

However, Lattimer \& Prakash \cite{Lattimer00} proposed a more accurate
representation of the binding energy:
\begin{equation}
BE/M \simeq 0.6\beta/(1-0.5\beta)\,, \label{newbind}
\end{equation}
which incorporates some radius dependence.  Thus, the observation of supernova
neutrinos, and the estimate of the total radiated neutrino energy, will yield
more accurate information about $M/R$ than about $M$ alone.

\subsection{Outlook for Radius Determinations} 

Any measurement of a radius will have some intrinsic uncertainty.  In
addition, the empirical relation we have determined between the
pressure and radius has a small uncertainty.  It is useful to display
how accurately the equation of state might be established from an
eventual radius measurement.  This can be done by inverting
Eq.~(\ref{correl}), which yields 
\begin{equation}
P(n) \simeq [R_M/C(n,M)]^4\,.
\label{correli}
\end{equation}
The inferred ranges of pressures, as a function of density and for
three possible values of $R_{1.4}$, are shown in the right panel of 
Fig.~\ref{fig:P-rho}.
It is assumed that the mass is 1.4 M$_\odot$, but the results are
relatively insensitive to the actual mass.  Note from
Table~\ref{cnmtab} that the 
differences between $C$ for 1 and 1.4 M$_\odot$ are typically less than the
errors in $C$ itself.  The light shaded areas show the pressures
including only errors associated with $C$.  The dark shaded areas show
the pressures when a hypothetical observational error of 0.5 km is
also taken into account.  These results suggest that a useful
restriction to the equation of state is possible if the radius of a
neutron star can be measured to an accuracy better than about 1 km.

The reason useful constraints might be obtained from just a single
measurement of a neutron star radius, rather than requiring a series
of simultaneous mass-radii measurements as Lindblom \cite{Lindblom92} proposed,
stems from the fact that we have been able to establish the empirical
correlation, Eq.~(\ref{correl}).  In turn, it appears that this
correlation exists because most equations of state have slopes $d \ln
P/d \ln n\simeq2$ near $n_s$.

\section{Tasks and Prospects}

There are several topics that will merit attention from theoretical
and experimental perspectives.  Among those dealing with $\nu$-matter
interactions are:

\subsection*{ Dynamic Structure Functions from Microscopic
Calculations}

Neutrino properties of astrophysical interest depend
crucially on the nature of the excitation spectrum of the nuclear
medium to spin and spin-isospin probes. The excited states
are very different depending upon whether or not interactions in the
medium conserve spin and spin-isospin. The importance of tensor
correlations in the medium is thus clear, since they break the
conservation laws.  Friman \& Maxwell \cite{frim79} 
 first emphasized the
importance of tensor correlations in the process
\begin{equation}
\nu_e + n + n \rightarrow e^- + p + n\,,
\end{equation}
and noted that their neglect underestimates the
rate of $\nu_e$ absorption by as much as an order of magnitude. In
their study, they used a hard core description of the short range
correlations and a one pion exchange model for the medium and long
range ones. Sawyer \& Soni \cite{sawy79}  
and Haensel \& Jerzak \cite{haen87}, who
used additional correlations based on a Reid soft core potential,
confirmed that large reductions were possible in degenerate
matter for non-degenerate neutrinos.

Since these earlier works, many-body calculations have vastly improved (e.g. 
\cite{Akmal}) 
and have been well-tested against data on light nuclei and nuclear
matter.  Much better tensor correlations are now available, so that we
may better pin down the rate of absorption due to the above process.
Detailed calculations to include
arbitrary matter and neutrino degeneracies encountered in 
 in many astrophysical applications are necessary. 

\subsection*{ Axial Charge Renormalization}

 In  dense matter,
the axial charge of the baryons is renormalized 
\cite{wilk73,rho74,brow91}, 
which alters the
neutrino-baryon couplings from their vacuum values. Since the
axial contribution to the scattering and absorption reactions is typically
three times larger than the vector contributions, small changes in the
axial vector coupling constants significantly affect the cross sections. 
The calculation of this renormalization requires 
a theoretical
approach which treats the pion and chiral symmetry breaking
explicitly.  So far, this has been done in isospin symmetric nuclear
matter \cite{cart96}, but not for neutron matter or
for beta-equilibrated neutron star matter. Substantial
reductions may be expected 
in the $\nu$-matter cross sections from this in-medium
effect. \\

\subsection*{ Multi-Pair Excitations}

Neutrinos can also excite many-particle states in an interacting
system, inverse bremsstrahlung being an example of a two-particle
excitation \cite{hann98}.  These excitations provide an efficient
means of transferring energy between the neutrinos and baryons which are
potentially significant in low-density matter.  However, multigroup neutrino
transport will be needed to  include this effect.  In addition, such
calculations require source terms for neutrino processes such as
bremsstrahlung and neutrino pair production.
The latter process has been accurately treated in \cite{pons98}. 

\subsection*{ $\pi^-$ and $K^-$ Dispersion Relations Through $\nu$-Nucleus
Reactions}

 The experimental program that would do the most to
illuminate theoretical issues permeating neutrino interactions in
dense matter would be studies of neutrino reactions on heavy nuclei,
the only direct way of probing the matrix elements of the axial
current in nuclear matter. Pioneering suggestions in this regard have
been put forth by Sawyer \& Soni \cite{sawy77,sawy78}, Ericson
\cite{eric90}, and Sawyer \cite{sawy94}. The basic idea is to detect
positively charged leptons ($\mu^+~or~e^+$) produced in inclusive
experiments
\begin{eqnarray}
\bar \nu + X &\rightarrow & \mu^+~(\rm~{or}~e^+)~ + \pi^- 
          ~(\rm~{or}~K^-) + X 
\end{eqnarray}
which is kinematically made possible when the in-medium $\pi^-$ or
$K^-$ dispersion relation finds support in space-like regions.  The
sharp peaks at forward angles in the differential cross section versus
lepton momentum survive the 100-200 MeV width in the incoming GeV or
so neutrinos from accelerator experiments.  Calculations
of the background from quasi-elastic reactions
indicate that the signal would be easily detectable.

\subsection*{ $\nu$-Matter Interactions at Sub-Nuclear Density}

Analgous to the effects of inhomogeneities for $\nu$-matter
interactions discussed in the case of a first-order kaon condensate or
quark matter transition is the case of coherent scattering of
neutrinos from closely-spaced nuclei at sub-nuclear densities.  The
sizes and separations of nuclei are similar to those of the droplets
discussed for the kaon and quark situations, so the range of neutrino
energies most affected will be similar.  These will be important in
reshaping the $\nu$ spectrum from PNSs, and are of potential
importance in the supernova mechanism itself due to the large energy
dependence of $\nu$-matter cross sections behind the shock.

In addition to these, several topics of interest from an astrophysical
perspective include:

\subsection*{ Improvements in PNS Simulations}

These include a) an adequate treatment of convection coupled with
neutrino transport appear to be necessary based upon large regions
that are potentially convectively unstable; b) the consideration of
other softening components in dense matter that might produce effects
dissimilar to those found when considering hyperons and kaons, i.e.,
quarks; c) improved transport calculations with many energy groups,
especially in the transparent regime; d) a self-consistent treatment
of accretion, which is known to significantly contribute to the early
$\nu$ emission.  The latter two items necessitate the coupling of 
a multi-group transport scheme with a hydrodynamical code of the type
generally used for supernova simulations.  

\subsection*{ Determination of the Neutron Skin of Neutron-rich Nuclei}

The Jefferson Lab experiment \cite{PREX} (PREX) is anticipated to
yield accurate measurements of the neutron-skin thickness of $^{208}$Pb.
This quantity, from a theoretical viewpoint, is the volume integral of
the inverse of the symmetry energy throughout a nucleus, and
represents how the nuclear symmetry energy is split between volume and
surface contributions.  Not coincidentally, the density dependence of
the symmetry energy is also implicated in the predicted neutron star
radius \cite{Lattimer00}.

\subsection*{ Determination of the Radius of a Neutron Star}

The best prospect for measuring a neutron star's radius may be the
nearby object RX J185635-3754.  Parallax information \cite{Walter01}
indicates its distance to be about 60 pc.  In addition, it may be
possible to identify spectral lines with the Chandra and XMM X-ray
facilities that would not only yield the gravitational redshift, but
would identify the atmospheric composition.  Not only would this
additional information reduce the uncertainty in the deduced value of
$R_\infty$, but, {\em both} the mass and radius for this object might
thereby be estimated.  It is also possible that an estimate of the
surface gravity of the star can be found from further comparisons of
observations with atmospheric modelling, and this would provide a
further check on the mass and radius.

\section*{Acknowledgements}

Research support from NSF grant INT-9802680 (for MP and JML) and DOE
grants FG02-88ER-40388 (for MP and AS) and FG02-87ER40317 (for JML and
JAP) and FG06-90ER40561 (for SR) is gratefully acknowledged. It is a
pleasure to acknowledge collaborations with 
George Bertsch, Greg Carter, Paul Ellis, Juan
Miralles, and Dany Page who 
have contributed significantly to the material presented in this
article. 


%


\begin{thebibliography}{8.}

\bibitem{burr86}A. Burrows, J.M. Lattimer:
Astrophys. J. \textbf{307}, 178 (1986) 

\bibitem{burr90}A. Burrows: Ann. Rev. Nucl. Sci. \textbf{40}, 181 (1990) 

\bibitem{burr88}A. Burrows: Astrophys. J. \textbf{334}, 891 (1988)

\bibitem{prak97a}M. Prakash, I. Bombaci, Manju Prakash, P.J. Ellis,
J.M. Lattimer, R. Knorren: Phys. Rep. \textbf{280}, 1 (1997)
P.J. Ellis, J.M. Lattimer, M. Prakash: Comm. Nucl. Part. Phys. 
\textbf{22}, 63 (1996)

\bibitem{Lattimer00}J.M. Lattimer, M. Prakash: Astrophys. J. {\it in press}
(2000)

\bibitem{AIP}{\it Next Generation Nucleon Decay and Neutrino
Dectector}, ed. by N. Diwan, C.K. Jung (AIP, New York 2000)

\bibitem{LvRPP94}
J.M. Lattimer, K.A. van Riper, M. Prakash, Manju Prakash: 
Astrophys. J. {\bf 425}, 802 (1994)

\bibitem{lind66}R.W. Lindquist: Ann. Phys. \textbf{37}, 478 (1966)

\bibitem{thor81}K.S. Thorne: Mon. Not. R. Astron. Soc. \textbf{194},
439 (1981)

\bibitem{pons99}J.A. Pons, S. Reddy, M. Prakash, J.M. Lattimer,
J.A. Miralles: Astrophys. J. \textbf{513}, 780 (1999)

\bibitem{Akmal} A. Akmal, V.R. Pandharipande: Phys. Rev. C {\bf 56},
2261 (1997)

\bibitem{sero92}
B.D. Serot, J.D. Walecka: In {\it Advances in Nuclear Physics}
Vol. 16, ed. by J.W. Negele, E. Vogt, (Plenum, New York 1986) p. 1

\bibitem{glen91}N.K. Glendenning, S. Moszkowski: Phys. Rev. Lett.
      \textbf{67}, 2414 (1991)


\bibitem{Pon00b} 
J.A. Pons, S. Reddy, P.J. Ellis, M. Prakash, 
J.M. Lattimer: Phys. Rev. C \textbf{62} 035803 (2000)

\bibitem{GS99} 
N.K. Glendenning, J. Schaffner-Bielich: 
Phys. Rev. C \textbf{60} 025803 (1999)

\bibitem{Gibbs} J.W. Gibbs: Tran. Conn. Acad. \textbf{III}, 108 (1876)

\bibitem{FGB94} 
E. Friedman, A. Gal, C.J. Batty: Nucl. Phys. A \textbf{579}, 578 (1994)

\bibitem{Fri99} 
E. Friedman, A. Gal, J. Mares, A. Cieply: Phys. Rev. C \textbf{60}, 
024314 (1999)

\bibitem{WW97} 
T. Waas, W. Weise: Nucl. Phys. A \textbf{625}, 287 (1997)

\bibitem{RO00} 
A. Ramos, E. Oset: Nucl. Phys. A \textbf{671}, 481 (2000)

\bibitem{BGN00} 
A. Baca, C. Garcia-Recio, J. Nieves: Nucl. Phys. A \textbf{673}, 335 (2000)

\bibitem{SPL00} A.W. Steiner, M. Prakash, J.M. Lattimer:
Phys. Lett. B \textbf{486} 239 (2000) 

\bibitem{PBP}
Manju Prakash, E. Baron, M. Prakash:
Phys. Lett. B, \textbf{243}, 175 (1990)

\bibitem{NJL}
Y. Nambu, G. Jona-Lasinio:
Phys. Rev. \textbf{122}, 345 (1961)

\bibitem{tHooft}
G. 't~Hooft:
Phys. Rep. \textbf{142}, 357 (1986)

\bibitem{Rehberg}
P. Rehberg, S.P. Klevansky, J. H\"ufner:
Phys. Rev. C \textbf{53}, 410 (1996)

\bibitem{parms2}
T. Kunihiro:
Phys. Lett. B \textbf{219}, 363 (1989)

\bibitem{Hatsuda}
T. Hatsuda, T. Kunihiro:
Phys. Rep. \textbf{247}, 221 (1994)

\bibitem{Buballa}
M. Buballa, M. Oertel:
Phys. Lett. B \textbf{457}, 261 (1999)

\bibitem{MS} H. M\"uller, B.D. Serot: Nucl. Phys. A {\bf 606}
(1996) 508

\bibitem{RBP}
S. Reddy, G.F. Bertsch, M. Prakash:
Phys. Lett. B \textbf{475}, 1 (2000)

\bibitem{glen1} N.K. Glendenning:
Phys. Rev. D \textbf{46}, 1274 (1992)

\bibitem{LR78}J.M. Lattimer, D.G. Ravenhall:
Astrophys. J. {\bf 223}, 314 (1978)

\bibitem{GS98} N.K. Glendenning, J. Schaffner-Bielich: 
Phys. Rev. Lett. \textbf{81}, 4564 (1998)


\bibitem{size} M. Christiansen, N.K. Glendenning, J. Schaffner-Bielich:
Phys. Rev. C \textbf{62}, 025804 (2000) T. Norsen, S. Reddy: 
nucl-th/0010075 (2000)

\bibitem{GP} N.K. Glendenning, S. Pei: 
Phys. Rev. C \textbf{52}, 2250 (1995)

\bibitem{PCL} M. Prakash, J.R. Cooke, J.M. Lattimer:
Phys. Rev. D \textbf{52}, 661 (1995)

\bibitem{brue85}S.W. Bruenn: Astrophys. J. Supp. \textbf{58}, 771 (1985)

\bibitem{mezz93}A. Mezzacappa, S.W. Bruenn:
Astrophys. J. \textbf{405}, 637 (1993)

\bibitem{wils89}J.R. Wilson, R.W. Mayle: In {\it The Nuclear
Equation of State} Part A, ed. W. Greiner, H. St\"ocker (Plenum, New
York 1989) p. 731

\bibitem{suzu92}H. Suzuki, K. Sato: In {\it The Structure and
Evolution of Neutron Stars}, ed. D. Pines, R. Tamagaki, S. Tsuruta
(Addison-Wesley, New York 1992) p. 276

\bibitem{keil95a}W. Keil, H.T. Janka: Astron. \&
Astrophys. \textbf{296}, 145 (1995)

\bibitem{redd97a}S. Reddy, J. Pons, M. Prakash, M., J.M. Lattimer:
In {\it Stellar Evolution, Stellar Explosions and Galactic
Chemical Evolution}, ed. T. Mezzacappa (IOP Publishing, Bristol 1997)
p. 585

\bibitem{redd97b}S. Reddy, M. Prakash: Astrophys. J. \textbf{423}, 
689 (1997)

\bibitem{prak97b}M. Prakash, S. Reddy: In {\it Nuclear
Astrophysics}, ed. M. Buballa, W. N\"orenberg, J. Wambach, A. Wirzba
(GSI: Darmstadt 1997), p. 187

\bibitem{redd98}S. Reddy, M. Prakash, J.M. Lattimer:
Phys. Rev. D \textbf{58}, 013009 (1998)

\bibitem{redd99a}S. Reddy, M. Prakash, J.M. Lattimer, J.A. Pons:
Phys. Rev. C \textbf{59}, 2888 (1999)

\bibitem{burr98}A. Burrows, R.F. Sawyer: Phys. Rev. C \textbf{58},
554 (1998)

\bibitem{burr99a}A. Burrows, R.F. Sawyer: Phys. Rev. C \textbf{59},
510 (1999)

\bibitem{sawy75}R.F. Sawyer: Phys. Rev. D \textbf{11}, 2740 (1975)

\bibitem{sawy89}R.F. Sawyer: Phys. Rev. C \textbf{40}, 865 (1989)

\bibitem{iwam82}N. Iwamoto, C.J. Pethick: Phys. Rev. D \textbf{25},
313 (1982)

\bibitem{horo91} C.J. Horowitz, K. Wehrberger: Nucl. Phys. A
\textbf{531}, 665 (1991)
Phys. Rev. Lett. \textbf{66}, 272  (1991)
Phys. Lett. B \textbf{226}, 236 (1992)

\bibitem{raff95}G. Raffelt, D. Seckel: Phys. Rev. D \textbf{52}, 1780 (1995)

\bibitem{sigl96}G. Sigl: Phys. Rev. Lett \textbf{76}, 2625 (1996)

\bibitem{FW} A.L. Fetter, J.D. Walecka: In {\it Quantum Theory of Many Particle
Systems} (McGraw-Hill, New York 1971)

\bibitem{DS} S. Doniach, E.H. Sondheimer: In {\it Green's Functions for Solid
State Physicists} (The Benjamin/Cummings Publishing Company, Inc.,
Reading 1974)

\bibitem{LPPH91}
J.M. Lattimer, C.J. Pethick, M. Prakash, P. Haensel:
Phys. Rev. Lett. \textbf{66}, 2701 (1991)

\bibitem{CJH}C.J. Horowitz: Phys. Rev. D \textbf{55}, 4577 (1997)

\bibitem{IP} N. Iwamoto, C.J. Pethick:
Phys. Rev. D \textbf{25}, 313 (1982)

\bibitem{gap}
B.C. Barrois: Nucl. Phys. B \textbf{129}, 390 (1977)
S.C. Frautschi: In {\it Workshop on Hadronic Matter at Extreme 
Energy Density} (Erice, Italy 1978)

\bibitem{qsf0}
M. Alford, K. Rajagopal, F. Wilczek:
Phys. Lett. B \textbf{422}, 247 (1998)
Nucl. Phys. B \textbf{357}, 443 (1999)
{\it ibid.} \textbf{558}, 219 (1999) 
R. Rapp, T. Sch\"{a}ffer, E.V. Shuryak, M. Velkovsky:
Phys. Rev. Lett. \textbf{81}, 53 (1998)
Ann. Phys. \textbf{280}, 35 (2000)

\bibitem{Bailin84}
D. Bailin, A. Love:
Phys. Rept. {\bf 107}, 325 (1984)

\bibitem{CR00}  
G. W. Carter, S. Reddy: Phys. Rev. D \textbf{62} 103002 (2000)

\bibitem{BCS}
J. Bardeen, L.N. Cooper, J.R. Schrieffer: Phys. Rev. {\bf 108},
1175 (1957)

\bibitem{tc}
R.D. Pisarski, D.H. Rischke:
Phys. Rev. D {\bf 61}, 051501 (2000)

\bibitem{propagators}
R.D. Pisarski, D.H. Rischke:
Phys. Rev. D {\bf 60}, 094013 (1999)

\bibitem{burr93c}
A. Burrows, B. Fryxell: Astrophys. J. Lett. \textbf{413}, L33 (1993)

\bibitem{hera94}
M. Herant, W. Benz, J. Hicks, C. Fryer, S.A. Colgate:
Astrophys. J. \textbf{425}, 339 (1994)

\bibitem{keil95b}W. Keil, T.H. Janka, E. M\"uller:
Astrophys. J. Lett. \textbf{473}, L111 (1995)

\bibitem{mezz98}
A. Mezzacappa, A.C. Calder, S.W. Bruenn, J.M. Blondin, M.W. Guidry,
M.R. Strayer, A.S. Umar:
Astrophys. J. \textbf{495}, 911 (1998)

\bibitem{mpu00}
J.A. Miralles, J.A. Pons, V.A. Urpin: 
Astrophys. J. \textbf{543}, 1001 (2000) 


\bibitem{Pon00a} 
J.A. Pons, J.A. Miralles, M. Prakash, J.M. Lattimer:
Astrophys. J. (2000) submitted; astro-ph/0008389

\bibitem{P98a}
S. Tsuruta: Phys. Rep. {\bf 292}, 1 (1998)
D.Page:
in {\it The Many Faces of Neutron Stars},
ed. by R. Bucheri, J. van Paradijs, M.A. Alpar
(Kluwer Academic Publishers, Dordrecht, 1998) p. 539

\bibitem{PA92}
D. Page, J.H. Applegate:
Astrophys. J. {\bf 394}, L17 (1992)

\bibitem{note} The relation between $T_e$ and $T$ is increased by
about 50\% if H or He dominates the atmosphere's composition.  Our qualitative
results concerning superfluidity, however, will be unaffected by the
atmospheric composition.

\bibitem{PPLP92}
M. Prakash, Manju Prakash, J.M. Lattimer, C.J. Pethick:
Astrophys. J. {\bf 390}, L77 (1992)

\bibitem{crevs}
C.J. Pethick: Rev. Mod. Phys. {\bf 64}, 1133 (1992)
M. Prakash: Phys. Rep. {\bf 242}, 297 (1994)
B. Friman, O.V. Maxwell: Astrophys. J. {\bf 232}, 541 (1979)
N. Iwamoto: Phys. Rev. Lett. {\bf 44}, 1637 (1980)

\bibitem{BEEHJS98}
M. Baldo, O. Elgaroy, L. Engvik, M. Hjorth-Jensen, H.-J. Schulze:
Phys. Rev. C {\bf 58}, 1921 (1998)

\bibitem{BB97}
S. Balberg, N. Barnea:
Phys. Rev. C {\bf 57}, 409 (1997)

\bibitem{qsf}
D.T. Son:
Phys. Rev. D {\bf 59}, 094019 (1999)
R. Pisarski, D. Rischke:
Phys. Rev. D {\bf 61}, 051501 (2000)
{\it ibid.} {\bf 61}, 074017 (2000)
J. Berges, K. Rajagopal:
Nucl. Phys. B {\bf 538}, 215 (1999)
M. Alford, J. Berges, K. Rajagopal:
Phys. Rev. D {\bf D60}, 074014 (1999)
T. Sch\"{a}ffer:
Nucl. Phys. A {\bf 642}, 45 (1998)
T. Sch\"{a}ffer, F. Wilczek:
Phys. Rev. Lett. {\bf 82}, 3956 (1999)
Phys. Rev. D {\bf 60}, 074014 (1999)
G.W. Carter, D. Diakonov:
Phys. Rev. D {\bf 60}, 16004 (1999)

\bibitem{B99}
P.F. Bedaque: hep-ph/9910247

\bibitem{Blaschke00}
D. Blaschke, T. Klahn, D.N. Voskresensky: Astrophys. J. 
\textbf{533}, 406 (2000)

\bibitem{PPLS}
D. Page, M. Prakash, J.M. Lattimer, A.W. Steiner: Phys. Rev. Lett.
\textbf{85}, 2048 (2000)

\bibitem{LY9496}
K.P. Levenfish, D.G. Yakovlev:
Astronomy Rep. {\bf 38}, 247 (1994)
Astronomy Lett. {\bf 20}, 43 (1994)
Astronomy Lett. {\bf 22}, 49 (1996)

\bibitem{FRS}
E. Flowers, M. Ruderman, P. Sutherland:
Astrophys. J. {\bf 205}, 541 (1976)

\bibitem{ZM90}
J. Zimanyi, S.A. Moszkowski:
Phys. Rev. C {\bf 42}, 416 (1990)

\bibitem{ARSW96}
T. Alm, G. R\"{o}pke, A. Sedrakian, F. Weber:
Nucl. Phys. A {\bf 604}, 491 (1996)

\bibitem{P97}
D. Page: Astrophys. J. Lett. {\bf 479}, L43 (1997)

\bibitem{APR98}
A. Akmal, V.R. Pandharipande, D.G. Ravenhall:
Phys. Rev. C {\bf 58}, 1804 (1998)

\bibitem{P98b}
D. Page:
in {\it Neutron Stars and Pulsars: Thirty Years after the Discovery},
ed. by N. Shibazaki, et al.
(Universal Academy Press, Tokyo 1998) p. 183

\bibitem{SBSB98}
C. Schaab, S. Balberg, J. Schaffner-Bielich:
Astrophys. J. Lett. {\bf 504}, L99 (1998)
\bibitem{LPMY} J.M. Lattimer, M. Prakash, D. Masak, A. Yahil:
Astrophys. J. {\bf 355} (1990) 241
\bibitem{Link} B. Link, R.I. Epstein, J.M. Lattimer:
Phys. Rev. Lett. {\bf 83} 3362 (1999)

\bibitem{PAL} M. Prakash, T.L. Ainsworth, J.M. Lattimer:
Phys. Rev. Lett. {\bf 61} 2518 (1988)

\bibitem{FP} B. Friedman, V.R. Pandharipande: Nucl. Phys. A
{\bf 361} 502 (1981)

\bibitem{PS} V.R. Pandharipande, R.A. Smith: Nucl. Phys. A {\bf 237} 
507 (1975)

\bibitem{WFF} R.B. Wiringa, V. Fiks, A. Fabrocine: Phys. Rev. C
{\bf 38} 1010 (1988)  

\bibitem{MPA} H. M\"uther, M. Prakash, T.L. Ainsworth: Phys. Lett. 
{\bf 199} 469 (1987)  

\bibitem{Engvik} L. Engvik, M. Hjorth-Jensen, E. Osnes, G. Bao, 
E. \O stgaard: Phys. Rev.  Lett. {\bf 73} 2650 (1994)

\bibitem{Witten84}E. Witten: Phys. Rev. \textbf{D30} 272 (1984)

\bibitem{Fahri84}E. Fahri, R. Jaffe: Phys. Rev. \textbf{D30} 2379 (1984)

\bibitem{Haensel86}P. Haensel, J.L. Zdunik, R. Schaefer: Astron. \&
Astrophys. \textbf{217} 137 (1986)

\bibitem{Alcock88}C. Alcock, A. Olinto: Ann. Rev. Nucl. Sci. \textbf{38}
161 (1988)

\bibitem{Prakash90}Manju Prakash, E. Baron, M. Prakash: Phys. Lett. B 
\textbf{243} 175 (1990)

\bibitem{Rhoades74}C.E. Rhoades, R. Ruffini: Phys. Rev. Lett. 
\textbf{32} 324 (1974)

\bibitem{Glendenning92}N.K. Glendenning, F. Weber:
Astrophys. J. \textbf{400} 672 (1992)

\bibitem{Lattimer01}J.M. Lattimer, M. Prakash: {\it to be published}

\bibitem{PREX}C.J. Horowitz, S.J. Pollock, P.A. Souder, R. Michaels: 
nucl-th/9912038

\bibitem{Tolman} R.C. Tolman: Phys. Rev. {\bf 55} 364 (1939)  

\bibitem{Buchdahl} H.A. Buchdahl: Astrophys. J. {\bf 147} 310 (1967)

\bibitem{RP} D.G. Ravenhall, C.J. Pethick: Astrophys. J. {\bf
424} 846 (1994)

\bibitem{glen} N.K. Glendenning: Phys. Rev. D {\bf 46} 4161 (1992) 

\bibitem{LY} J.M. Lattimer, A. Yahil: Astrophys. J. {\bf 340} 426 (1989)

\bibitem{Lindblom92}L. Lindblom: Astrophys. J. \textbf{398} 569 (1992)


\bibitem{frim79}B. Friman, O. Maxwell: Astrophys. J. 232, 541 (1979)

\bibitem{sawy79}R.F. Sawyer, A. Soni: Astrophys. J. \textbf{230}, 859 (1979)

\bibitem{haen87}P. Haensel, A.J. Jerzak: Astron. \& Astrophys. \textbf{179}, 127 (1987)

\bibitem{wilk73}D.H. Wilkinson: Phys. Rev. C \textbf{7}, 930 (1973)

\bibitem{rho74}M. Rho: Nucl. Phys. A \textbf{231}, 493 (1974)

\bibitem{brow91}G.E. Brown, M. Rho: Phys. Rev. Lett. \textbf{66}, 2720 (1991)

\bibitem{cart96}G. Carter, P.J. Ellis, S. Rudaz:
Nucl. Phys. A \textbf{603}, 367 (1996)

\bibitem{hann98}S. Hannestad, G. Raffelt:
Astrophys. J. \textbf{507}, 339 (1998)

\bibitem{pons98}J.A. Pons, J.A. Miralles, J.-M. Iba\~nez:
Astron. \& Astrophys. Supp. \textbf{129}, 343 (1998)

\bibitem{sawy77}R.F. Sawyer, A. Soni: Phys. Rev. Lett. \textbf{38}, 1383 (1977)

\bibitem{sawy78}R.F. Sawyer, A. Soni: Phys. Rev. C \textbf{18}, 898 (1978)

\bibitem{eric90}M. Ericson: Nucl. Phys. A \textbf{518}, 116 (1990)

\bibitem{sawy94}R.F. Sawyer: Phys. Rev. Lett. \textbf{73}, 3363 (1994)

\bibitem{Walter01}F.M. Walter: Astrophys. J. {\it in press} 
(2001)


\end{thebibliography}
\end{document}